\documentclass[lettersize,journal]{IEEEtran}
\DeclareUnicodeCharacter{FB01}{fi}
\DeclareUnicodeCharacter{FB02}{fl}
\usepackage{fancyhdr}
\usepackage{epsfig}
\usepackage{threeparttable}
\usepackage{epsf,epsfig}
\usepackage{amsthm}
\usepackage[cmex10]{amsmath}
\usepackage{amssymb, amsfonts}
\usepackage[noadjust]{cite}
\usepackage{dsfont}
\usepackage{subfigure}
\usepackage{color}
\usepackage{soul}
\usepackage{amssymb}
\usepackage{accents}
\usepackage{algorithm}
\usepackage{diagbox}
\usepackage[english]{babel}
\usepackage{url}
\usepackage{framed}
\usepackage{multirow}
\usepackage{graphicx}
\usepackage{indentfirst}
\usepackage{url}
\usepackage{framed}
\usepackage{booktabs}
\usepackage[colorlinks,
linkcolor=black,
anchorcolor=black,
citecolor=black]{hyperref}

\newtheorem{remark}{Remark}
\def\phi{\varphi}

\def\({\left(}
\def\){\right)}

\def\b0{{\mathbf{0}}}

\usepackage{xcolor}
\usepackage{hyperref}
\hypersetup{
    colorlinks=true,
    linkcolor=blue,
    filecolor=blue,      
    urlcolor=blue,
    citecolor=blue,
}
\definecolor{mygold}{RGB}{247,173,69}

\allowdisplaybreaks[2]

\def\bf#1{\mathbf{#1}}

\def\rv#1{\textcolor{black}{#1}}

\usepackage{mathrsfs}
\usepackage{algorithm}
\usepackage{algorithmic}
\usepackage[justification=centering]{caption}
\begin{document}	
	\pagestyle{empty}
	\title{\huge Semantics-Guided Diffusion for Deep Joint Source-Channel Coding in Wireless Image Transmission}
	\author{Maojun Zhang, {\it Student Member, IEEE}, Haotian Wu, {\it Member, IEEE}, Guangxu Zhu, {\it Member, IEEE}, \\Richeng Jin, {\it Member, IEEE}, Xiaoming Chen, {\it Senior Member, IEEE}, Deniz Gündüz, {\it Fellow, IEEE}
	\thanks{

	M. Zhang, and X. Chen are with the College of Information Science and Electronic Engineering, Zhejiang University, Hangzhou, China (Email: $\{$zhmj, chen\_xiaoming$\}$@zju.edu.cn.). 
	H. Wu and D. Gündüz is with the Department of Electrical and Electronic Engineering, Imperial College London (Email: $\{$haotian.wu17, d.gunduz$\}$@imperial.ac.uk). 
	G. Zhu is with Shenzhen Research Institute of Big Data, Shenzhen, China (Email: gxzhu@sribd.cn). 
	R. Jin are with the Department of Information and Communication Engineering, Zhejiang University, and also with Zhejiang Provincial Key Laboratory of Multi-Modal Communication Networks and Intelligent Information Processing, Hangzhou, China, 310007 (e-mail: richengjin@zju.edu.cn). 
	This work was carried out when M. Zhang was a visiting student at the Information Processing Laboratory (IPC Lab) at Imperial College London.

	}	
	\vspace{-6mm}
	}
	\maketitle
	\thispagestyle{empty}
	\vspace{-20mm}
	\begin{abstract}
 Joint source-channel coding (JSCC)
 offers a promising avenue for enhancing transmission efficiency by jointly incorporating source and channel statistics into the system design. {A key advancement in this area is the deep joint source and channel coding (DeepJSCC) technique that designs a direct mapping of input signals to channel symbols parameterized by a neural network, which can be trained for arbitrary channel models and semantic quality metrics.}  
{This paper advances the DeepJSCC framework toward a semantics-aligned, high-fidelity transmission approach,} 
called semantics-guided diffusion DeepJSCC (SGD-JSCC). 
Existing schemes that integrate diffusion models (DMs) with JSCC face challenges in transforming random generation into accurate reconstruction and adapting to varying channel conditions.  
{SGD-JSCC incorporates two key innovations: 
(1) {utilizing some inherent information that contributes to the semantics of an image, such as text description or edge map, to guide the diffusion denoising process;} 
and (2) enabling seamless adaptability to varying channel conditions with the help of a semantics-guided DM for channel denoising. 
The DM is guided by diverse semantic information and integrates seamlessly with DeepJSCC. 
In a slow fading channel, 
SGD-JSCC dynamically adapts to the instantaneous channel state information (CSI) directly estimated from the channel output, thereby eliminating the need for additional pilot transmissions for channel estimation. 
In a fast fading channel, we introduce a training-free denoising strategy, allowing SGD-JSCC to effectively adjust to fluctuations in channel gains.} 
   {Numerical results demonstrate that,} guided by semantic information and leveraging the powerful DM, our method outperforms existing DeepJSCC schemes, delivering satisfactory reconstruction performance even at extremely poor channel conditions. 
   The proposed scheme highlights the potential of incorporating diffusion models in future communication systems. 
	{The code and pretrained checkpoints will be publicly available at \emph{\url{https://github.com/MauroZMJ/SGDJSCC}}, allowing integration of this scheme with existing DeepJSCC models, without the need for retraining from scratch.}
	\end{abstract}
	
	\vspace{-2mm}
	
	\section{Introduction}

{Over the past decades, wireless communication has undergone significant evolution, progressing from 1G to 5G. Advanced coding techniques, such as polar codes and low-density parity-check (LDPC) codes, have pushed performance closer to theoretical limits. However,  rapidly growing communication demands, driven by applications like autonomous driving and advanced artificial intelligence, pose a risk of saturating network capacity. Addressing this challenge requires a shift from viewing communication systems as passive bit-pipes to developing semantic-aware communication frameworks with intrinsic intelligence. This paradigm shift aims to bridge the gap between escalating demands and the theoretical performance bottleneck of conventional communication systems, leading researchers to explore new paradigms at the semantic level, known as semantic communication (SemCom) \cite{gunduz2022beyond}.}  
Emerging SemCom systems leverage neural networks to extract and utilize semantic information, which essentially refers to the information that is most relevant for the task desired to be carried out by the receiver. 
{Shifting the transmission objective from bit-level accuracy to semantic level, or end-to-end accuracy, allows SemCom to outperform traditional separation-based approaches.
} 

\subsection{{Deep Joint Source and Channel Coding (DeepJSCC)}}\label{subsec: deepjscc}
 {
{One of the most promising advancements for SemCom is the DeepJSCC approach \cite{bourtsoulatze2019deep}, which combines source compression and error correction into a unified encoder parameterized by a neural network. Unlike separate source and channel coding, DeepJSCC allows end-to-end optimization and promises significant improvements in 
 the practical finite-blocklength regime \cite{gunduz2024joint}.} 
DeepJSCC for wireless image transmission was initially proposed in \cite{bourtsoulatze2019deep}, where a convolutional neural network (CNN)-based JSCC architecture is proposed, outperforming standard separation-based schemes over additive white Gaussian noise (AWGN) and Rayleigh fading channels. Subsequent works, such as \cite{dai2022nonlinear, wu2024transformer}, have enhanced DeepJSCC approach by incorporating advanced vision transformer architectures. DeepJSCC has also been extended to various channel models and scenarios with superior performance, including multiple-input multiple-output (MIMO) channels \cite{wu2024deep}, orthogonal frequency division multiplexing (OFDM) \cite{yang2021deep,wu2022channel}, relay channels \cite{bian2024process}, multi-user transmission \cite{yilmaz2023distributed, zhang2023model}, and transmission using a finite constellation \cite{tung2022deepjscc, bo2022learning}.}
However, as we will show in this paper, it is possible to further push the limits of DeepJSCC. DeepJSCC maps the original data directly to a latent feature vector as channel symbols. The global distribution of the image itself or its latent features constitutes a new dimension of prior knowledge in DeepJSCC. This motivates the integration of generative models, which can efficiently capture the underlying data distribution, to enhance DeepJSCC \cite{erdemir2023generative}.  However, merging generative models with DeepJSCC presents new challenges. Specifically, generative models aim to randomly generate realistic data while DeepJSCC focuses on accurately transmitting data from the transmitter to the receiver. Moreover, there are concerns about whether the generative models-aided DeepJSCC can remain effective under varying channel conditions. These challenges are crucial for the integration of generative models with DeepJSCC, yet they have not been fully addressed.
\subsection{Motivations}
{\textbf{Towards an accurate diffusion model (DM)-aided DeepJSCC framework:} 
We utilize the DM, a powerful generative model for visual data. Speciﬁcally, we consider employing DM for channel denoising, leveraging the resemblance between the diffusion process and the wireless channel, as demonstrated in \cite{wu2024cddm}. As discussed earlier, while randomly denoising the channel output can generate realistic data, it risks compromising key semantics, especially under high channel noise. To address this, we consider transmitting semantics as side information to guide the diffusion denoising process towards a semantics-aligned direction. This framework enhances performance by transforming unconditional denoising into conditional denoising. Moreover, as the definition and type of the underlying side information are ﬂexible, the transmitter can adjust the semantic side information based on the instantaneous transmission objective. This flexibility enables effective adaptation to various semantic quality metrics without the need for specific training or fine-tuning. 
}

\textbf{Adapting DM to varying channels:}
{Another essential challenge in advancing DeepJSCC is improving model adaptability across diverse channel conditions, {which becomes especially critical when adopting large models containing billions of parameters}. 
Typically, a separate JSCC model must be trained for each specific channel environment to ensure {best performance is achieved in that environment}. However, this approach demands substantial computational resources and storage capacity, making it both impractical and costly. Currently, there are two primary approaches to address this problem. The first approach involves acquiring the channel state information (CSI), and providing it as side information to both the transmitter and receiver, typically utilizing an attention mechanism to introduce the CSI into the coding process \cite{xu2021wireless}, \cite{wu2022channel}. 
{This network is then trained over a wide variety of channel conditions, and learns to adapt to each channel state dynamically.}
The second approach uses DM and 
employs the current CSI to select an appropriate matching step to start the denoising process \cite{wu2024cddm}. 
{Both approaches require accurate CSI information which necessitates pilot transmission and explicit channel estimation, as well as CSI feedback if we want the encoder to adapt to the CSI.}
{
In this paper, the proposed SGD-JSCC method addresses this challenge by dynamically adapting to the channel state directly from the channel output, eliminating the need for pilot transmissions for channel estimation. 
Furthermore, 
in fast fading channels, where noise levels vary across different symbols, we propose a denoising scheme inspired by the water-filling principle. This approach allows us to directly leverage the DM trained on slow fading channels for denoising in fast fading channels. 
	\vspace{-3mm}
	\subsection{Contribution and Organization}
	\vspace{-1mm}
	In this paper, we {advance} DeepJSCC towards a semantics-aligned, high-fidelity transmission approach.
	Specifically, we propose transmitting some underlying semantics as side information 
	alongside JSCC latent features. These semantics serve as guidance for DM denoising, improving the denoising performance while preserving key semantic information. Furthermore, to fully leverage the advantages of DM within the DeepJSCC system, we design a DM tailored for the wireless channel, making it {adaptive to varying channel conditions.} The key contributions of this paper are summarized as follows. 
 \begin{itemize}
     \item {\textbf{Semantics-guided DM for semantics-aligned denoising:} We propose transmitting inherent information that contributes to the semantics of an image as side information alongside JSCC latent features. At the receiver side, a transmission-tailored DM is introduced to conduct channel denoising under the guidance of semantics, thereby accurately reconstructing key semantic content.}
     \item {\textbf{Adaptive DM for time-varying channel conditions:} 
	 Considering a slow fading channel, we propose to estimate the instantaneous CSI directly from the normalized channel output, alleviating the need for dedicated pilot transmission. A continuous timestep matching approach is proposed to mitigate matching errors common in previous discrete-timestep DM-DeepJSCC methods \cite{wu2024cddm}. In a fast fading channel, we introduce a training-free denoising strategy, 
     which enables SGD-JSCC to dynamically adapt to variations in channel gains. 
	 }
     
     \item {\textbf{Performance evaluation:} Our numerical results demonstrate that the proposed method achieves superior performance compared to other {DeepJSCC schemes} in the literature and can achieve satisfactory performance even at extremely poor channel conditions. i.e., ${\rm SNR} = -15$ dB. Moreover, with the integration of semantic side information, the proposed method effectively preserves relevant semantics in the reconstructed images, resulting in a more satisfactory perceptual reconstruction quality.}
 \end{itemize}
	The rest of this paper is organized as follows. 
    The related works are discussed in Section \ref{sec: related works}.
    Section \ref{sec: system model} introduces the semantics-guided transmission framework. 
	Section \ref{sec: seamtics extraction and transmission} presents two types of semantic {guidance} and the corresponding transmission scheme. 
	DeepJSCC and DM design for slow fading channels are detailed in Section \ref{sec: JSCC and diffusion design}, followed by the extension to fast fading channels in Section \ref{sec: extension to rayleigh fading case}. 
	Numerical simulation results are provided in Section \ref{sec: numerical results}, followed by the concluding remarks in Section \ref{sec: conclusion}.  
\section{Related works}\label{sec: related works}
\subsection{Generative Models: Foundations and Control Principles}
	Generative models aim to generate realistic samples by learning the underlying data distribution and sampling from it. Among various generative models, DM has demonstrated remarkable results, particularly in visual generation tasks. During training, DMs learn the conditional distribution of data over a progressively noisier latent space. In the inference stage, DMs iteratively remove noise and finally obtain a generated data instance \cite{ho2020denoising}.  Nevertheless, the original DM in \cite{ho2020denoising} operates under an unconditional generation framework, which introduces randomness in the generated results. 
{Our goal in this work, however, is to convey the input image with the highest fidelity, rather than generating an arbitrary realistic image sample at the receiver. 
}
	Hence, we want to design a DM that is able to generate data that is aligned with the semantics of the input image,  
leading to semantics-guided generation. 
The authors in \cite{dhariwal2021diffusion} first introduced the concept of incorporating class labels into the DM, employing a classifier to guide the generation process and improve alignment between the generated image and the desired class. This was further extended to text-based semantics, leading to the development of popular text-to-image DMs like stable diffusion \cite{rombach2022high}. Stable diffusion shifts the diffusion process from the pixel level to the latent feature space and integrates contrastive language-image pretraining (CLIP) models to better align the generated images with their corresponding text descriptions. This method has been further enhanced by advanced diffusion transformer models \cite{chen2024pixart}. Additionally, the authors in \cite{zhang2023adding} introduced structural semantics, adding spatial control over the generation process, and allowing for fine-grained regulation of the output. 

\vspace{-2mm}
\subsection{DMs for Source Coding}
Pretrained semantics-guided DMs have also been applied in the field of data compression. The authors in \cite{lei2023text+} initially explored compressing an image by representing it through its key semantic features. During the decoding process, these semantics are served as guidance for the generation process. Compared to directly compressing the image itself, these semantics are lightweight and incur lower encoding costs. This method has been extended in \cite{qiao2024latency, wang2024fast,wijesinghe2023diff}. However, due to the inherent randomness of the generative process, the decoded image may differ significantly from the original image, particularly in color accuracy. To address this, the authors in \cite{yang2024lossy} utilized the compressed image from existing compression schemes as additional guidance, which is shown to improve the consistency of the received images. 
{While the benefits of DM in compression has been shown, 
}
further investigation is needed to develop DM-based transmission schemes that can simultaneously optimize data compression and noise resilience over wireless channels. 
\vspace{-2mm}
\subsection{DMs for DeepJSCC}
By effectively capturing data distributions, generative models can help in addressing the semantic distortion issue present in DeepJSCC. In the context of DM, a hybrid JSCC scheme was first proposed in \cite{Niu:SPAWC:23}, which conveyed a low-resolution image using separate source and channel coding, followed by a refinement layer that exploited diffusion. A fully joint scheme was later presented in \cite{yilmaz2023high} that employ DM for post-processing, specifically to {refine} the reconstructed images from DeepJSCC. 
{An alternative scheme relying on invertible neural networks was proposed in \cite{chen2024commin}. 
}
\rv{Besides, the authors in \cite{yang2024diffusion} proposed incorporating signal-to-noise ratio (SNR)  information into DMs to accommodate varying channel conditions. 
}Using DM for post-processing requires modeling the distortion function from the reconstructed image to the original one. However, this distortion is highly non-linear as it arises from {both} the encoder/decoder neural networks and {the dynamic nature} of the wireless channels, which either involve complex computations or {fluctuate unpredictably, making accurate distortion characterization challenging.} To address this, the authors in \cite{wu2024cddm} proposed using DM for preprocessing the channel output, called channel denoising diffusion models (CDDM). Thanks to the similarity between the diffusion process and the noise added over the wireless channel, DM can naturally serve as denoisers for removing channel noise. Given the high inference latency of CDDM, the authors in \cite{pei2024latent} proposed a consistency distillation strategy to reduce the number of diffusion steps. 
\rv{In addition, the authors in \cite{xie2024hybrid} proposed a hybrid analog-digital transmission scheme, where the received analog and digital signals are first combined and then denoised using DM. 
}The authors in \cite{guo2024diffusion} considered using a variational autoencoder (VAE) for downsampling and transforming the original feature distribution to a Gaussian distribution, which then can be processed by DMs.  
\rv{
	Subsequently, the authors in \cite{wang2024diffcom} proposed to directly utilize the noisy latent feature as prior knowledge, and developed a gradient-based guidance scheme for the DM to enhance perceptual performance.}
{Despite significant advancements, several critical challenges in DeepJSCC, particularly in terms of practicality and interpretability, remain unresolved. Additionally, the full potential of diffusion-based generative models for wireless communication remains underexplored. Current approaches, such as CDDM, overlook the explicit utilization of inherent data semantics. This restricts their denoising efficiency and increases computational complexity. 
}
	\vspace{-6mm}
    \section{System Model}\label{sec: system model}
\subsection{Problem Statement}\label{subsec: problem formulation}
We consider the problem of transmitting an image $\mathbf{x}\in \mathbb{R}^{h\times w\times 3}$ over a point-to-point wireless channel, where $h$, $w$, and $3$ denote the height, width, and color channels of an RGB image, respectively. The transmitter maps the source image $\mathbf{x}$ into a vector of complex-valued channel input symbols $\mathbf{z}\in \mathbb{C}^M$. An average transmit power constraint is imposed on $\mathbf{z}$, such that 
\begin{align}\label{eq: power constraint on z}
\frac{1}{M}\mathbb{E}_{\mathbf{z}}{[\|\mathbf{z}\|_2^2]}\leq 2. 
\end{align}
$\mathbf{z}$ is then transmitted over a noisy channel. 
\rv{We consider a slow fading channel with perfect synchronization, with the channel output denoted by $\mathbf{y}$. 
}The $i$-th element of $\mathbf{y}$ is given by 
\begin{align}\label{eq: channel model}
	y_i = h z_i + n_{c,i},
\end{align}
where $h \in \mathbb{C}$ denotes the random channel gain\footnote{Note that we primarily consider slow fading channels, where the channel gain remains constant during the transmission of each $\mathbf{z}$. In Section \ref{sec: extension to rayleigh fading case}, we will also discuss how to extend the transmission method from slow fading channels to fast fading channels, where the channel gain may vary with each transmitted symbol $z_i$.}, and $n_{c,i} \in \mathcal{CN}(0,2\sigma^2)$ denotes the independent additive complex Gaussian noise. 
The receiver reconstructs the image from the channel output $\mathbf{y}$. 
We assume that the transmitter has no access to CSI. 
\rv{The receiver estimates CSI either by utilizing transmitted pilot symbols or, as we discuss later, by directly estimating it from the channel output $\mathbf{y}$. }

The transmission objective is to minimize the distortion between the source and the reconstructed image, measured by various semantic quality metrics, subject to a given bandwidth compression ratio (BCR). 
The BCR is defined as $R\triangleq\frac{M}{3hw}$, representing the average number of available channel symbols per source dimension. 
\vspace{-3mm}
\subsection{Semantics Guided Transmission Framework}\label{subsec: semantics guided transmission framework}
    \vspace{-1mm}
	{
	DeepJSCC is a promising paradigm for addressing the wireless image transmission problem described in Section \ref{subsec: problem formulation}. 
	As illustrated in Fig. \ref{Fig:previous jscc transmission paradigm}, a standard DeepJSCC scheme directly maps the original data into channel symbols, and the decoder reconstructs the input image from the noisy channel output. 
	In this paper, we consider an enhanced DeepJSCC scheme that incorporates inherent information contributing to the semantics of an image as side information to improve transmission performance, as illustrated in Fig. \ref{Fig:semantics guided deep jscc transmission paradigm}. 
	The transmission model of the proposed SGD-JSCC scheme is detailed in the following subsection. 
	}
	\subsubsection{Transmitter}
	As depicted in Fig. \ref{Fig:semantics guided deep jscc transmission paradigm}, the transmitter directly encodes the {source} image $\mathbf{x}$  to its latent representation. The encoding process is described as follows: 
	\begin{align}\label{eq: jscc encoding process}
		{\mathbf{f}}=\mathcal{F}(\mathbf{x};\boldsymbol{\Theta}),
	\end{align}
	where $\mathbf{x}$ and $\boldsymbol{\Theta}$ denote the original image $\mathbf{x}$ and the trainable parameters of the JSCC encoder, respectively. 
	$\mathcal{F}(\cdot)$ is the encoding function, which generates the latent representation ${\mathbf{f}}\in \mathbb{R}^{N}$. 
	The latent representation $\mathbf{f}$ satisfies
	the power constraint $\frac{1}{N}\mathbb{E}_\mathbf{f}[\|\mathbf{f}\|_2^2]\leq 1$. 


	Besides, a semantic extractor is employed at the encoder to directly extract the semantic features of $\mathbf{x}$, yielding $\mathbf{s}$. 
	The exact form of $\mathbf{s}$ is flexible and can be customized at the transmitter for a specific semantic quality metric. 
	$\mathbf{s}$ is subsequently encoded into its latent representation, leading to:
	\begin{align}\label{eq: semantics extracting process}
		\rv{{\mathbf{o}}= \mathcal{H}(\mathbf{s}, \boldsymbol{\Pi}),}
	\end{align}
	\rv{where $\mathcal{H}(\cdot)$ denotes the encoding function of $\mathbf{s}$ and $\boldsymbol{\Pi}$ denotes the trainable parameters of the semantic extractor.}  
	The latent feature of $\mathbf{s}$ is denoted by $\mathbf{o}\in \mathbb{R}^{K}$, which satisfies the power constraint $\frac{1}{K}\mathbb{E}_{\mathbf{o}}[\|\mathbf{o}\|_2^2]\leq 1$.

\rv{We define two mappings between a ``real'' vector $\mathbf{v}_r\in \mathbb{R}^{2L}$ and its complex counterpart $\mathbf{v}\in\mathbb{C}^{L}$: \\$\mathcal{C}:\mathbb{R}^{2L}\rightarrow \mathbb{C}^{L}$, $[\mathcal{C}(\mathbf{v}_r)]_i=[\mathbf{v}_r]_i+j[\mathbf{v}_r]_{i+L}$; \\$\mathcal{R}:\mathbb{C}^L\rightarrow\mathbb{R}^{2L}: [\mathcal{R}(\mathbf{v})]_i=\Re([\mathbf{v}]_i),[\mathcal{R}(\mathbf{v})]_{i+L}=\Im([\mathbf{v}]_i)$, \\
where $\Re(\cdot)$ and $\Im(\cdot)$ denote the real and imaginary parts of a complex number, respectively. 
	Then, $\mathbf{f}$ and $\mathbf{o}$ are transformed into its complex form and then concatenated, yielding $\mathbf{z}=[\mathcal{C}(\mathbf{f});\mathcal{C}(\mathbf{o})]\in \mathbb{C}^{M}$.   
}
Since both $\mathbf{f}$ and $\mathbf{o}$ satisfy their respective power constraints, $\mathbf{z}$ inherently satisfies the power constraint in (\ref{eq: power constraint on z}). 
Finally, $\mathbf{z}$ is transmitted through the wireless channel in (\ref{eq: channel model}).

\subsubsection{Receiver}
\rv{After transmitting $\mathbf{z}$ over the wireless channel, the receiver observes the noisy output $\mathbf{y}$. Writing the complex channel coefficient as $h=|h|e^{j\phi}$, the receiver performs equalization in two steps: 
First, it compensates for the phase rotation by 
\begin{align}\label{eq: phase remove with gt phase}
	y_{eq,i}'=e^{-j\phi}y_i,
	\end{align}
which removes the phase term and prevents interference between the real and imaginary components. Next, it normalizes the signal power via
\begin{align}
	y_{eq,i}=\frac{y_{eq,i}'}{\sqrt{|h|^2+\sigma^2}}. 
	\end{align}
Given the channel model described in (\ref{eq: channel model}), we further have:
\begin{align}\label{eq: equalized channel output}
y_{eq,i}=\frac{|h|}{\sqrt{|h|^2+\sigma^2}}z_i + \frac{e^{-j\phi}}{\sqrt{|h|^2+\sigma^2}}n_{c,i}. 
\end{align}
} 
\rv{After equalization, $\mathbf{y}_{eq}$ undergoes the complex-to-real operation, i.e., $\mathbf{y}_r=\mathcal{R}(\mathbf{y}_{eq})$.  
}Let $y_{r,i}$ denote the $i$-th element of $\mathbf{y}_{r}$, 
based on the equalization result in (\ref{eq: equalized channel output}), 
the conditional distribution of $y_{r,i}$ given $z_{r,i}$ and $h$ is given by 
\begin{align}\label{eq: conditional distribution of y}
p(y_{r,i}|z_{r,i},h)\sim\mathcal{N}\bigg(\frac{|h|}{\sqrt{|h|^2+\sigma^2}}z_{r,i},\frac{\sigma^2}{|h|^2+\sigma^2}\bigg).
\end{align}
Let $\mathbf{\tilde{f}}$ and $\tilde{\mathbf{o}}$ denote the received latent features of the image $\mathbf{x}$ and the semantics $\mathbf{s}$, respectively. 
According to (\ref{eq: conditional distribution of y}), 
the conditional distribution is given by
\begin{align}
p(\tilde{\mathbf{f}}|\mathbf{f},h) \sim \mathcal{N}\bigg(\frac{|h|}{\sqrt{|h|^2+\sigma^2}}\mathbf{f},\frac{\sigma^2}{|h|^2+\sigma^2}\mathbf{I}\bigg),\\
p(\tilde{\mathbf{o}}|\mathbf{o},h) \sim \mathcal{N}\bigg(\frac{|h|}{\sqrt{|h|^2+\sigma^2}}\mathbf{o},\frac{\sigma^2}{|h|^2+\sigma^2}\mathbf{I}\bigg),
\end{align}
where $\mathbf{I}$ denotes the identity matrix.

Then, the receiver reconstructs the semantic side information, as follows: 
\begin{align}
	\rv{\tilde{\mathbf{s}} = \mathcal{G}(\tilde{\mathbf{o}},\boldsymbol{\Lambda}),}
\end{align}
\rv{where $\boldsymbol{\Lambda}$ and $\mathcal{G}(\cdot)$ denote the trainable parameters in the semantic decoder and the decoding function, respectively.
$\tilde{\mathbf{s}}$ denotes the reconstructed semantic side information.  
}\begin{figure}[t]
	\centering 
	\subfigure[Standard DeepJSCC scheme.]{
		\label{Fig:previous jscc transmission paradigm} 
		\hspace*{-2.5mm}
		\includegraphics[width=1\linewidth]{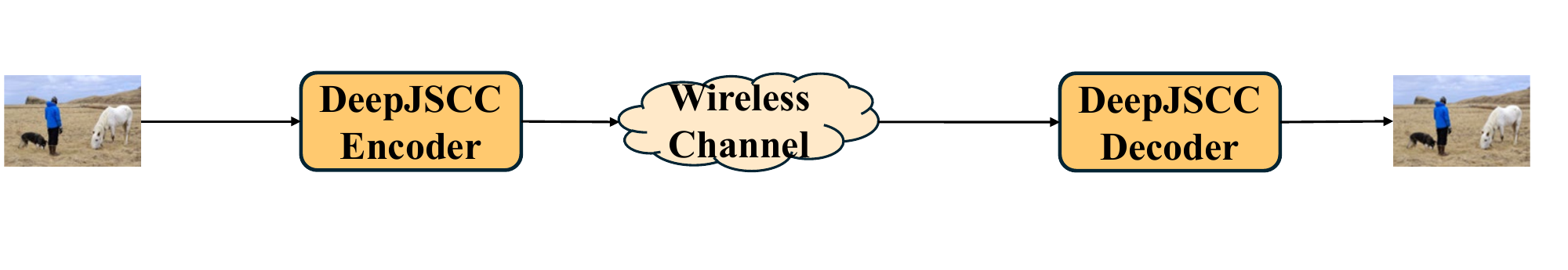}
		\vspace{-15mm} 
		}
	\subfigure[Proposed semantics guided DeepJSCC scheme.]{
		\label{Fig:semantics guided deep jscc transmission paradigm} 
		\includegraphics[width=1\linewidth]{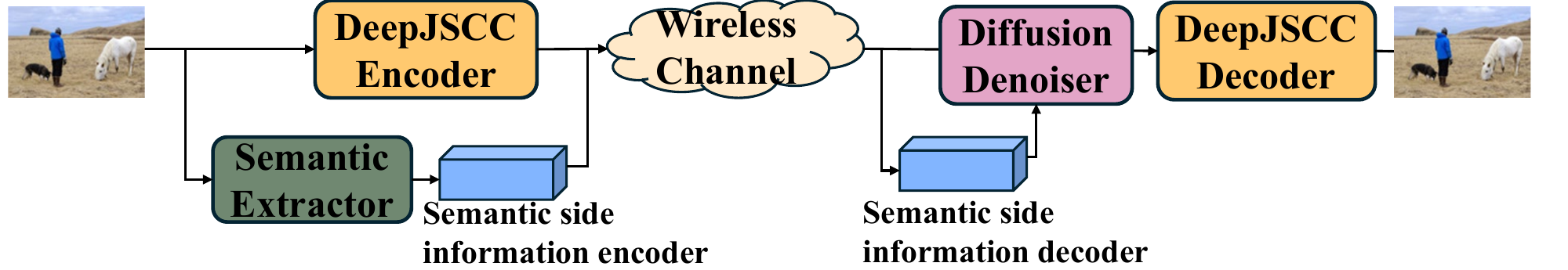}}
	\vspace{3mm}
		\caption{{Different DeepJSCC transmission paradigms.}}\label{transmission paradigm}
	\vspace{-7mm}
\end{figure}

With the noisy feature $\tilde{\mathbf{f}}$ and semantic side information $\tilde{\mathbf{s}}$ in hand, 
we consider employing DM to preprocess the noisy feature $\tilde{\mathbf{f}}$ under the guidance of $\tilde{\mathbf{s}}$. 
DM is a powerful generator and denoiser capable of refining content and generating outputs aligned with semantic guidance. The denoising process is given as follows: 
\begin{align}
	\mathbf{\hat{f}}=\mathcal{D}(\tilde{\mathbf{f}}|\tilde{\mathbf{s}};\boldsymbol{\Omega}),
\end{align}
where $\boldsymbol{\Omega}$ and $\mathcal{D}$ denote the trainable parameters in DM and the denoising operation, respectively. 

Consequently, the denoised feature $\hat{\mathbf{f}}$ is fed into the JSCC decoder to reconstruct the image in the pixel domain, as follows: 
\begin{align}
	\hat{\mathbf{x}}=\mathcal{G}(\hat{\mathbf{f}};\boldsymbol{\Psi}),
\end{align}
where $\boldsymbol{\Psi}$ and $\mathcal{G}$ are the trainable parameters in DeepJSCC decoder and the decoding function, respectively.

\section{Semantic Guidance Extraction and Transmission}\label{sec: seamtics extraction and transmission}
{To facilitate the reliable transmission of critical semantic information, we explore two types of semantic guidance:} text descriptions, which convey coarse semantic information, and edge maps, which offer finer semantic details.
 
 \vspace{-4mm}
	\subsection{Coarse Semantics: Text Description}
 \vspace{-1mm}
	In human-level communication, we can often imagine a reasonable reconstruction of an image based on the descriptions provided by other people. Inspired by this, text descriptions serve as suitable semantic {guidance} that encapsulate the general information behind images. 
	Utilizing text descriptions in DM-aided image compression has been considered in \cite{pan2022extreme,lei2023text+,careil2023towards}. 
	We use BLIP2 \cite{li2023blip}, an off-the-shelf state-of-the-art image captioning model, {to extract the text descriptions for the input images}. 
	The transmission cost of text descriptions is negligible compared to that of images. Therefore, we neglect the transmission cost of text and assume the text description can be transmitted to the receiver perfectly.\footnote{\rv{This assumption holds reasonable for most transmission scenarios. 
	However, in some extreme cases (e.g., ${\rm SNR} =-15$dB), transmitting text accurately becomes challenging. 
		Nevertheless, as the goal shifts from word-level accuracy to preserving the semantic meaning for guidance, specialized DeepJSCC techniques can help to address such challenges. 
		Due to limited paper space, we have opted not to discuss it in detail. 
		}}
    


 	\begin{figure}[t] 
		\centering
		\includegraphics[width=1\linewidth]{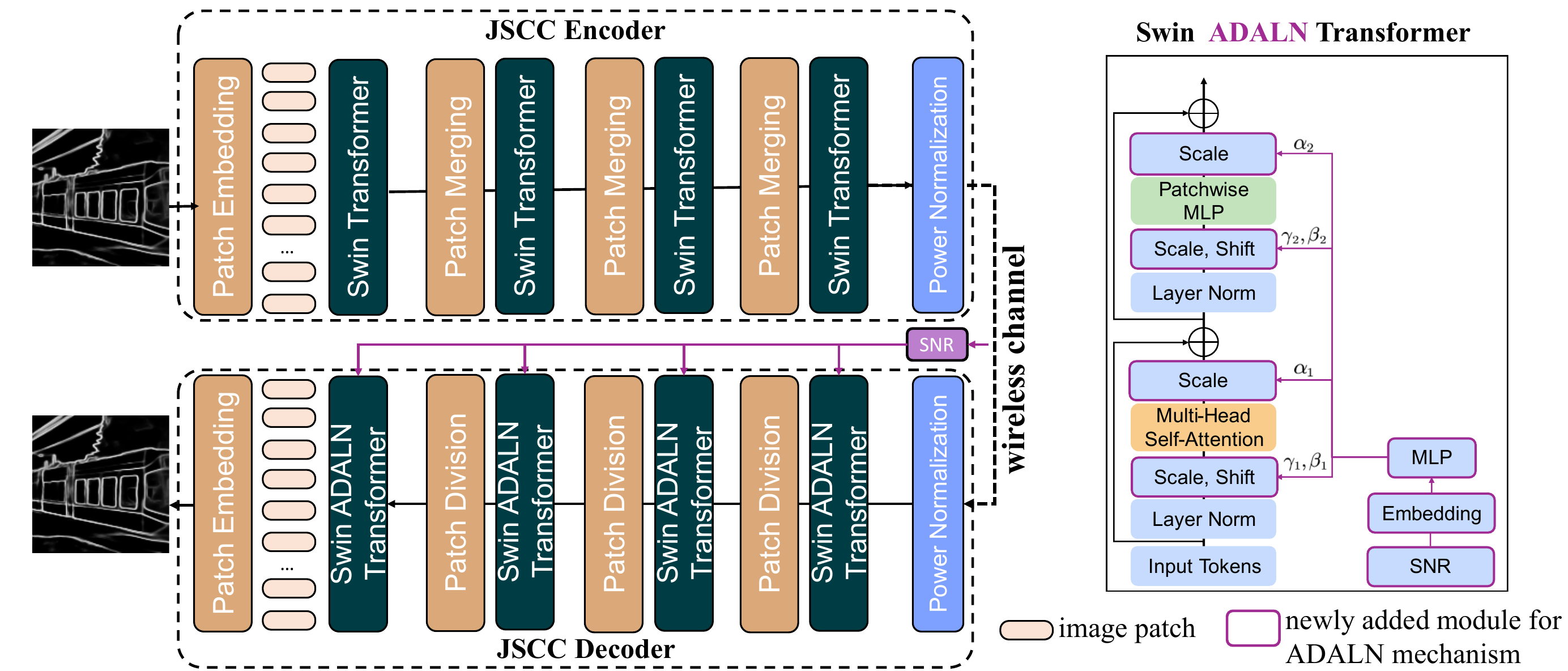}
	\vspace{2mm}
        \caption{{DeepJSCC transmission for the image edge map.}}
\vspace{-4mm} 
		\label{Fig:edge transmission}
	\end{figure}
	\subsection{Fine Semantics: Edge Map}\label{subsec: edge map extraction and transmission}
	Text descriptions provide lightweight semantics but only allow for coarse guidance during the denoising process. To address this limitation, we explore the use of edge maps as an additional semantic guidance with structural details. An edge map is a grayscale image that highlights the edges of objects in an image. 
    We use MuGE \cite{zhou2024muge}, a deep learning-based edge extractor, to extract edge information from input images.

\rv{
Specifically, the semantic information in edge maps resides in the global structural lines, we consider a vision transformer (ViT) as the architecture of the JSCC model for edge map transmission, given its effectiveness for capturing these long-range dependencies. Moreover, 
since each pixel is represented by a single scalar and most of the pixels are set to zero in an edge map,  we use a small embedding size for each patch to reduce computation complexity. 
As illustrated in Section \ref{subsec: problem formulation}, the SNR information is available at the decoder side,  
	we project the SNR values as side information to the transformer blocks at the decoder side. 
}	The output of the JSCC model represents the probability of each pixel being part of the foreground, for which we adopt binary cross-entropy (BCE) as the first part of the loss function. Moreover, the edge map is transmitted to provide structural information that guides the DM denoising process. In this context, the focus is on accurately identifying foreground information in the non-zero pixels rather than achieving overall pixel-level accuracy. Therefore, we incorporate  Dice loss as the second part of the loss function, encouraging the model to prioritize foreground information. Specifically, the DeepJSCC model in Fig. \ref{Fig:edge transmission} is trained in an end-to-end manner to minimize the weighted sum of BCE loss and Dice loss. 
\section{SGD-JSCC over Slow Fading Channels}\label{sec: JSCC and diffusion design}
In this section, we {consider slow fading channels, where the fading state remains constant during the transmission of $\mathbf{f}$, i.e., $h_{c,i}=h, \forall i$.} {Under this scenario, the received {channel output} can be equivalently treated as the channel output of an additive white Gaussian noise (AWGN) channel with ${\rm SNR}=\frac{|h|^2}{\sigma^2}$.} 
We begin by detailing the design of the DeepJSCC model, followed by the specific architecture of the DM, and proceed to the training strategies and pilot-free channel estimation design. 
We then extend the proposed approach to a more challenging fast-fading scenario in Section \ref{sec: extension to rayleigh fading case}. 

\begin{figure}[t] 
	\centering
		\captionsetup{skip=2pt}
	\includegraphics[width=1\linewidth]{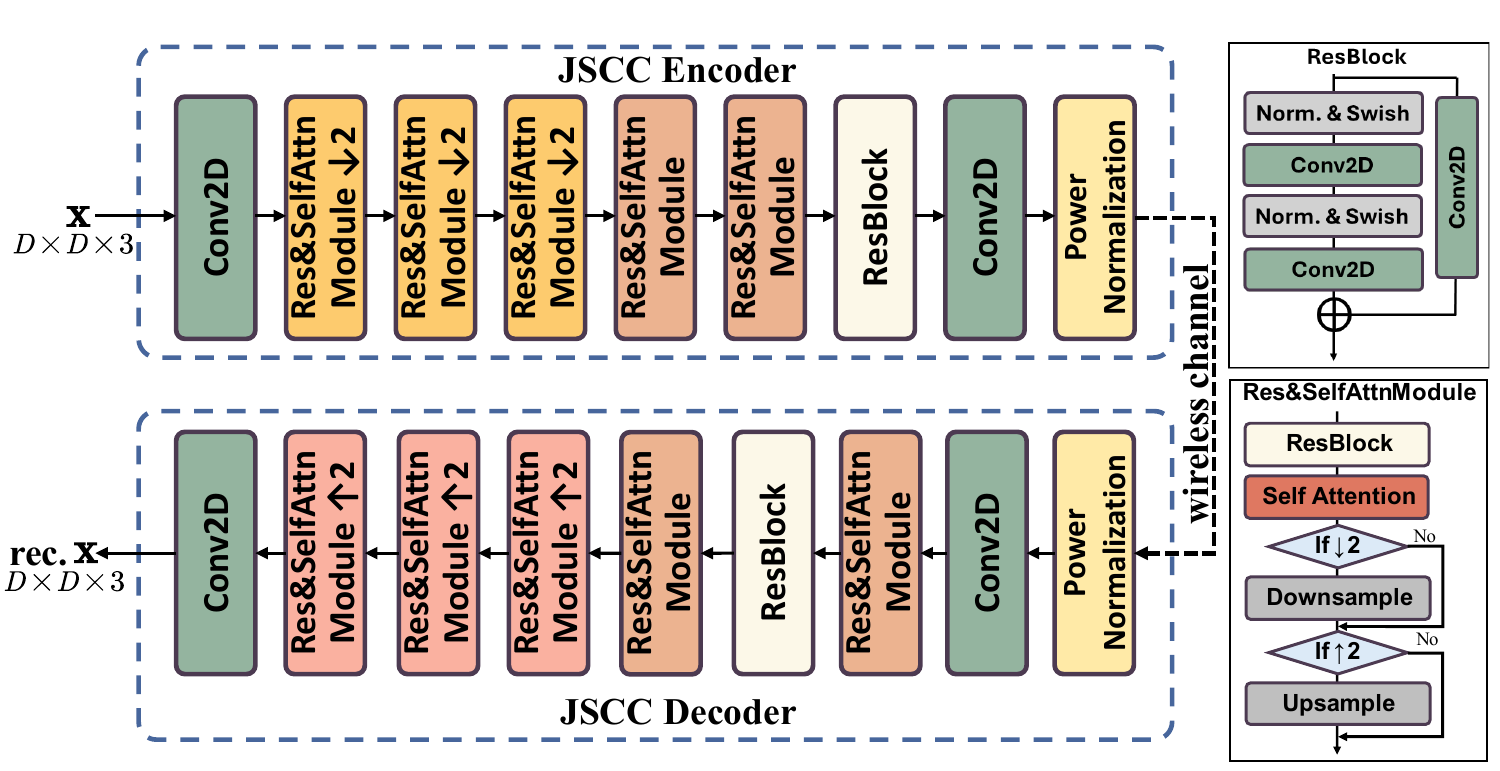}
	\vspace{-2mm}
	\caption{{Architecture of the DeepJSCC.}}
	\vspace{-6mm}
	\label{Fig: jscc transmission paradigm}
\end{figure} 
\subsection{DeepJSCC architecture}
	As the DeepJSCC architecture for the extraction and transmission of the input image features, 
	we adopt the architecture proposed in \cite{ho2020denoising}. 
	Specifically, as depicted in Fig. \ref{Fig: jscc transmission paradigm}, residual blocks are utilized as the basic feature extraction modules. 
	Additionally, a self-attention module is incorporated after each residual block to further enhance representation and reconstruction capabilities of the DeepJSCC model. 
	Within each self-attention module, three convolutional layers are employed to obtain the query, key, and value sequences from the corresponding intermediate feature map, 
	which are subsequently processed through the self-attention mechanism. 
 
	We train {this autoencoder pair} in an end-to-end manner under a fixed noisy channel setting {(i.e., AWGN channel with SNR = $10$dB)\footnote{Note that, the adaptability to various SNRs can be achieved by DM module as detailed in Section \ref{subsec: diffusion denoiser}, thereby we consider a fixed SNR setting for the training of the JSCC model.}.} 
	The total training objective is divided into two parts as follows.
	\vspace{-2mm}
	\begin{align}\label{eq: training loss function of JSCC autoencoder}
	\mathcal{L}_{\rm JSCC}=\min_{\boldsymbol{\Theta},\boldsymbol{\Psi}}\|\mathbf{x}-\hat{\mathbf{x}}\|_2^2+\lambda&\min_{\boldsymbol{\Theta},\boldsymbol{\Psi}}\max_{\mathscr{D}}(L_{GAN})
	\end{align}
    The mean squared error (MSE) between the reconstructed and original images serves as the first part of the loss function. 
    However, merely minimizing the pixel-level distortion can significantly degrade the semantic information. To address this, we incorporate a patch-based discriminator to enhance perceptual performance. {The discriminator model is denoted by $\mathscr{D}$, and the corresponding discriminator loss is denoted by $L_{GAN}$.} 
	$\lambda$ is the weighting factor. 
	\vspace{-4mm}
	\subsection{Diffusion Denoiser}\label{subsec: diffusion denoiser}
	\vspace{-2mm}
	DMs are natural denoisers, adept at iteratively learning to remove additive noise from data \cite{ho2020denoising}. 
	Given the similar effects of wireless channels on transmitted signals, we employ DMs to mitigate channel noise. 
	{Notably, conventional and state-of-the-art} DMs are specifically designed for generative tasks, where visual quality is the primary focus. 
	However, when applying DMs to DeepJSCC, it is crucial to consider {the transmission distortion} in addition to visual quality. 
{In the following section, we will elaborate on the diffusion algorithm and the design of the transmission-oriented diffusion denoising model. }
 
	\subsubsection{{Training Strategy}}
	Given a data point sampled from a real data distribution $\mathbf{f}_0 \sim \mathbf{q}(\mathbf{f})$, the forward trajectory of the diffusion process involves iteratively adding Gaussian noise with a specific variance to the data sample, ultimately resulting in a standard Gaussian noise $\mathbf{f}_1 \sim \mathcal{N}(0, \mathbf{I})$. 
	{Let $t$ represent the timestep corresponding to a specific noise level, where $t=0$ corresponds to the clean sample $\mathbf{f}_0$.} We have
	\begin{align}\label{eq: forward trajectory of diffusion process}
		\mathbf{f}_t = \sqrt{1-\bar{\beta}_t} \mathbf{f}_{0} + \sqrt{\bar{\beta}_t} \mathbf{n},
	\end{align}
where $\bar{\beta}_t$ is the variance of the noise at timestep $t$, and $\mathbf{n}\sim \mathcal{N}(\mathbf{0},\mathbf{I})$ denote the additive Gaussian noise. 
Note that in most DMs, $t$ takes discrete values ranging from $0$ to $T$. 
However, the wireless channel noise can take continuous values, which means that a discrete noise schedule cannot accurately characterize its state. 
We consider $t$ as a continuous value, i.e., $t\in \mathbb{R}$ and $0\leq t\leq 1$. 
The continuous setting of $t$ is helpful for mitigating the step-matching error that will be discussed later. 

We consider a variance-preserving forward trajectory, that is, $\mathbb{E}[\|\mathbf{f}_t\|^2]=1$. 
Based on (\ref{eq: forward trajectory of diffusion process}), 
it can be found that the conditional distribution $q(\mathbf{f}_t|\mathbf{f}_s)$ for any $t>s$ is Gaussian distribution as well, which is given by
\begin{align}
	\mathbf{f}_t = \sqrt{\frac{{1-\bar{\beta}_t}}{1-\bar{\beta}_{s}}}\mathbf{f}_s + \sqrt{\bar{\beta}_{t}-\bar{\beta}_s\frac{1-\bar{\beta}_t}{1-\bar{\beta}_s}}~\mathbf{n}.
\end{align} 
Given the forward trajectory described in (\ref{eq: forward trajectory of diffusion process}), the objective of the reverse trajectory in diffusion process is to recover $\mathbf{f}_0$ from $\mathbf{f}_1$ in $T$ steps. To achieve this, a DM is introduced to learn the conditional distribution, i.e., $p_{\boldsymbol{\Omega}}(\mathbf{f}_{s}|\mathbf{f}_t)$, $s<t$. 
For an intermediate timestep $t$, we have $\mathbf{f}_t=\sqrt{1-\bar{\beta}_t}\mathbf{f}_0+\sqrt{\bar{\beta}_t}\boldsymbol{\epsilon}$, where $\boldsymbol{\epsilon}$ is a instance of $\mathbf{n}$. 
Given the forward distribution $q(\mathbf{f_t|\mathbf{f}_0})\sim\mathcal{N}(\sqrt{1-\bar{\beta}_t}\mathbf{f}_0,\bar{\beta}_t\mathbf{I})$, 
following the non-Markovian sampling distribution proposed in \cite{song2020denoising}, 
the distribution of $\mathbf{f}_{s}$ conditioned on $\mathbf{f}_0$ and $\mathbf{f}_{t}$ can be built by
\begin{align}\label{eq: conditional distribution of reverse trajectory}
q(\mathbf{f}_{s}|\mathbf{f}_t,\mathbf{f}_0)&=\mathcal{N}(\sqrt{1-\bar{\beta}_{s}}\mathbf{f}_0\notag\\ &+\sqrt{\bar{\beta}_{s}-\sigma_{s,t}^2}\frac{\mathbf{f}_t-\sqrt{1-\bar{\beta}_t}\mathbf{f}_0}{\sqrt{\bar{\beta}_t}},\sigma_{s,t}^2\mathbf{I}), 
\end{align}
 where $\sigma_{s,t}^2$ is the variance for the reverse distribution. To note, DM serves as a denoising module as shown in Fig. \ref{Fig:semantics guided deep jscc transmission paradigm}. The objective is to recover the transmitted feature from a {noisy one rather than generating an arbitrary new sample. Consequently, unlike the original DM in \cite{ho2020denoising}, introducing additional noise or randomness is undesirable.} 
 Therefore, we adopt a deterministic reverse process {with} $\sigma_{s,t}^2=0, \forall s,t\sim [0,1]$. 

 \begin{figure}[!]
	\label{alg:LSB}
	\vspace{-2mm}
	\begin{algorithm}[H]
	  \caption{Training algorithm for the denoising DM}\label{algo:training algorithm for diffusion}
	  \begin{algorithmic}[1]
		\renewcommand{\algorithmicrequire}{ \textbf{Input:}}
		\REQUIRE Training dataset, $\bar{\beta}_t$
		\renewcommand{\algorithmicrequire}{ \textbf{Output:}}
		\REQUIRE DM $\boldsymbol{\Omega}$ after training
		\WHILE{$\boldsymbol{\Omega}$ not converged} 
		\STATE $\mathbf{f}_0\sim q(\mathbf{f}_0)$.
		\STATE $t\sim {\rm Uniform}(0,1)$.
		\STATE $\bar{\beta}_t=\mathcal{S}(t)$ defined in (\ref{eq: noise scheduler definition}).
		\STATE $\mathbf{n}\sim \mathcal{N}(\mathbf{0},\mathbf{I})$.
		\STATE {Take gradient descent step:}\\ $\nabla_{\boldsymbol{\Omega}}\|\mathbf{f}_0-\boldsymbol{\epsilon}_{\boldsymbol{\Omega}}(\sqrt{1-\bar{\beta}_t}\mathbf{f}_0+\sqrt{\bar{\beta}_t}\mathbf{n},\bar{\beta}_{t})\|^2$. 
		\ENDWHILE
	  \end{algorithmic}
	\end{algorithm}
	\vspace{-10mm}
\end{figure}
\begin{figure*}[t] 
	\centering
	\includegraphics[width=1.0\linewidth]{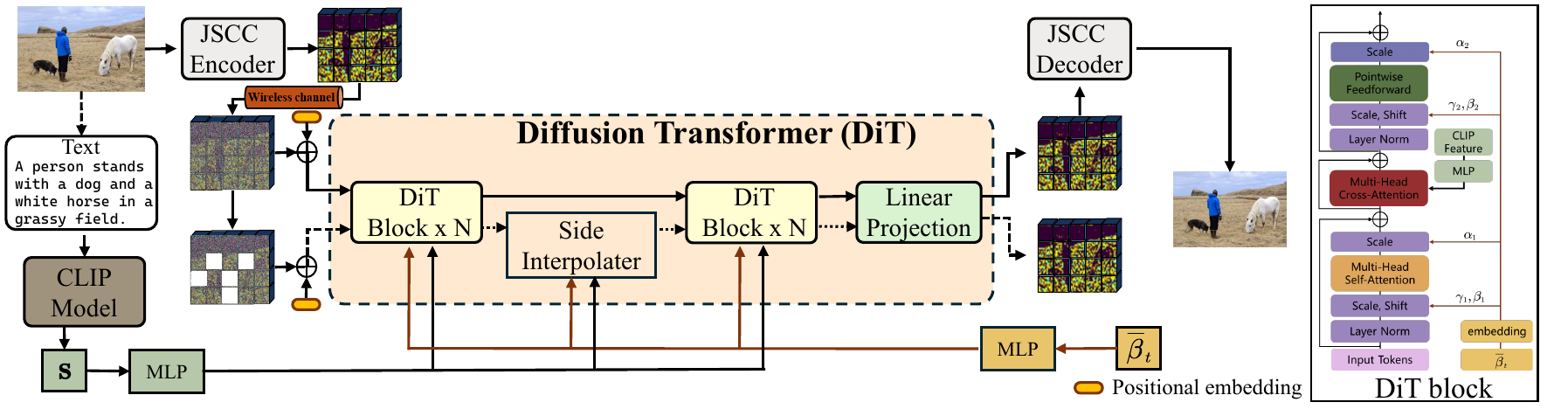}
	\vspace{0mm}
	\caption{{Network architecture of diffusion transformer (DiT) model.}}\label{fig: network architecture of diffusion model}
	\label{Fig:diffusion model}
	\vspace{-5mm}
	\end{figure*} 
From (\ref{eq: conditional distribution of reverse trajectory}), {it is evident that obtaining $\mathbf{f}_{s}$ requires both $\mathbf{f}_t$ and $\mathbf{f}_0$, where $\mathbf{f}_0$ is the desired result of reverse trajectory and is unknown at the $t$-th step of reverse trajectory.} To address this, a neural network $\boldsymbol{\Omega}$ is {introduced} to predict  $\mathbf{f}_0$ from $\mathbf{f}_t$, yielding $\boldsymbol{\epsilon}_{\boldsymbol{\Omega}}(\mathbf{f}_t,\bar{\beta}_{t})$. 
 Moreover, as revealed in \cite{chen2023importance}, the noise scheduling function (i.e., $\bar{\beta}_{t}$ over $t$) plays a critical role to the performance of DMs. 
 We adopt a sigmoid scheduling function, which is given by
 \vspace{-2mm}
 \begin{align}\label{eq: noise scheduler definition}
	\mathcal{S}(t)=\frac{{\rm sigmoid}(\frac{t(e-g)+g}{\tau})-{\rm sigmoid}(\frac{g}{\tau})}{{\rm sigmoid}(\frac{e}{\tau})-{\rm sigmoid}(\frac{g}{\tau})},
\end{align}
\rv{where ${\rm sigmoid}(x)=\frac{1}{1+\exp(-x)}$. $e$, $g$, and $\tau$ are hyperparameters that determine the shape of the scheduling function.  
}
The training algorithm is concluded in Algorithm \ref{algo:training algorithm for diffusion}. 

\subsubsection{DM for Channel Denoising}
With the well-trained DM for predicting $\mathbf{f}_0$ from $\mathbf{f}_t$, we are able to perform the denoising operation, also known as the sampling operation in diffusion theory \cite{ho2020denoising}. 
Let $s$ be the next target timestep, where $s<t$.  
Building on the conditional distribution $q(\mathbf{f}_{s}|{\mathbf{f}_t,\mathbf{f}_0})$ in (\ref{eq: conditional distribution of reverse trajectory}) with $\sigma_{s,t}^2=0$, 
the optimal sample of $\mathbf{f}_{s}$ is given by
\vspace{-2mm}
\begin{align}
	\mathbf{f}_{s}=\sqrt{\frac{\bar{\beta}_{s}}{\bar{\beta}_t}}\mathbf{f}_t+\bigg(\sqrt{\bar{\beta}_{s}}-\sqrt{\frac{\bar{\beta}_{s}(1-\bar{\beta}_{t})}{\bar{\beta}_t}}\bigg)\boldsymbol{\epsilon}_{\boldsymbol{\Omega}}(\mathbf{f}_t,\bar{\beta}_{t}). 
\end{align}

\rv{The overall procedure of diffusion denoising is depicted in Fig. \ref{Fig:diffusion denoising}. 
Unlike the generation-oriented reverse diffusion process which denoises from pure Gaussian noise $\mathbf{f}_1$, our DM starts from the equalized channel output $\tilde{\mathbf{f}}$, a noisy latent feature vector whose noise variance depends on the channel state. Therefore, determining the appropriate starting point is essential. We refer to this process as step matching, which calculates the corresponding timestep based on the SNR of $\tilde{\mathbf{f}}$. 
While the original approach in \cite{wu2024cddm} maps SNR values to a discrete timestep, it introduce a matching error because the SNR typically spans a continuous range. 
To address this, we treat the timestep as a continuous value and directly feed the noise level $\bar{\beta}_t$ into the DM instead of the discrete timestep, effectively mitigating the matching discrepancy.} 
Specifically, let $\gamma$ denote the SNR value, 
the current timestep is given by
\begin{align}\label{eq: step matching}
	m=\mathcal{S}^{-1}\bigg(\frac{1}{1+\gamma}\bigg)=\mathcal{S}^{-1}\bigg(\frac{\sigma^2}{\sigma^2+|h|^2}\bigg), 
\end{align}
\rv{where $\mathcal{S}^{-1}(\cdot)$ denotes the inverse function of the scheduling function in (\ref{eq: noise scheduler definition}).
}\rv{Once $m$ is determined, we can perform iterative denoising with DM, as summarized in Algorithm \ref{algo:sampling algorithm for diffusion}.}
\vspace{-2mm}
\begin{figure}[H] 
	\centering
		\captionsetup{skip=2pt}
	\includegraphics[width=0.9\linewidth]{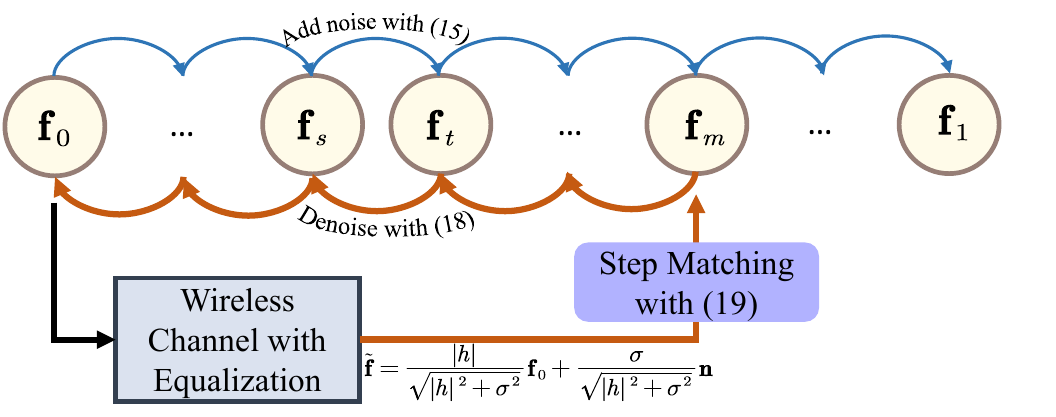}
	\vspace{3mm}
	\caption{{Diffusion Denoising Procedure.}}
	\vspace{-4mm}
	\label{Fig:diffusion denoising}
\end{figure}
\begin{remark}
\emph{Using diffusion for preprocessing the channel output, rather than post-processing the DeepJSCC output as in \cite{chen2024commin,yilmaz2023high}, offers some new advantages. While DM for post-processing can leverage the prior knowledge from the learned data distribution to counteract distortion, it struggles to accurately and efficiently characterize the distortion caused by JSCC decoder and wireless channel. The preprocessing approach, however, moves the DM model before the JSCC decoder, requiring the DM to handle only the distortion introduced by the wireless channel. The channel-induced distortion can be naturally interpreted as an intermediate state within the diffusion process. This alignment with the diffusion process highlights the inherent suitability of using DM for preprocessing. Moreover, by adopting a preprocessing strategy, channel adaptation can be delegated entirely to the DM model. As a result, the DeepJSCC model itself can be fixed and does not need further fine-tuning for specific channel environments or semantic quality metrics, while only the DM needs to be personalized and tailored. Once the JSCC model is trained, various DMs can be trained on it and flexibly integrated into the DeepJSCC framework in a plug-in manner. Moreover, compared to the existing preprocessing methods \cite{wu2024cddm}, we introduce semantics guidance for denoising and address the step-matching errors by adopting a continuous timestep setting, which further improves its applicability in DeepJSCC.}
\end{remark}
\begin{figure}[t]
	\label{alg:LSB_}
	\vspace{-4mm}
	\begin{algorithm}[H]
	  \caption{Sampling algorithm of the denoising DM}\label{algo:sampling algorithm for diffusion}
	  \begin{algorithmic}[1]
		\renewcommand{\algorithmicrequire}{ \textbf{Input:}}
		\REQUIRE {Channel output $\hat{\mathbf{y}}$,} $\bar{\beta}_t$
		\renewcommand{\algorithmicrequire}{ \textbf{Output:}}
		\REQUIRE The denoised latent feature $\mathbf{f}_0$
		\STATE {\textbf{Step matching:}} \\
  Calculate the current timestep $m$ with (\ref{eq: step matching}). 
		\STATE $\mathbf{f}_m=\hat{\mathbf{y}}$. 
		\STATE {\textbf{Initialize:}} $t=m$. 
		\WHILE{$t>0$} 
		\STATE $s=t-\frac{m}{T}$. 
		\STATE Calculate $\bar{\beta}_t$, $\bar{\beta}_t$ with (\ref{eq: noise scheduler definition}). 
		\STATE $\mathbf{f}_{s}=\sqrt{\frac{\bar{\beta}_{s}}{\bar{\beta}_t}}\mathbf{f}_t+\bigg(\sqrt{1-\bar{\beta}_{s}}-\sqrt{\frac{\bar{\beta}_{s}(1-\bar{\beta}_{t})}{\bar{\beta}_t}}\bigg)\boldsymbol{\epsilon}_{\boldsymbol{\Omega}}(\mathbf{f}_t,\bar{\beta}_{t}).$
		\STATE $t=s$. 
		\ENDWHILE
	  \end{algorithmic}
	\end{algorithm}
	\vspace{-10mm}
\end{figure}
	\begin{figure*}[t] 
		\centering
		\includegraphics[width=1.0\linewidth]{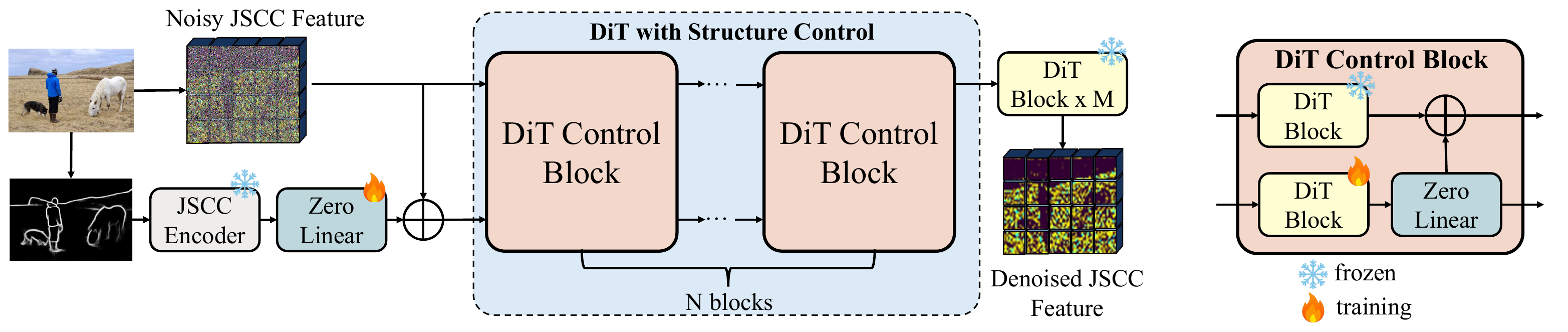}
		\vspace{0mm} 
		\caption{{Block diagram of the proposed semantics-guided diffusion framework.}}\label{fig: structural semantics guided diffusion framework}
	\vspace{-6mm}
	\end{figure*}

\vspace{-2mm} 
 \subsubsection{Network Design}
	\rv{We employ the diffusion transformer (DiT) model as the main architecture, which has been widely adopted in visual generation tasks and demonstrated its superior scalability and training efﬁciency compared to CNNs \cite{peebles2023scalable}.} As shown in Fig. \ref{fig: network architecture of diffusion model}, the base DM facilitates guidance through text descriptions. In each denoising iteration, the noisy feature is passed through a sequence of DiT blocks, yielding a ``cleaner’’ JSCC feature. The detailed architecture of DiT block is depicted on the right hand side of Fig. \ref{fig: network architecture of diffusion model}, where we adopt the same DiT block as in \cite{chen2023pixart}.  \rv{The timestep information, represented by $\bar{\beta}_t$, is projected onto each DiT block through the adaptive layer norm (ADALN) mechanism \cite{peebles2023scalable}, which performs modulation operations like scale $y=\alpha x$ and scale plus shift $y=\gamma x + \beta$.} The text information is first encoded with the CLIP model and then serves as the key and value in the cross-attention layer. 

	Moreover, as revealed in \cite{gao2023mdtv2}, incorporating masked data enhances the denoising and generation capabilities of the DM. Given this, we introduce masked data as an auxiliary input to boost performance. Specifically, during the training process, given the latent feature $\mathbf{f}_t$ and timestep $t$, we randomly mask some patches of the latent feature, yielding $\hat{\mathbf{f}}_t$. Both $\mathbf{f}_t$ and $\hat{\mathbf{f}}_t$ are then fed into the diffusion transformer. 
\rv{	Since some patches are dropped in $\hat{\mathbf{f}}_t$, the unmasked patches are first flattened and undergo the position embedding operation. 
}	After passing through $N_1$ diffusion transformer (DiT) blocks, a side interpolator, which has the same structure as a DiT block, is introduced to predict the masked tokens. Finally, the loss function is given by
\begin{align}
	\mathcal{L}_{\rm DM}(\mathbf{f}_t,\mathbf{f}_0,\bar{\beta}_t)=\|\boldsymbol{\epsilon}_{\boldsymbol{\Omega}}(\mathbf{f}_t,\bar{\beta}_{t})-\mathbf{f}_0\|^2+\|\boldsymbol{\epsilon}_{\boldsymbol{\Omega}}(\hat{\mathbf{f}}_t,\bar{\beta}_{t})-\mathbf{f}_0\|^2.
	\end{align}
\vspace{-5mm} 

	\textbf{Controlnet for semantics guided denoising:}
	As discussed in Section \ref{subsec: edge map extraction and transmission}, semantics can also be in a more structural form. The text-guided DM illustrated in Fig. \ref{Fig:diffusion model} does not readily support edge maps as a finer form of guidance. 
	To address this limitation, we integrate ControlNet \cite{zhang2023adding,chen2024pixart} to utilize edge maps as guidance during the denoising process. Specifically, as depicted in Fig. \ref{fig: structural semantics guided diffusion framework}, the first $N$ DiT blocks are replaced with DiT control blocks, which independently process the image features and structural semantic features, subsequently combining them using a weighted sum operation. 
	Each DiT control block consists of two DiT blocks and a ``zero'' linear layer. 
	The two DiT blocks are exact copies of those in the well-trained text-guided DM in Fig. \ref{Fig:diffusion model}, ensuring the preservation of the prior knowledge embedded in the original model. 
	The zero linear layer is a linear layer initialized with parameters set to zero, allowing it to adaptively learn how to fuse the structural guidance with the image features. 
	During the training of the DiT control blocks, 
	the parameters of the original text-guided DM are frozen, and only the parameters of the DiT blocks handling the structural semantic features are updated. This approach preserves the prior knowledge embedded in the original text-guided DM while enabling the model to incorporate structural information.
	\begin{figure}[b]
		\centering
		\vspace{-6mm}
			\captionsetup{skip=2pt}
		\subfigure[previous scheme \cite{wu2024cddm}]{
			\label{Fig:step matching:a} 
			\includegraphics[width=0.7\linewidth]{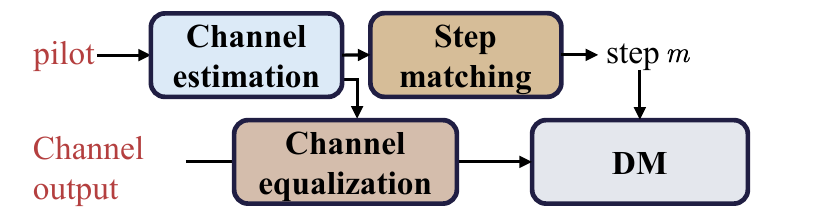}}\\
		\subfigure[proposed scheme]{
			\label{Fig:step matching:b} 
	\includegraphics[width=0.7\linewidth]{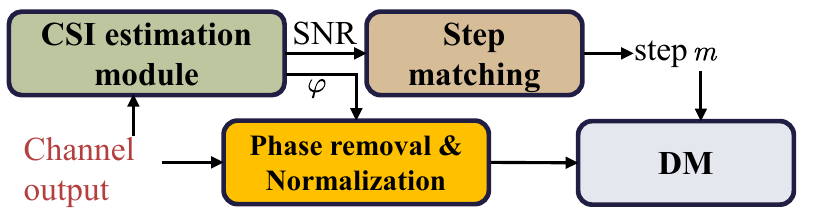}}
		\caption{{Comparison of step matching method.}}\label{fig: step matching}
	\vspace{-4mm}
	\end{figure}
	\vspace{-5mm}
	\subsection{Pilot-free step matching}\label{subsec: snr-estimation free step matching}
	Using DM for preprocessing of the noisy channel output can naturally adapt to dynamic wireless environments with the assistance of step matching (\ref{eq: step matching}).
	However, it necessitates the signals to be the form of (14) and the corresponding instantaneous SNR, which requires the knowledge of 
	the phase term $\phi$ and the equivalent SNR $\frac{|h|^2}{\sigma^2}$. 
	This, however, usually requires pilot transmission and introduce additional overhead as in Fig. \ref{Fig:step matching:a}. 
	Instead, we propose a pilot-free scheme for the slow-fading channel scenarios, that is, we directly estimate $\phi$ and the equivalent SNR from the channel output. 
	As shown in Fig. \ref{Fig:step matching:b}, 
\rv{	The noisy latent feature is first normalized with its L2 norm, yielding the resulting normalized noisy feature, which can be expressed as follows: 
\vspace{-2mm}	
\begin{align}
		\bar{\mathbf{f}}=\sqrt{\alpha}\mathcal{R}(e^{j\phi}\mathcal{C}(\mathbf{f}))+\sqrt{1-\alpha}\mathbf{n},
	\end{align}
	where $\alpha=\frac{|h|^2}{{|h^2|+\sigma^2}}$ denotes the signal level of $\bar{\mathbf{f}}$. 
	Then we utilize neural networks to estimate the $\phi$ and $\alpha$ separately. 
}	\subsubsection{SNR Estimation}For SNR estimation, we assume the perfect phase is available. In this case, we can conduct phase removal as in (\ref{eq: phase remove with gt phase}), and we have $e^{-j\phi}\bar{\mathbf{f}}\sim \mathcal{N}(\sqrt{\alpha}\mathbf{f},(1-\alpha)\mathbf{I})$. The training objective for SNR estimation is given by 
\vspace{-1mm}	
\begin{align}\label{eq: snr estimation}
		\min_{\mathcal{P}}~~\mathbb{E}[\|\zeta_{\mathcal{P}}(\sqrt{\alpha}\mathbf{f}+\sqrt{1-\alpha}\mathbf{n})-\alpha\|_2^2], 
	\end{align}
	where $\zeta_{\mathcal{P}}(\cdot)$ denotes the estimation function, with $\mathcal{P}$ being the trainable parameters in the estimation module. 
	
	\begin{table}[t]
		\centering  %
		\caption{Parameters of the SNR 
		estimation module.}
		\vspace{3mm}
		\setlength{\tabcolsep}{1.5mm}{\begin{tabular}{ll}
		\hline
		Layer& Parameter\\
		\hline
		Layer 1 (input)& \begin{minipage}{4cm} \vspace{0.1cm}Input of size $c\times h \times w$, batch  of size 128, 100 epochs \vspace{0.1cm}\end{minipage} \\
		Layer 2 (residual block)&  \begin{minipage}{4cm} \vspace{0cm}output of size $32\times\frac{h}{2}\times\frac{w}{2}$\vspace{0.1cm}\end{minipage} \\
		Layer 3 (residual block)&  \begin{minipage}{4cm} \vspace{0cm}output of size $64\times\frac{h}{4}\times\frac{w}{4}$\vspace{0.1cm}\end{minipage} \\
		Layer 4 (residual block)&  \begin{minipage}{4cm} \vspace{0cm}output of size $128\times\frac{h}{8}\times\frac{w}{8}$\vspace{0.1cm}\end{minipage} \\
		Layer 5 (residual block)&  \begin{minipage}{4cm} \vspace{0cm}output of size $256\times\frac{h}{16}\times\frac{w}{16}$\vspace{0cm}\end{minipage} \\
		Layer 6 (average pool)& output of size $256\times1 \times 1$\\
		Layer 7 (flatten) \\
		Layer 9 (fully-connected)& $1$ output neuron\\
		Layer 10 (activation)  & Sigmoid\\
		Layer 11 output layer &  \begin{minipage}{4cm} \vspace{0cm}Adam optimizer, learning rate of 0.001, MSE metric\vspace{0cm}\end{minipage} \\
		\hline
		\end{tabular}}
		\label{model structure of prediction model}
		\vspace{-6mm}
		\end{table}
	Then we detail the model architecture of the SNR estimation module. 
	We note that \rv{direct SNR estimation from the received signal has been explored in previous work \cite{yang2019snr}, where the quadrature amplitude modulation (QAM) signal is first flattened into a vector and then processed using convolutional neural networks (CNNs) to estimate the SNR. In our framework, we take a different approach by leveraging the distributional discrepancy between the entire latent representation and the Gaussian noise, rather than focusing on the per-symbol distribution differences between the QAM signal and Gaussian noise. Consequently, as shown in Table \ref{model structure of prediction model}, we preserve the latent features in their original two-dimensional form and utilize 2D residual blocks to extract features from the noisy latent representation. This is followed by a fully connected layer, and the final output is produced by a sigmoid activation function.
	}
	\subsubsection{Phase Estimation}
\rv{For phase estimation, similarly, we assume the equivalent SNR ($\frac{\alpha}{1-\alpha}$) is known. The training objective is given by
\vspace{-2mm}
	\begin{align}
		\min_{\mathcal{Q}}\mathbb{E}\left[\left\|\xi_{\mathcal{Q}}(\sqrt{\alpha}\mathcal{R}(e^{j\phi}\mathcal{C}(\mathbf{f}))+\sqrt{1-\alpha}\mathbf{n},\alpha)-\frac{\phi}{\pi}\right\|_2^2\right],
		\end{align}
		where $\xi_{\mathcal{Q}}(\cdot)$ denotes the phase estimation function, with $\mathcal{Q}$ being the trainable parameters in the estimation module. 
		We adopt a similar network structure as the SNR estimation module, while integrating a attention feature modules \cite{xu2021wireless} after each each residual block to project SNR information into the phase estimation module.
}		
	\subsubsection{Joint Estimation}
			\begin{figure}[t]
				\label{alg:LSB}
				\begin{algorithm}[H]
				  \caption{Joint phase and SNR estimation}\label{algo:CSI estimation}
				  \begin{algorithmic}[1]
					\renewcommand{\algorithmicrequire}{ \textbf{Input:}}
					\REQUIRE the normalized noisy feature $\bar{\mathbf{f}}$.
					\renewcommand{\algorithmicrequire}{ \textbf{Output:}}
					\REQUIRE equalized channel output $\tilde{\mathbf{f}}$, $\alpha$.
					\STATE $\phi=0$
					\WHILE{not converged or maximum iterations not reached} 
					\STATE Remove phase: $\bar{\mathbf{f}}\leftarrow e^{-j\phi}\bar{\mathbf{f}}$. 
					\STATE SNR estimation: $\alpha=\zeta_{\mathcal{P}}(\bar{\mathbf{f}})$. 
					\STATE Phase estimation: $\phi=\xi_{\mathcal{Q}}(\bar{\mathbf{f}},\alpha)$. 
					\ENDWHILE
					\STATE $\tilde{\mathbf{f}}\leftarrow \bar{\mathbf{f}}$.
				  \end{algorithmic}
				\end{algorithm}
				\vspace{-11mm}
			\end{figure} 
\rv{The SNR and phase estimation modules described above operate under the assumption that one parameter is known while estimating the other. In this part, we investigate the joint estimation approach. Specifically, we begin by initializing $\varphi$ to 0 and then alternate between phase removal, signal-to-noise ratio estimation, and phase estimation until convergence is achieved or the maximum number of iterations is exceeded. The detailed procedure is outlined in Algorithm \ref{algo:CSI estimation}.
}	
\vspace{-4mm}
\section{Extension to Fast Fading Case}\label{sec: extension to rayleigh fading case}
Up to this point, semantics-guided DMs have shown to be a promising solution for slow fading and AWGN scenarios. 
The next question is \emph{whether the DM trained under an AWGN channel can be directly applied to a fast fading scenario, without further training or specific fine-tuning}. 
In a fast fading scenario, the $i$-th element of the channel output $\mathbf{y}$ is given by $y_i=h_iz_i + n_i$. \rv{We assume perfect CSI at the receiver, i.e., $\mathbf{h}=[h_1,...,h_M]$.}
At the receiver side, $\mathbf{y}$ is first processed by the MMSE channel equalization and normalization (i.e., $\frac{h_i^*}{|h_i|\sqrt{|h_i|^2+\sigma^2}}y_i$), then transformed from a complex vector into a real vector, yielding the resulting latent feature $\tilde{\mathbf{f}}$. 
Let $[\mathbf{e}]_i$ denote the $i$-th element of  the vector $\mathbf{e}$, and $\mathbf{f}_0=\mathbf{f}$ be the desired latent feature.  
Then $[\tilde{\mathbf{f}}]_i$ can be expressed as 
\begin{align}\label{eq: received signal in fading channel}
[\tilde{\mathbf{f}}]_i=\sqrt{1-d_i}{[\mathbf{f}_0]_i}+\sqrt{d_i}[\mathbf{n}]_i, 
\end{align}
where $d_i = \frac{\sigma^2}{{|h_{c,i}|^2+\sigma^2}}$, and $h_{c,i} = \left\{\begin{array}{l}
	h_i, i\leq N/2 \\
	h_{i-N/2}, {\rm else}
\end{array}
\right.$.
\begin{figure}[t]
	\centering
	\vspace{-4mm}
		\captionsetup{skip=2pt}
\subfigure[Signal and noise levels in channels]{
		\label{Fig:fading case:a} 
		\includegraphics[width=0.48\linewidth]{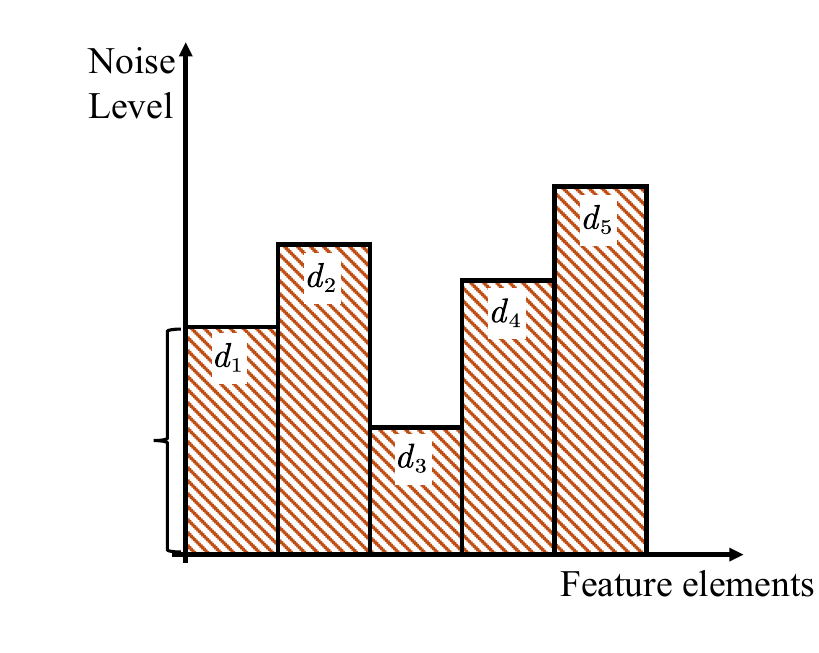}}
	\subfigure[The proposed denoising scheme]{
		\label{Fig:fading case:b} 
		\includegraphics[width=0.48\linewidth]{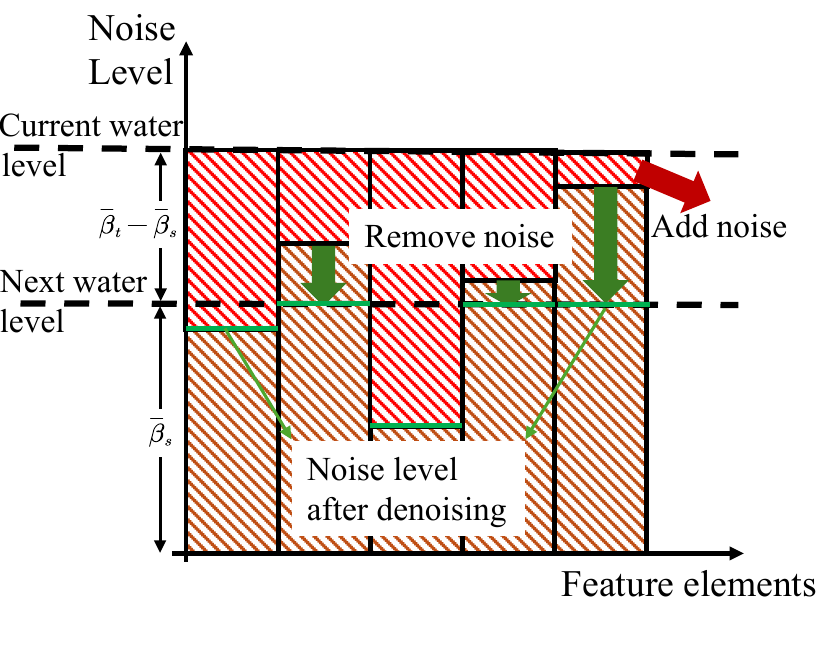}}
	\caption{{Adaptation to the fast fading channel.}}\label{fig: fading case}
	\vspace{-5mm}
\end{figure} 

As illustrated in Fig. \ref{Fig:fading case:a}, no direct intermediate state as described in (\ref{eq: forward trajectory of diffusion process})  can be used to facilitate step matching for $\tilde{\mathbf{f}}$, since different elements experience varying SNR levels due to the distinct fading coefficients. To address this, We draw inspiration from water-filling: \rv{We manually add Gaussian noise with a carefully chosen variance to elements which have lower noise level than the current target noise level.} Specifically,  let $t$ and $s$ represent the current and the next target timesteps ($s<t$), respectively. The diffusion denoising step refers to inferring $\mathbf{f}_s$ from $\mathbf{f}_t$ using DM. 
\rv{Denote $\mathbf{b}_t$ as the exact noise level vectors of $\mathbf{f}_t$, which indicates that $p([\mathbf{f}_t]_i|[\mathbf{f}_0]_i)\sim\mathcal{N}(\sqrt{1-[\mathbf{b_t}]_i}[\mathbf{f}_0]_i,[\mathbf{b_t}]_i)$. For the initial time step $m$, $\mathbf{b}_m$ is given by $[\mathbf{b}_m]_i=d_i$. Denote $\bar{\beta}_t$ as the target noise level at the time step $t$, which can be calculated by (\ref{eq: noise scheduler definition}).}
We add noise to each $[\mathbf{f}_t]_i$, equalizing the noise level of all elements to $\bar{\beta}_t$, as follows. 
\begin{align}\label{eq: wather filling}
[\mathbf{g}_t]_i=\sqrt{\frac{1-\bar{\beta}_t}{1-[\mathbf{b}_t]_i}}[\mathbf{f}_t]_i+\sqrt{\bar{\beta}_t-[\mathbf{b}_t]_i\frac{1-\bar{\beta}_t}{1-[\mathbf{b}_t]_i}}\epsilon,  
\end{align}
where $\epsilon \sim \mathcal{N}(0,1)$ denotes an instance of standard Gaussian noise.  
Therefore, the conditional distribution of $\mathbf{g}_t$ given $\mathbf{f}_0$ is given by $p(\mathbf{g}_t|\mathbf{f}_0)\sim \mathcal{N}(\sqrt{1-\bar{\beta}_t}\mathbf{f}_0,\bar{\beta}_t\mathbf{I})$, which indicates that all $[\mathbf{g}_t]_i$ have the same noise level. 
 $\mathbf{g}_t$ thus can be fed into DM trained in slow fading channels for denoising, 
yielding 
\vspace{-2mm}
\begin{align}\label{eq: denoising in fast fading channel}
\hat{\mathbf{g}}_s=\sqrt{\frac{\bar{\beta}_s}{\bar{\beta}_t}}\mathbf{g}_t+\left(\sqrt{1-\bar{\beta}_s}-\sqrt{\frac{\bar{\beta}_s(1-\bar{\beta}_t)}{\bar{\beta}_t}}\right)\boldsymbol{\epsilon}_{\boldsymbol{\Omega}}(\mathbf{g}_t,\bar{\beta}_t). 
\end{align}
As for the updating process, for a specific element $[\mathbf{f}_t]_i$ with its noise level $[\mathbf{b}_t]_i$ and the current noise level $\bar{\beta}_t$, there are two cases\footnote{Note that, throughout all denoising iterations, the condition $[\mathbf{b}_t]_i\leq \bar{\beta}_t, \forall i$ is always satisfied. Based on the updating method, we observe that $[\mathbf{b}_s]_i \leq \bar{\beta}_s$ if $[\mathbf{b}_t]_i \leq \bar{\beta}_t$. This condition can be ensured by setting the initial noise level $\bar{\beta}_1$ to be the maximum noise level in the channel.}.
\begin{itemize}
\item $\bar{\beta}_s\leq[\mathbf{b}_t]_i\leq \bar{\beta}_t$, which means that the current noise level is higher than the next target noise level. Therefore, $[\mathbf{f}_s]_i$ and its corresponding $[\mathbf{b}_s]_i $needs to be updated, i.e., $[\mathbf{f}_s]_i=\hat{\mathbf{g}}_s$, and $[\mathbf{b}_t]_i =\bar{\beta}_s$. 

\item $[\mathbf{b}_t]_i < \bar{\beta}_s$, which means that the current noise level is already lower than the next target level, indicating that $[\mathbf{f}_t]_i$ is ``cleaner'' than $[\hat{\mathbf{g}}_t]_i$. Therefore, we leave these elements unchanged, i.e., $[\mathbf{f}_s]_i=[\mathbf{f}_t]_i$, and $[\mathbf{b}_s]_i=[\mathbf{b}_t]_i$. 
\end{itemize}
The detailed denoising algorithm for fast fading channels is outlined in Algorithm \ref{alg: fading case}. 
\begin{figure}[h]
	\vspace{-8mm}
	\begin{algorithm}[H]
	  \caption{The proposed denoising algorithm for fast fading channels.}\label{alg: fading case}
	  \begin{algorithmic}[1]
		\STATE {\textbf{Input:}} the noisy latent feature $\tilde{\mathbf{f}}$ in (\ref{eq: received signal in fading channel}).  
		\STATE {\textbf{Output:}} the denoised latent feature $\hat{\mathbf{f}}_0$.
		\STATE {\textbf{Initialize:}} $t=\mathcal{S}^{-1}(\max\{d_i,\forall i\})$. $\mathbf{f}_t=\tilde{\mathbf{f}}$, $[\mathbf{b}_t]_i=d_i, \forall i$.
		\WHILE {$t>0$}
		\STATE $s=t-\frac{1}{T}$
		\STATE Calculate $\bar{\beta}_s$, $\bar{\beta}_t$ with (\ref{eq: noise scheduler definition}).
		\STATE Conduct water filling, and obtain $\mathbf{g}_t$ with (\ref{eq: wather filling}). 
		\STATE Conduct denoising with DM, and obtain $\hat{\mathbf{g}}_s$ with (\ref{eq: denoising in fast fading channel}).
		\STATE \# Update the latent feature 
		\FOR {$i=1,..., N$}
		\IF {$[\mathbf{b}_t]_i\geq {\bar{\beta}_s}$}
		\STATE $[\mathbf{f}_s]_i= [\hat{\mathbf{g}}_s]_i$, $[\mathbf{b}_s]_i={\bar{\beta}_s}$. 
		\ELSE
		\STATE $[\mathbf{f}_s]_i=[\mathbf{f}_t]_i$, $[\mathbf{b}_s]_i=[\mathbf{b}_t]_i$. 
		\ENDIF
		\ENDFOR
		\STATE $t=s$.
		\ENDWHILE
	  \end{algorithmic}
	\end{algorithm}
    \vspace{-9mm}
\end{figure}
\section{Numerical Results}\label{sec: numerical results}
In this section, we conduct a series of experiments to evaluate the performance of the proposed scheme, providing a comprehensive demonstration of its effectiveness across various scenarios.
\vspace{-2mm}
\subsection{Simulation Setup}
\textbf{Training Details:} The proposed SGD-JSCC scheme is trained in three stages. In the first stage, the JSCC encoder and decoder are jointly trained using the loss function in (\ref{eq: training loss function of JSCC autoencoder}) on the Imagenet dataset under a fixed channel setting (AWGN channel with SNR=$10$dB in our simulations). 
The JSCC model is fixed after this training stage. 
Then, in the second stage, we train the text-guided DM shown in Fig. \ref{fig: network architecture of diffusion model}, following the steps outlined in Algorithm \ref{algo:training algorithm for diffusion}. We collect approximately 14 million text-image pairs from various open datasets, including SA-1B \cite{kirillov2023segment}, JourneyDB \cite{sun2024journeydb}, CC3M \cite{changpinyo2021conceptual}, Datacomp \cite{gadre2024datacomp}, and CelebA-HQ \cite{zhu2022celebv}. With this diverse dataset, our DM is capable of understanding open-domain text descriptions and generating the corresponding visual data. All the images are center-cropped and resized to 
$128\times 128$. 
In the third stage, we incorporate edge maps as structural guidance for the well-trained text-guided DM obtained from stage two. The DMs in stage two and stage three are both trained with about $250,000$ gradient descent steps on a single NVIDIA A100 GPU, requiring about 2 GPU days. 
The training parameters and dataset composition for the second and third training stage are detailed in Table \ref{table:dataset parameters} and Table \ref{table:training parameters}, respectively. 

\textbf{Benchmark Schemes:} We compare the proposed SGD-JSCC scheme with three DeepJSCC-based schemes: ADJSCC \cite{xu2021wireless}, JSCCformer \cite{wu2024transformer}, DeepJSCC-Diff \cite{yilmaz2023high}, JSCCDiff \cite{yang2024diffusion}, and VAEJSCC. The ADJSCC scheme refers to the DeepJSCC architecture in \cite{erdemir2023generative}, which iteratively downsamples and upsamples image data using residual and attention blocks. The AF modules are integrated after each upsample block to incorporate  SNR information into the DeepJSCC network. 
\rv{Additionally, we also compare our method with DeepJSCC-Diff \cite{yilmaz2023high} and JSCCDiff \cite{yang2024diffusion}, which are two diffusion-based schemes that aim to improve the perceptual performance of DeepJSCC through post-processing.
}
The JSCCformer scheme refers to the JSCC architecture with vision transformer, which can also achieve SNR-adaptivity using a single model. 
Furthermore, we compare our SGD-JSCC method with VAEJSCC, a variation of the proposed SGD-JSCC scheme that does not use diffusion for denoising. VAEJSCC serves as a baseline to validate the effectiveness of our semantic-guided DM. 
\rv{{All schemes set their hyperparameters to ensure a CBR of $R=\frac{1}{20}$, and trained with the loss function in (\ref{eq: training loss function of JSCC autoencoder}) for a fair comparison \footnote{\rv{We note that incorporating either a discriminator or the LPIPS metric can enhance the perceptual performance. In this paper, to ensure a fair comparison, we adopt a consistent approach aligned with the training scheme of the SGD-JSCC model, utilizing the discriminator to improve the perceptual performance of the benchmark schemes. 
}}}. 
}For the proposed SGD-JSCC scheme, the transmission cost of the edge map and JSCC features in terms of CBR are set as $\frac{1}{24}$, $\frac{1}{120}$, respectively, resulting in a total CBR of $R=\frac{1}{20}$. 
\rv{The hyperparameters of scheduling function in (\ref{eq: noise scheduler definition}) are set to $e=3$, $s=0$, $\tau=0.7$.
}
\begin{table}[t]
	\centering
	\caption{Dataset and model parameters used in the second and third stage of training of SGD-JSCC.}
\begin{subtable}[Dataset composition]{
	\begin{tabular}{|l|l|}
		\hline
		training dataset         &  samples \\ \hline
		SA-1B & 7M                \\ \hline
		JourneyDB                & 3M                \\ \hline
		CC3M                     & 2M                \\ \hline
		Datacomp                 & 2M                \\ \hline
		Celeba-HQ                 & 30K                \\ \hline
		\end{tabular}
	\label{table:dataset parameters}}
\end{subtable}
\qquad
\begin{subtable}[Training parameters]{
	\begin{tabular}{|l|l|}
		\hline
		Parameters      & value \\ \hline
		number of channels $c$             & $16$    \\ \hline
		batch size      & $64$    \\ \hline
		embedding size   & $256$   \\ \hline
		CFG scalar & $4.5$   \\ \hline
		Guidance scalar & $0.3$   \\ \hline
		\end{tabular}
	\label{table:training parameters}}
	\end{subtable}
\vspace{-8.5mm}
\end{table}

\begin{figure*}[!]
	\centering
	\subfigure[PSNR, AWGN w/ SNR]{
		\label{Fig:awgn_wcsi:a} 
		\includegraphics[width=0.235\linewidth]{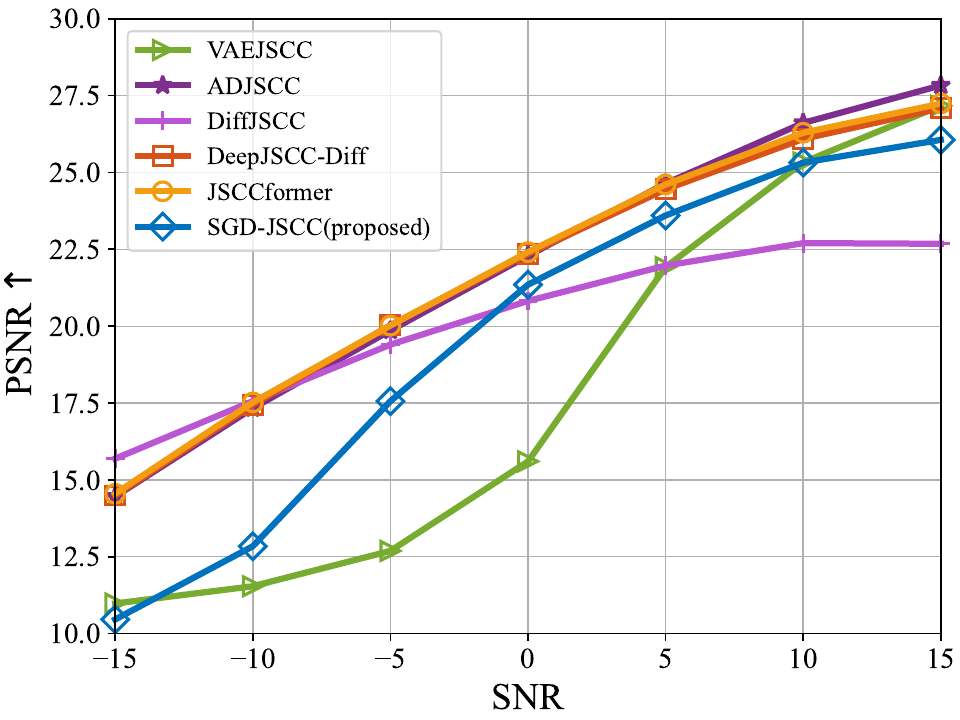}}
	\subfigure[LPIPS, AWGN w/ SNR]{
		\label{Fig:awgn_wcsi:b} 
		\includegraphics[width=0.235\linewidth]{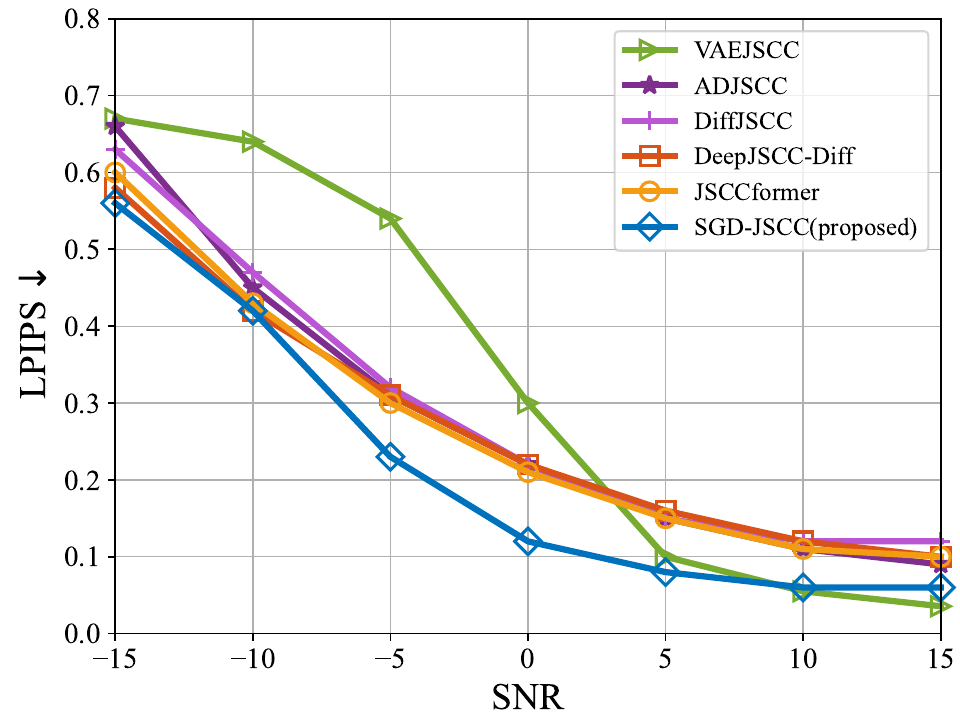}}
		\subfigure[CLIP Score, AWGN w/ SNR]{
			\label{Fig:awgn_wcsi:c} 
			\includegraphics[width=0.235\linewidth]{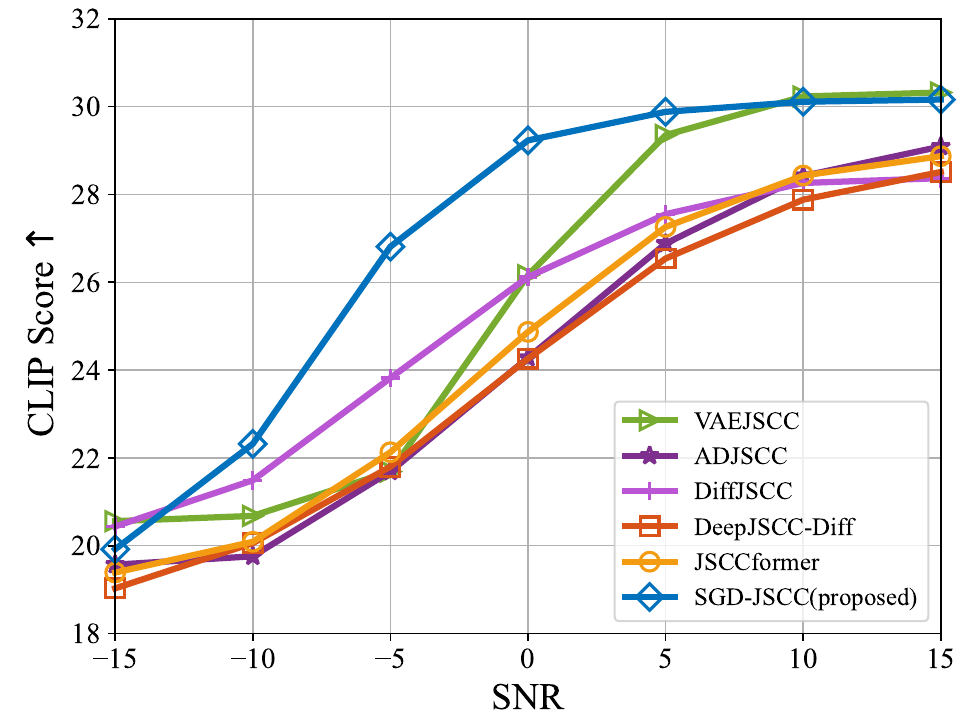}}						    
		\subfigure[FID, AWGN w/ SNR]{
		\label{Fig:awgn_wcsi:d} 
		\includegraphics[width=0.235\linewidth]{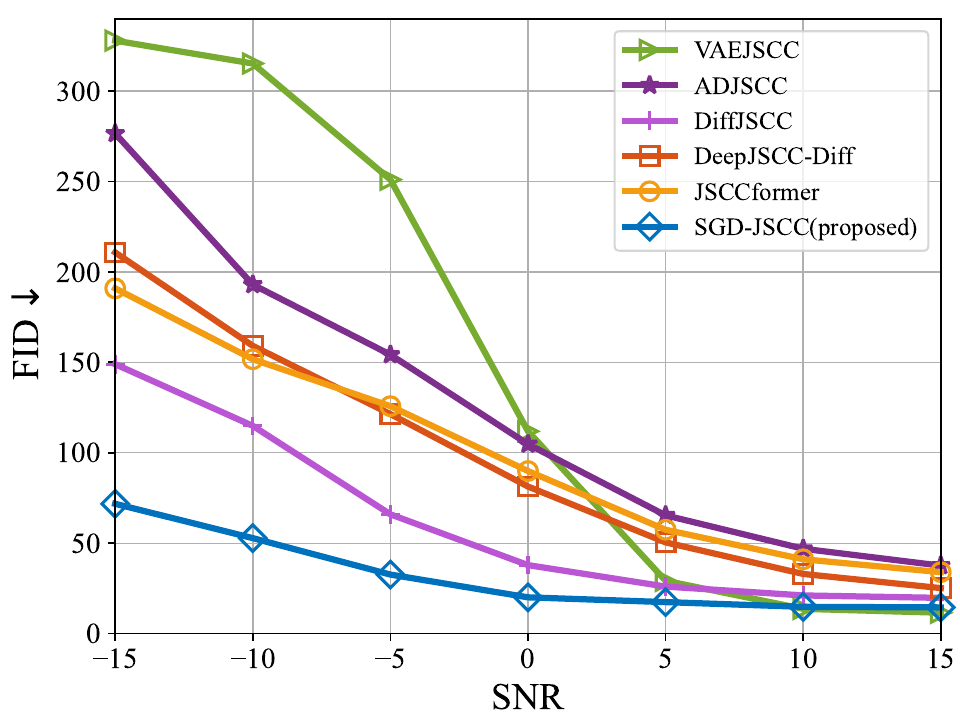}}						
	\caption{\rv{Average reconstruction performance over AWGN channels with receiver-side SNR information.}}\label{Fig:performance evaluation of awgn with csi setting}
\vspace{-2mm}
\end{figure*}
\begin{figure*}[!]
	\centering
	\subfigure[PSNR, AWGN w/o SNR]{
		\label{Fig:awgn_wocsi:a} 
		\includegraphics[width=0.235\linewidth]{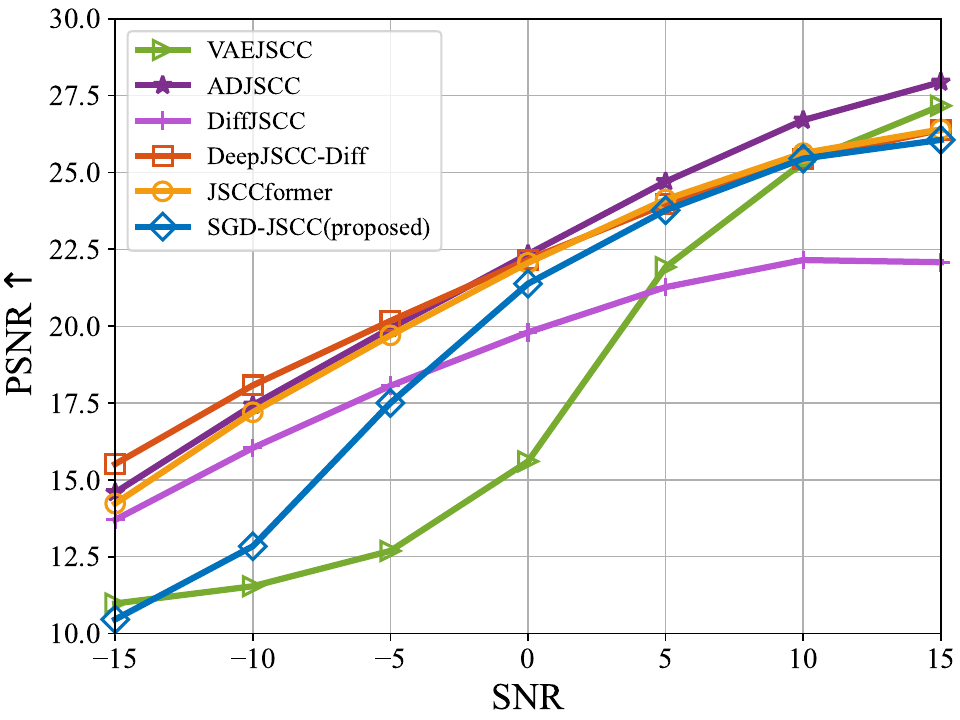}}
	\subfigure[LPIPS, AWGN w/o SNR]{
		\label{Fig:awgn_wocsi:b} 
		\includegraphics[width=0.235\linewidth]{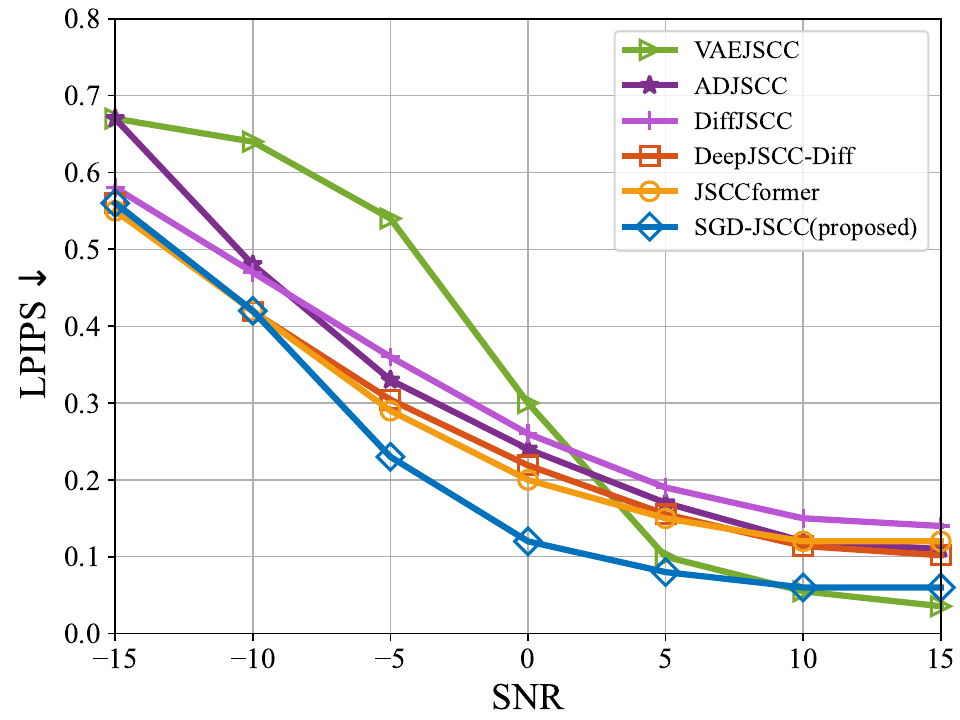}}
		\subfigure[CLIP Score, AWGN w/o SNR]{
			\label{Fig:awgn_wocsi:c} 
			\includegraphics[width=0.235\linewidth]{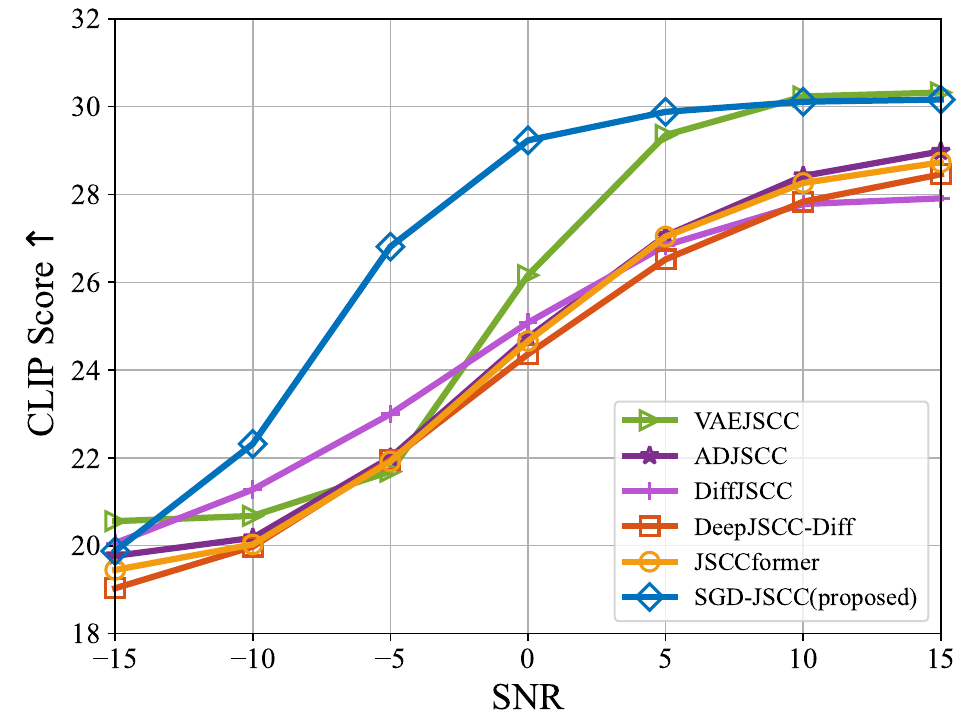}}						    
		\subfigure[FID, AWGN w/o SNR]{
		\label{Fig:awgn_wocsi:d} 
		\includegraphics[width=0.235\linewidth]{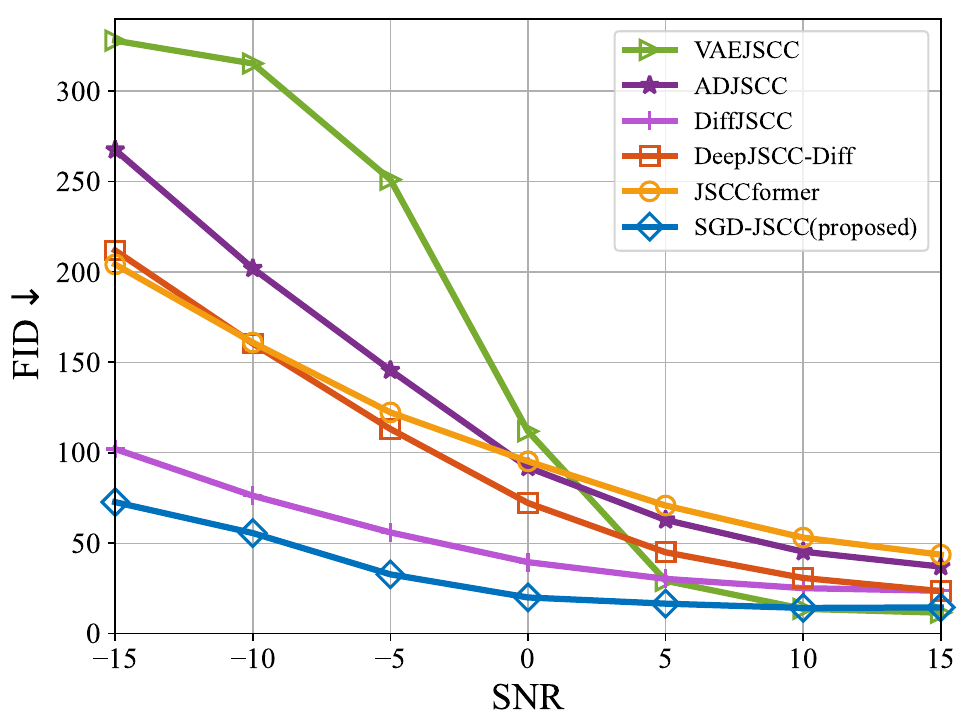}}			
	\caption{\rv{Average reconstruction performance over AWGN channels without SNR information.}}\label{Fig:awgn without csi}
	\vspace{-2mm}
\end{figure*}
\begin{figure*}[!]
	\centering
	\subfigure[PSNR, Fading w/o CSI]{
		\label{Fig:awgn_wocsi:a} 
		\includegraphics[width=0.235\linewidth]{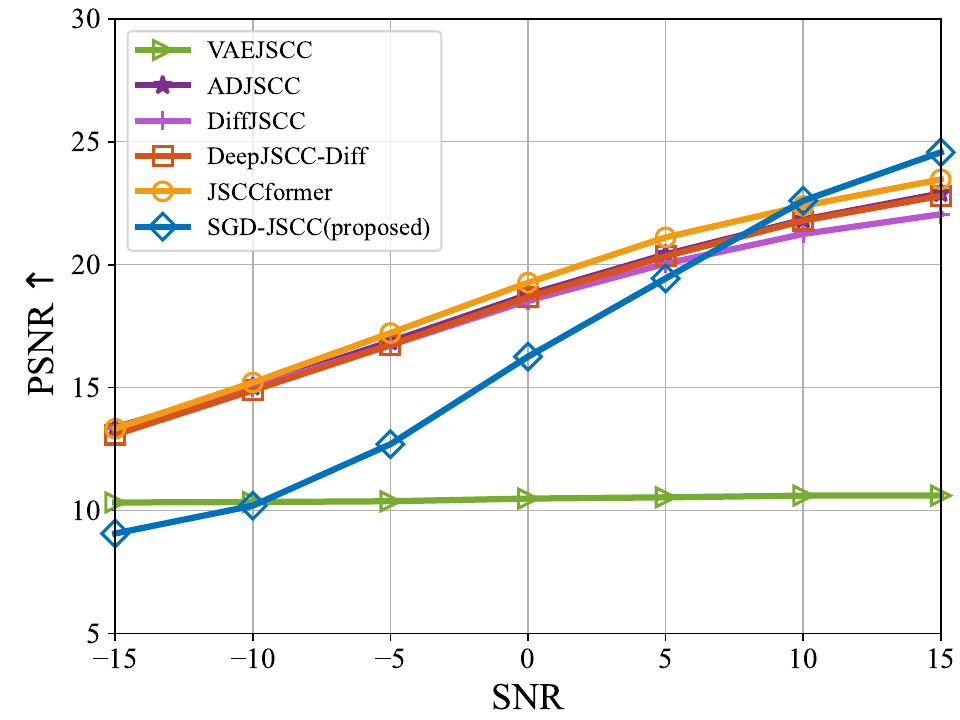}}
	\subfigure[LPIPS, Fading w/o CSI]{
		\label{Fig:awgn_wocsi:b} 
		\includegraphics[width=0.235\linewidth]{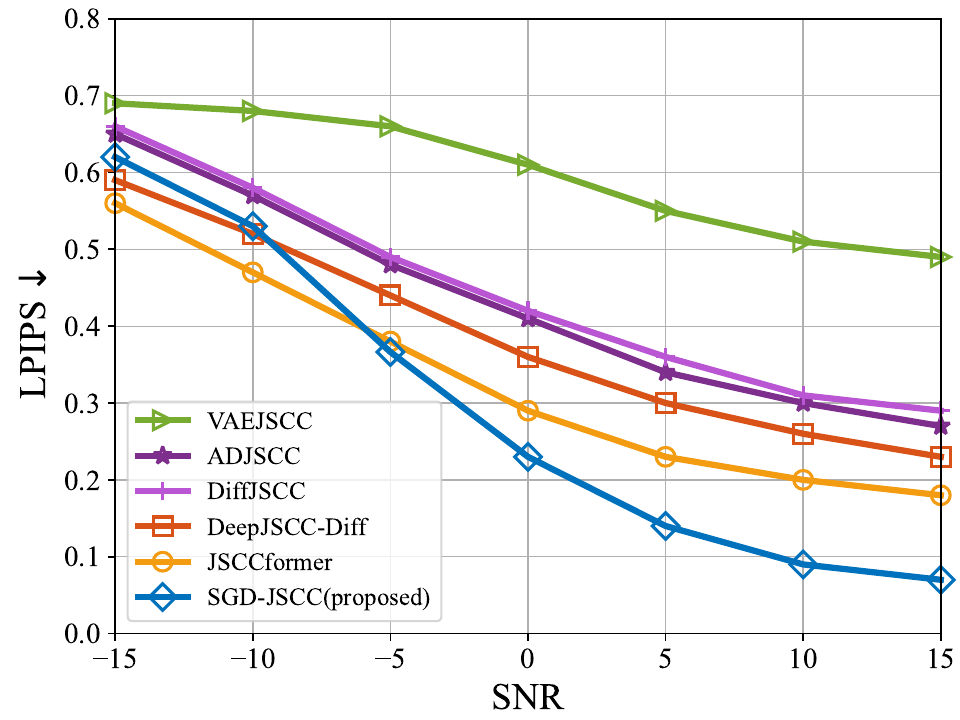}}
		\subfigure[CLIP Score, Fading w/o CSI]{
			\label{Fig:awgn_wocsi:c} 
			\includegraphics[width=0.235\linewidth]{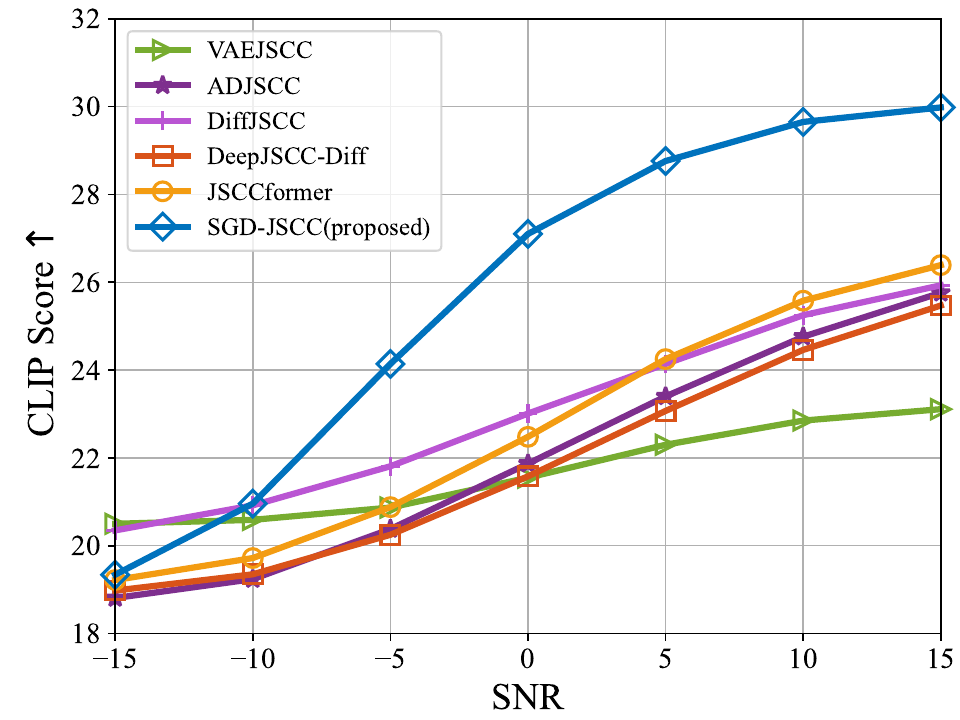}}						    
		\subfigure[FID, Fading w/o CSI]{
		\label{Fig:awgn_wocsi:d} 
		\includegraphics[width=0.235\linewidth]{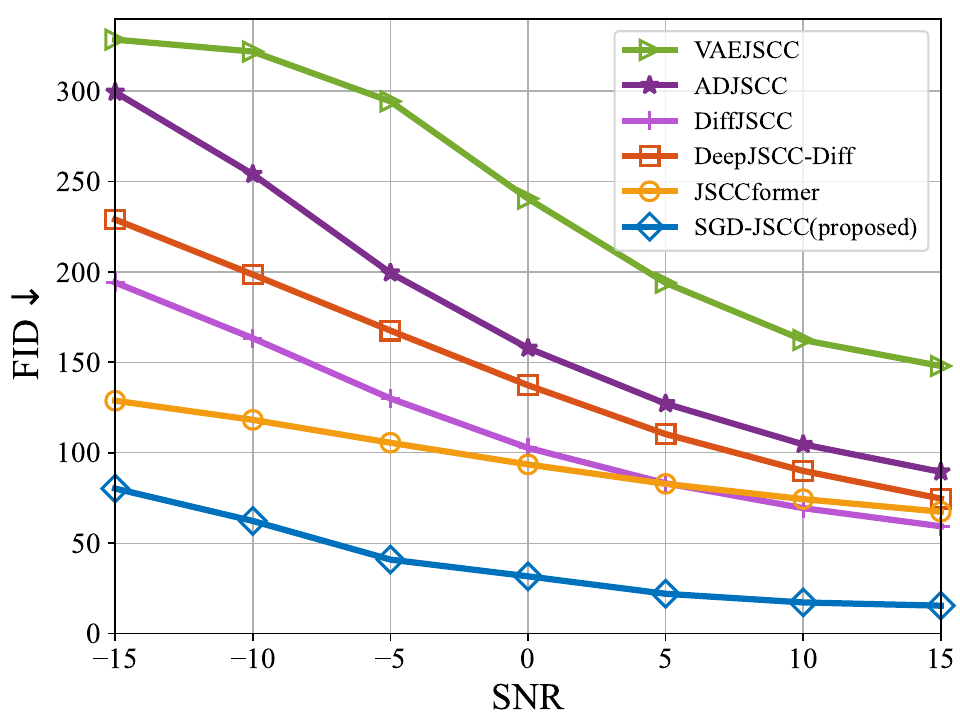}}			
	\caption{\rv{Average reconstruction performance over slow Rayleigh fading channels without CSI information.}}\label{Fig:fading without csi}
	\vspace{-4mm}
\end{figure*}

\textbf{Evaluation Dataset:} We adopt the COCO2017 dataset \cite{lin2014microsoft} for evaluation.  Specifically, for DeepJSCC and DeepJSCC-Diff, the COCO training set is used for training the JSCC models. 
We use the Imagenet dataset for training the DM used in the DeepJSCC-Diff scheme. 
Similarly, all the images are center-cropped and resized to $128\times 128$. The COCO validation set, consisting of $5,000$ images and their corresponding text descriptions, is used for evaluation. 
  
\textbf{Performance Metrics:} We employ the commonly used peak signal-to-noise ratio (PSNR) and learned perceptual image patch similarity (LPIPS) to evaluate the reconstruction performance. 
Additionally, perception is also a crucial aspect of image transmission, which the aforementioned metrics may not fully capture.
To address this, we introduce two additional metrics: CLIP score and Frechet inception distance (FID). The CLIP score measures the similarity between image and text descriptions. Since COCO2017 dataset already includes text descriptions for each image, we can evaluate the consistency between the reconstructed image and its corresponding ground-truth text description. FID assesses visual quality by calculating the statistical similarity between the original image set and the reconstructed image set.

\subsection{{Performance Evaluation over Slow Fading Channels}}
In this subsection, we compare the proposed scheme\footnote{For a fair comparison, we incorporate edge map guidance into the proposed SGD-JSCC scheme in this subsection. The effectiveness of text guidance will be evaluated in Section \ref{subsec: ablation study}.} with three benchmarks 
\rv{under three channel conditions: 1) AWGN channels with receiver-side SNR information ($h=1$, $\sigma^2$ known), 2) AWGN channels without SNR information ($h=1$, $\sigma^2$ unknown.), and 3) Rayleigh fading channels without CSI ($h\in \mathcal{CN}(0,1)$, $h$ and $\sigma^2$ unknown.).}
\subsubsection{AWGN channels with receiver-side SNR information}
\begin{figure*}[t]
	\begin{center}
		\resizebox{\textwidth}{!}
		{\begin{tabular}{ccccccc}
  {\large Original} & {\large VAEJSCC} & {\large ADJSCC(w. SNR)} & {\large ADJSCC(w/o SNR)} & {\large JSCC-Diff(w. SNR)}& {\large JSCC-Diff(w/o SNR)}& {\large SGD-JSCC}\\
  \includegraphics[width=0.2\textwidth]{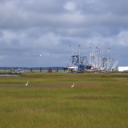}  &
  \includegraphics[width=0.2\textwidth]{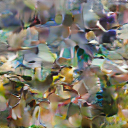}   &
  \includegraphics[width=0.2\textwidth]{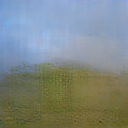} & 
  \includegraphics[width=0.2\textwidth]{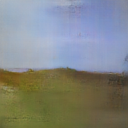} & 
  \includegraphics[width=0.2\textwidth]{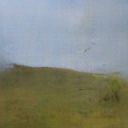} &
  \includegraphics[width=0.2\textwidth]{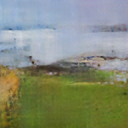} &
  \includegraphics[width=0.2\textwidth]{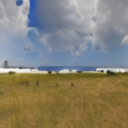} 
  \\
  \includegraphics[width=0.2\textwidth]{./Figures/selected/_15dB/proposed/origin_37.png}  &
  \includegraphics[width=0.2\textwidth]{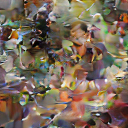}  &
  \includegraphics[width=0.2\textwidth]{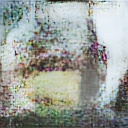}& 
  \includegraphics[width=0.2\textwidth]{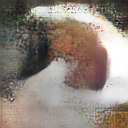}& 
  \includegraphics[width=0.2\textwidth]{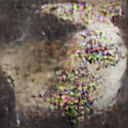} &
  \includegraphics[width=0.2\textwidth]{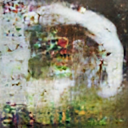} &
  \includegraphics[width=0.2\textwidth]{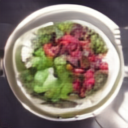} 
  \\
\end{tabular}}
   \end{center}
	  \caption{{Examples of reconstructed images under AWGN channel with SNR $=-15$ dB.}}
	  \label{fig:example of awgn reconstruction images snr=-15db}
	  \vspace{-1mm}
  \end{figure*}
  \begin{figure*}[htb]
	\begin{center} 
		\resizebox{\textwidth}{!}
		{\begin{tabular}{ccccccc}
  {\large Original} & {\large VAEJSCC} & {\large ADJSCC(w. SNR)} & {\large ADJSCC(w/o SNR)} & {\large JSCC-Diff(w. SNR)}& {\large JSCC-Diff(w/o SNR)}& {\large SGD-JSCC}\\
  \includegraphics[width=0.2\textwidth]{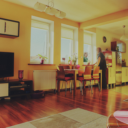}  &
  \includegraphics[width=0.2\textwidth]{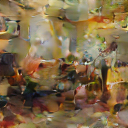}   &
  \includegraphics[width=0.2\textwidth]{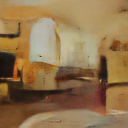} & 
  \includegraphics[width=0.2\textwidth]{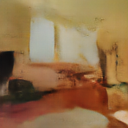} & 
  \includegraphics[width=0.2\textwidth]{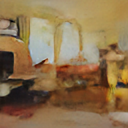} &
  \includegraphics[width=0.2\textwidth]{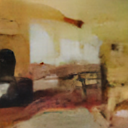} &
  \includegraphics[width=0.2\textwidth]{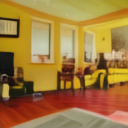} 
  \\
  \includegraphics[width=0.2\textwidth]{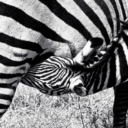}  &
  \includegraphics[width=0.2\textwidth]{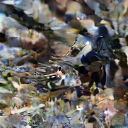}  &
  \includegraphics[width=0.2\textwidth]{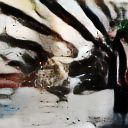}& 
  \includegraphics[width=0.2\textwidth]{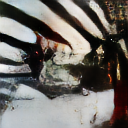}& 
  \includegraphics[width=0.2\textwidth]{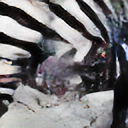} &
  \includegraphics[width=0.2\textwidth]{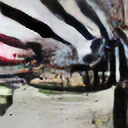} &
  \includegraphics[width=0.2\textwidth]{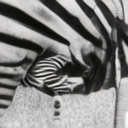} 
  \\
\end{tabular}}
   \end{center}
	  \caption{{Examples of reconstructed images under AWGN channel with SNR $=-5$ dB.}}
	  \label{fig:example of awgn reconstruction images snr=-5db}
	 \vspace{-1mm}
  \end{figure*}
  
  \begin{figure*}[htb]
	\begin{center}
		\resizebox{\textwidth}{!}
		{\begin{tabular}{ccccccc}
  {\large Original} & {\large VAEJSCC} & {\large ADJSCC(w. SNR)} & {\large ADJSCC(w/o SNR)} & {\large JSCC-Diff(w. SNR)}& {\large JSCC-Diff(w/o SNR)}& {\large SGD-JSCC}\\
  \includegraphics[width=0.2\textwidth]{./Figures/selected/5dB/proposed/origin_1.png}  &
  \includegraphics[width=0.2\textwidth]{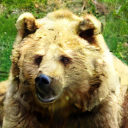}   &
  \includegraphics[width=0.2\textwidth]{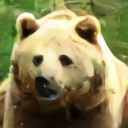} & 
  \includegraphics[width=0.2\textwidth]{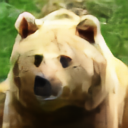} & 
  \includegraphics[width=0.2\textwidth]{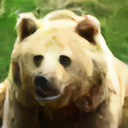} &
  \includegraphics[width=0.2\textwidth]{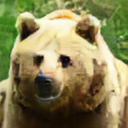} &
  \includegraphics[width=0.2\textwidth]{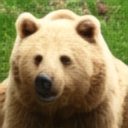} 
  \\
  \includegraphics[width=0.2\textwidth]{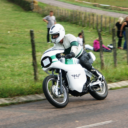}  &
  \includegraphics[width=0.2\textwidth]{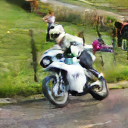}  &
  \includegraphics[width=0.2\textwidth]{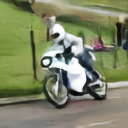}& 
  \includegraphics[width=0.2\textwidth]{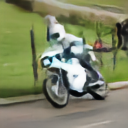}& 
  \includegraphics[width=0.2\textwidth]{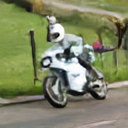} &
  \includegraphics[width=0.2\textwidth]{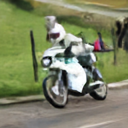} &
  \includegraphics[width=0.2\textwidth]{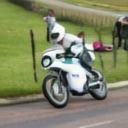} 
  \\
\end{tabular}}
   \end{center}
	  \caption{{Examples of reconstructed images under AWGN channel with SNR $=5$ dB.}}
	  \label{fig:example of awgn reconstruction images snr=5db}
	  \vspace{-3mm}
  \end{figure*}

We first evaluate the performance of the proposed scheme in a AWGN channel scenario where 
where only the receiver has the SNR information. 
The reconstruction performance is depicted in Fig. \ref{Fig:performance evaluation of awgn with csi setting}. First, it can be observed that the ADJSCC scheme has better PSNR performance. The proposed SGD-JSCC scheme improves the PSNR performance compared to VAEJSCC, with a gap of only 2.5dB compared to the state-of-the-art JSCC scheme. Second, in terms of LPIPS metric, the proposed SGD-JSCC scheme outperforms both the ADJSCC and JSCCformer schemes. It also achieves much better performance than VAEJSCC when SNR $\leq 5$ dB, validating the effectiveness of the semantics-guided DM in denoising. 
Third, generative models are naturally beneficial for improving perceptual performance, as measured by CLIP score and FID shown in Fig. \ref{Fig:awgn_wcsi:c} and Fig. \ref{Fig:awgn_wcsi:d}. The DeepJSCC-Diff scheme aims to first reconstruct a lower-resolution image, followed by a super-resolution process using DM. This approach results in better perceptual performance in the low SNR regime (i.e., SNR $\leq 5$ dB) compared to ADJSCC, especially in terms of the FID metric. 
\rv{For DiffJSCC, which refines the image with the fine-tuned DM, the perceptual performance is significantly improved compared to the ADJSCC scheme as measured by CLIP score and FID.  
However, the performance of DeepJSCC-Diff and DiffJSCC is limited by the preceding JSCC model. 
}
As as result, a performance floor occurs when SNR $>5$ dB. In contrast, by transforming the paradigm from post-processing the DeepJSCC output into preprocessing the channel output, the proposed SGD-JSCC scheme benefits from dynamically translating the eq. SNR to a specific intermediate state of diffusion process. SGD-JSCC significantly improves the perceptual quality of VAEJSCC in low SNR regime and retains performance in the high SNR regime. Moreover, by embracing semantics guidance and open-world text-image datasets, SGD-JSCC outperforms the DeepJSCC-Diff in terms of both reconstruction and perceptual performance. The demonstrated improvements highlight the superiority of the proposed SGD-JSCC scheme. 

\vspace{-3mm} 

	\subsubsection{AWGN channels without SNR information}\label{subsubsec: awgn without snr}
Next, we consider the AWGN scenario where neither the transmitter nor the receiver has access to SNR.
The results are presented in Fig. \ref{Fig:awgn without csi}. In this case, the AF modules in the benchmark JSCC models are removed, leading to a performance drop compared to the corresponding setups with CSI available at the receiver. 
{Interestingly, the performance of the proposed SGD-JSCC method remains nearly identical, whether the SNR is available or not. 
This robustness arises because our scheme does not require SNR at the transmitter for adaptive design, and it bypasses the need for precise SNR information at the receiver by directly estimating a sufficiently accurate from the noisy channel output and leveraging this estimate for step matching in the diffusion denoising process. This indicates that the proposed scheme is a promising solution in scenarios where SNR is unavailable or challenging to measure accurately. 
}
\begin{figure*}[!]
	\centering
	\captionsetup{skip=2pt}
	\subfigure[PSNR]{
		\label{Fig:awgn_wocsi:a} 
		\includegraphics[width=0.235\linewidth]{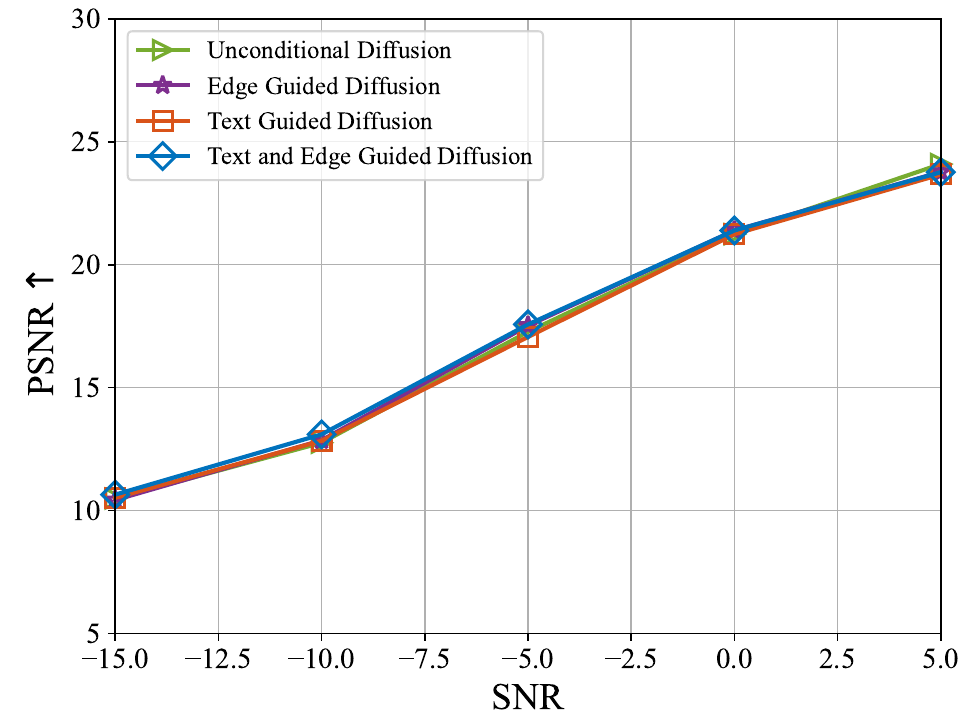}}
	\subfigure[LPIPS]{
		\label{Fig:awgn_wocsi:b} 
		\includegraphics[width=0.235\linewidth]{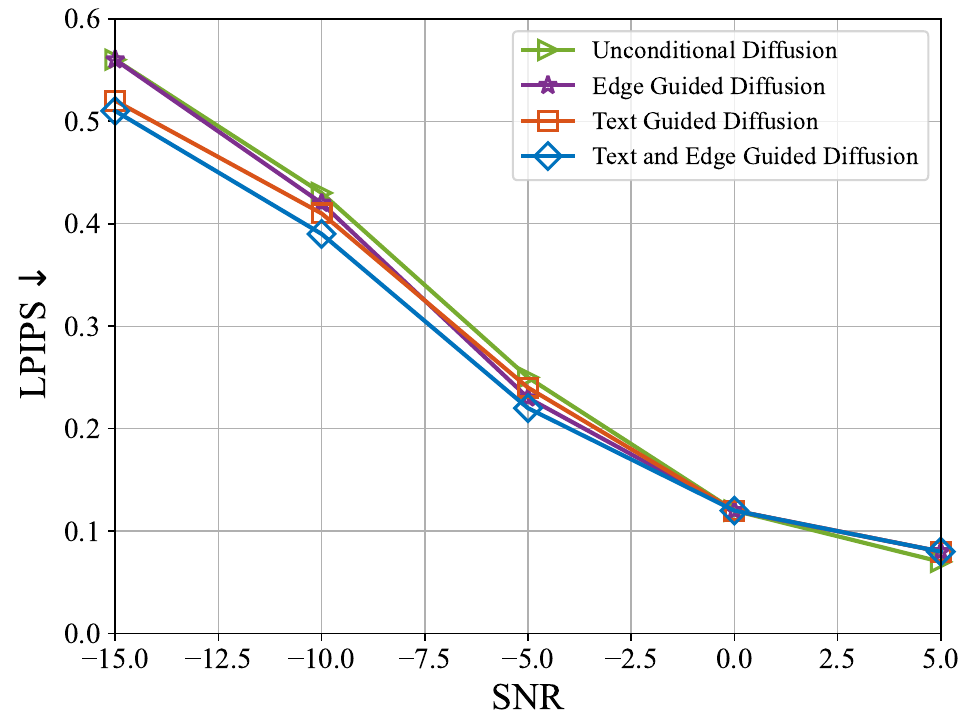}}
		\subfigure[CLIP Score]{
			\label{Fig:awgn_wocsi:c} 
			\includegraphics[width=0.235\linewidth]{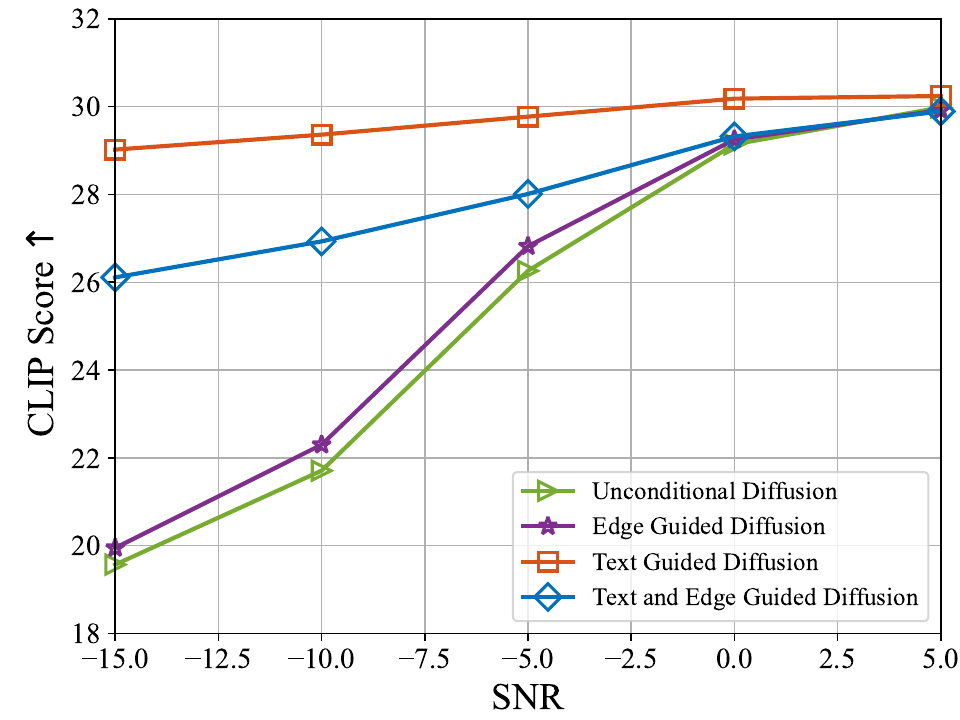}}						    
		\subfigure[FID]{
		\label{Fig:awgn_wocsi:d} 
		\includegraphics[width=0.235\linewidth]{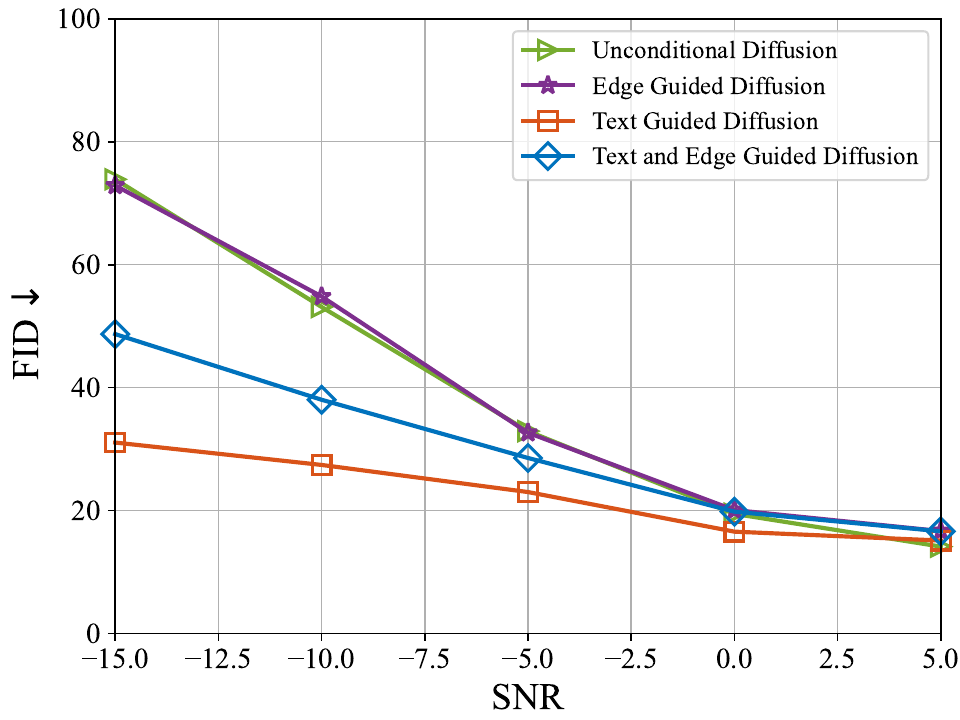}}			
	\caption{\rv{Ablation study of different guidance schemes.}}\label{Fig: semantics guided DM ablation study}
	\vspace{-2mm}
\end{figure*}
  \begin{figure*}[t]
	\begin{center}
		\resizebox{\textwidth}{!}
		{\begin{tabular}{ccccccc}
  {\large Original} & {\large Edge Map} & {\large Rec. Edge Map}& {\large Unconditional} & {\large Text Guided} & {\large Edge Guided}& {\large Text and Edge Guided}\\
  \includegraphics[width=0.2\textwidth]{./Figures/selected/ablation_study/_10dB/text_edge_map_guided/origin_1.png}  &
  \includegraphics[width=0.2\textwidth]{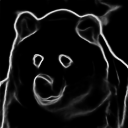}   &
  \includegraphics[width=0.2\textwidth]{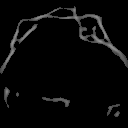} & 
  \includegraphics[width=0.2\textwidth]{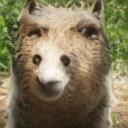} & 
  \includegraphics[width=0.2\textwidth]{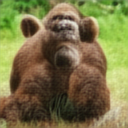} &
  \includegraphics[width=0.2\textwidth]{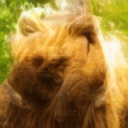} &
  \includegraphics[width=0.2\textwidth]{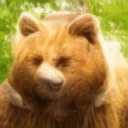} 
  \\
  \includegraphics[width=0.2\textwidth]{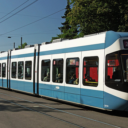}  &
  \includegraphics[width=0.2\textwidth]{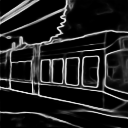}  &
  \includegraphics[width=0.2\textwidth]{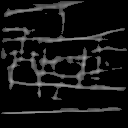}& 
  \includegraphics[width=0.2\textwidth]{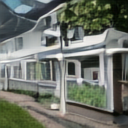}& 
  \includegraphics[width=0.2\textwidth]{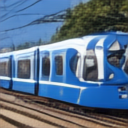} &
  \includegraphics[width=0.2\textwidth]{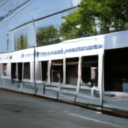} &
  \includegraphics[width=0.2\textwidth]{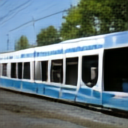} 
  \\
\end{tabular}}
   \end{center}
   \caption{{Examples of reconstructed images under different guidance methods with SNR $=-10$ dB. The extracted text description for the first row images: {\itshape A brown bear is sitting in the grass.} The extracted text description for the second-row images: {\itshape A blue and white train on the tracks.}}}
	  \label{fig:ablation study}
	  \vspace{-4mm}
  \end{figure*}

  We provide examples of reconstructed images in Fig. \ref{fig:example of awgn reconstruction images snr=-15db}, Fig. \ref{fig:example of awgn reconstruction images snr=-5db}, and Fig. \ref{fig:example of awgn reconstruction images snr=5db} for SNR values of  $-15$ dB, $-5$ dB, and $5$ dB, respectively. As shown in Fig. \ref{fig:example of awgn reconstruction images snr=-15db}, under SNR $=-15$ dB, although ADJSCC and JSCC-Diff exhibit better PSNR performance, their reconstructed images degrade significantly due to the high noise levels, resulting in the loss of key semantic information. In contrast, under the guidance of semantic side information, the proposed SGD-JSCC preserves these key semantics and delivers better perceptual performance. This also indicates that LPIPS and CLIP score are better performance metrics for evaluating reconstructed images under extremely low SNR conditions. Similarly, as shown in Fig. \ref{fig:example of awgn reconstruction images snr=-5db}, the proposed SGD-JSCC reconstructs the images with the best visual quality, consistent with the FID performance in Fig. \ref{Fig:awgn_wcsi:d}. When SNR $=5$ dB, the benchmark schemes are able to reconstruct images that capture some of the semantic information, while the advantage of the proposed SGD-JSCC algorithm lies in providing more detailed reconstructions.

\rv{\subsubsection{Rayleigh fading channels without CSI}\label{subsubsec: fading without csi}
	We then examine the  scenario of Rayleigh fading channels,  where neither the transmitter nor the receiver has access to channel state information ($h$, $\sigma^2$). This setting proves to be far more challenging than the AWGN scenarios because there exists severe inter-symbol interference, in addition to the additive Gaussian noise. The benchmark schemes are trained from scratch under this channel configuration, while our approach only involves specific training of the phase estimation module. As shown in Fig. \ref{Fig:fading without csi}, the performance of the benchmark schemes decreases markedly in this scenario, with the ViT-based JSCCFormer outperforming the CNN-based ADJSCC. Our proposed SGD-JSCC scheme also experiences a performance drop at very low SNR (i.e., SNR $\leq -10$dB), mainly due to large phase estimation errors. However, as SNR increases and the phase estimation becomes more accurate, our method achieves significantly better results than the baselines. These findings underscore the advantages of the SGD-JSCC approach in cases where channel state information is unavailable or severely impaired by channel estimation errors.}

\subsection{Performance Evaluation of Semantics Guided DM}\label{subsec: ablation study}
\begin{figure*}[!]
	\centering
	\captionsetup{skip=2pt}
	\subfigure[SNR Estimation]{
		\label{Fig:estimation :a} 
		\includegraphics[width=0.235\linewidth]{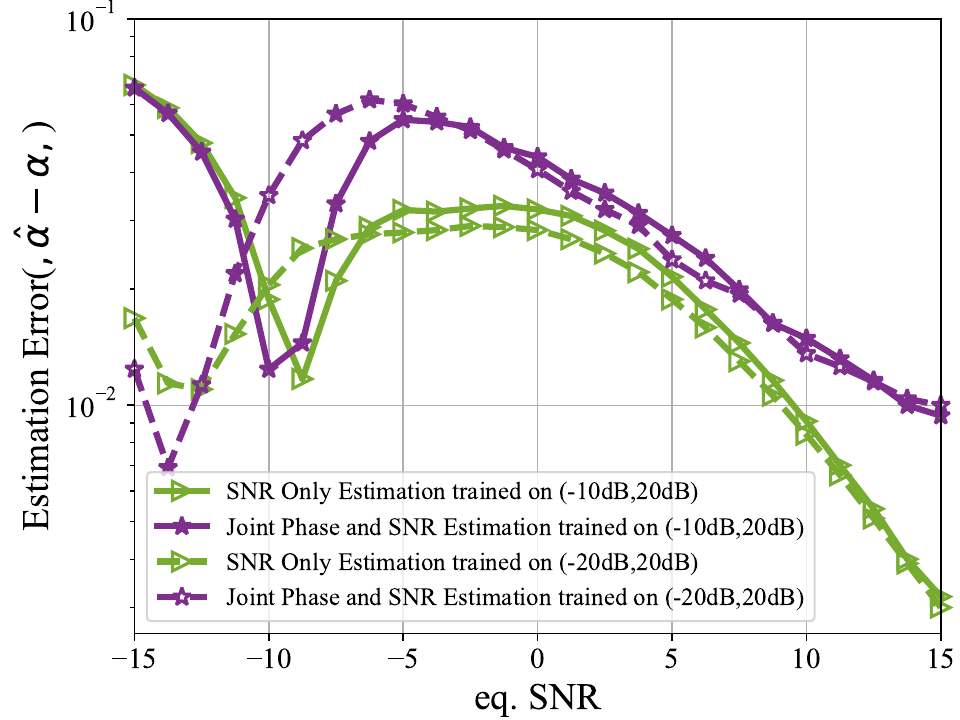}}
	\subfigure[Phase Estimation]{
		\label{Fig:estimation :b} 
		\includegraphics[width=0.235\linewidth]{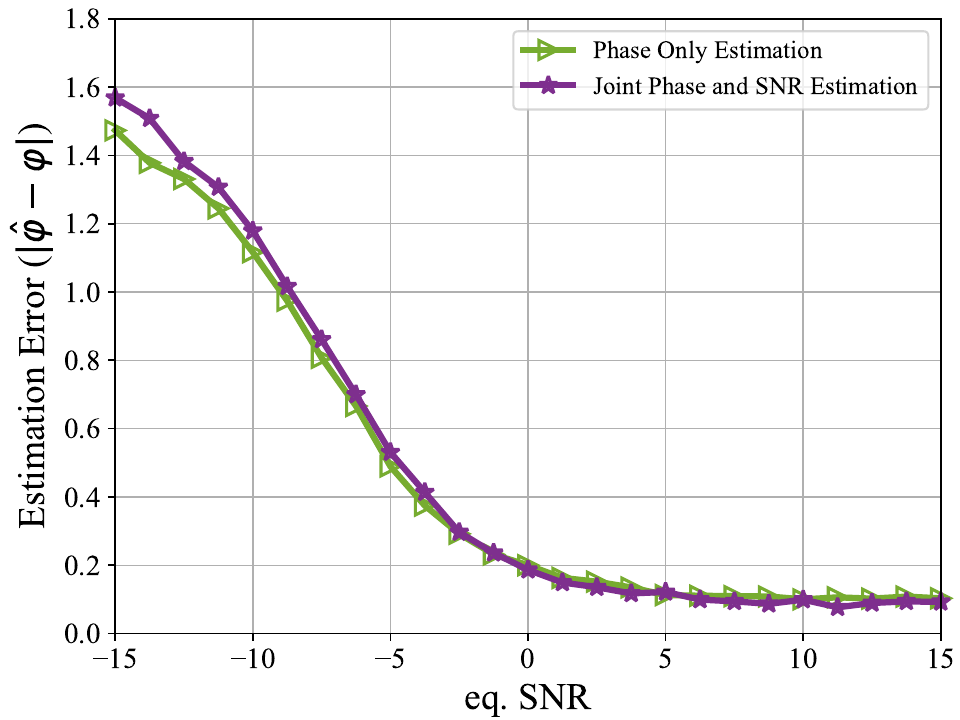}}
		\subfigure[Sensitivity to SNR Est. Error]{
			\label{Fig:estimation :c} 
			\includegraphics[width=0.235\linewidth]{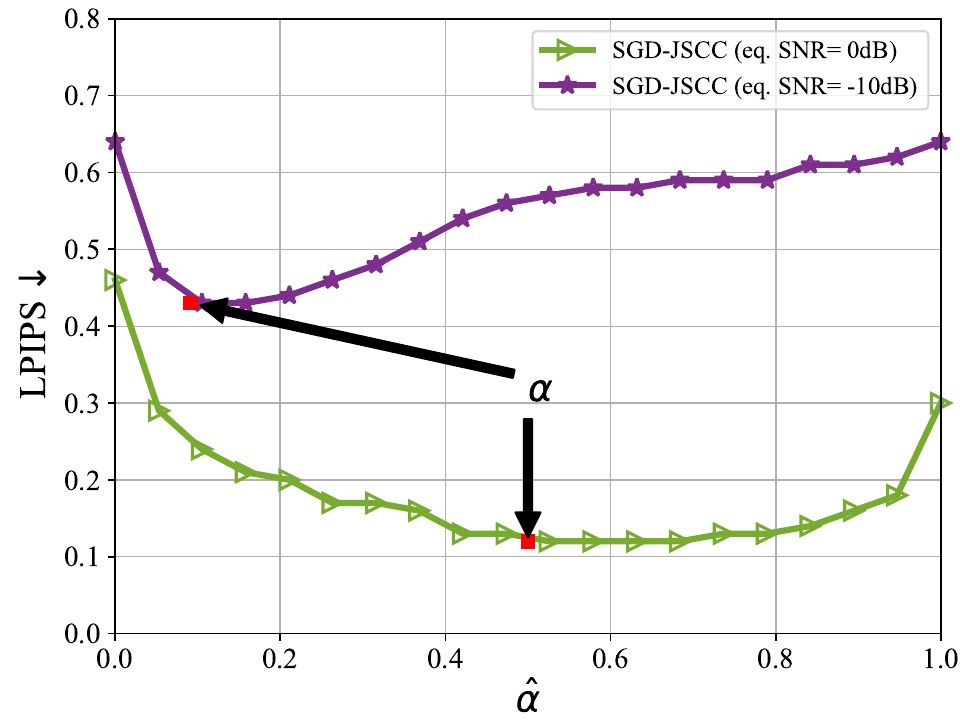}}						    
		\subfigure[Sensitivity to Phase Est. Error]{
		\label{Fig:estimation :d} 
		\includegraphics[width=0.235\linewidth]{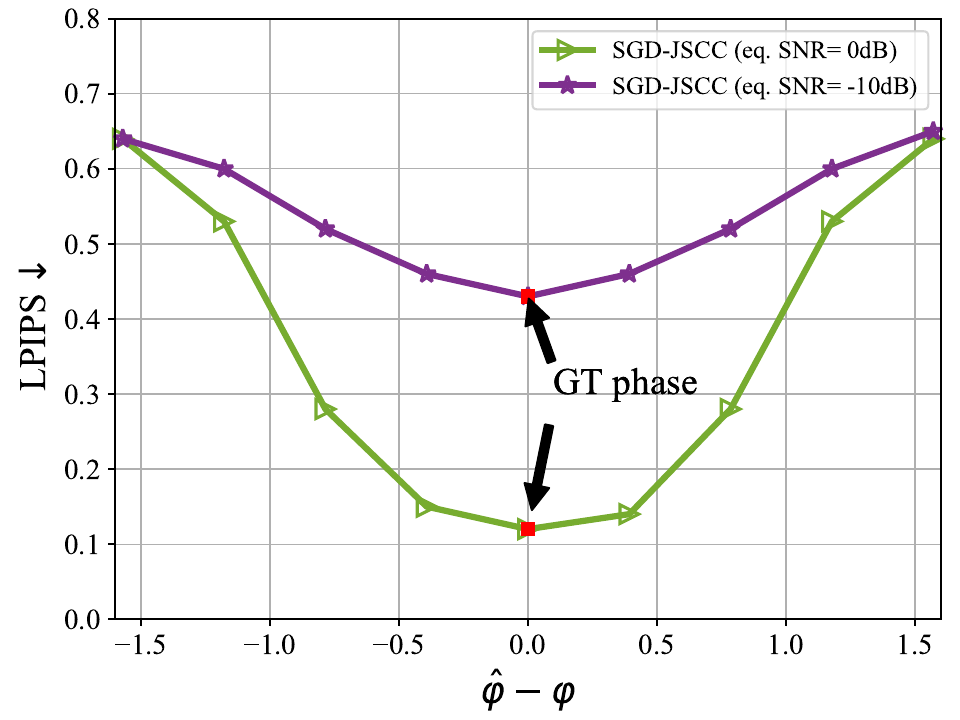}}			
	\caption{{Performance Evaluation of Channel Estimation and Sensitivity Analysis.}}\label{Fig: estimation} 
\end{figure*}
\begin{figure*}[htb]
	\begin{center}
		\resizebox{\textwidth}{!}
		{\begin{tabular}{ccccc}
  {\large Original} & {\large L2 norm (MR=0.5)} & {\large L2 norm (MR=0.4)}& {\large L2 norm (MR=0.3)} & {\large L2 norm (MR=0.2)} \\
  \includegraphics[width=0.28\textwidth]{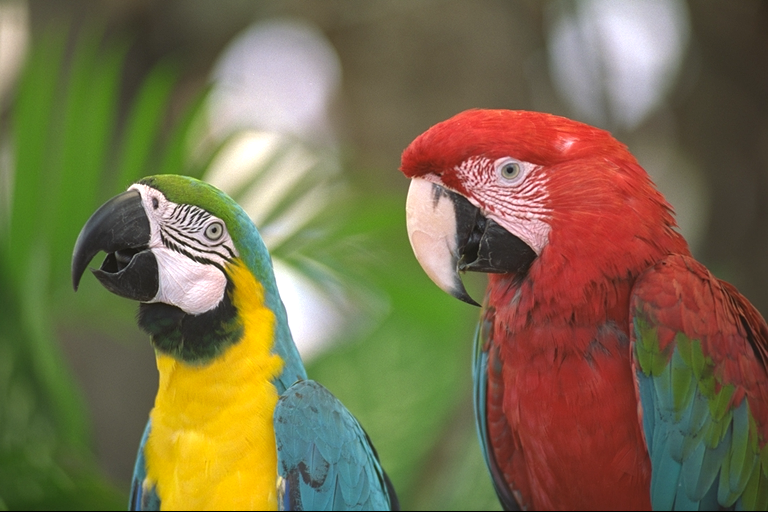}  &
  \includegraphics[width=0.28\textwidth]{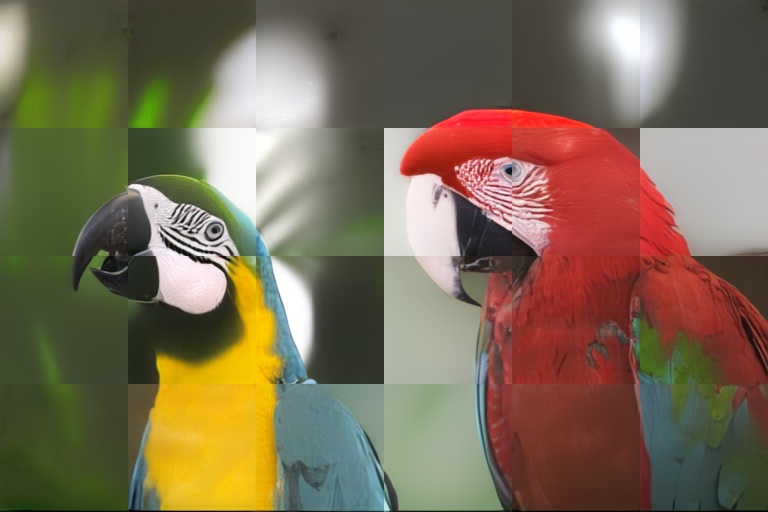}  &
  \includegraphics[width=0.28\textwidth]{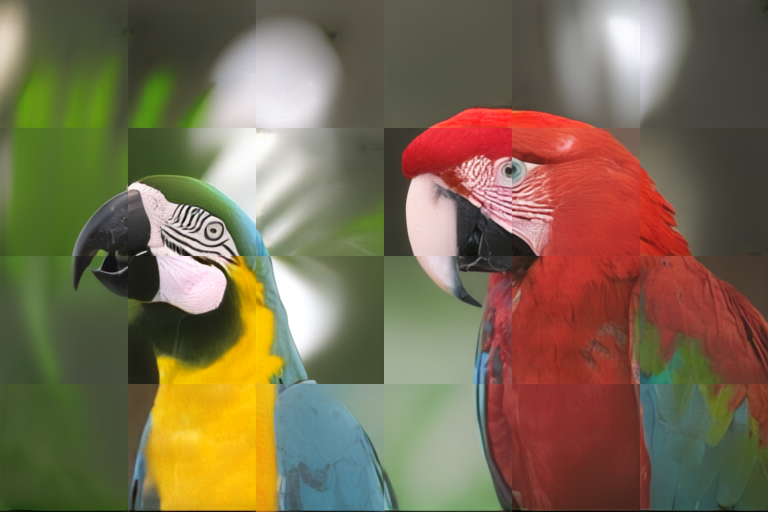}  &
  \includegraphics[width=0.28\textwidth]{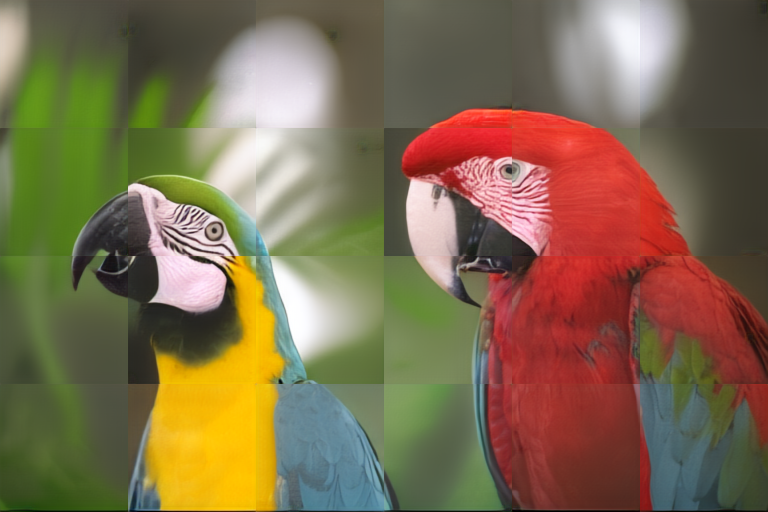}   &
  \includegraphics[width=0.28\textwidth]{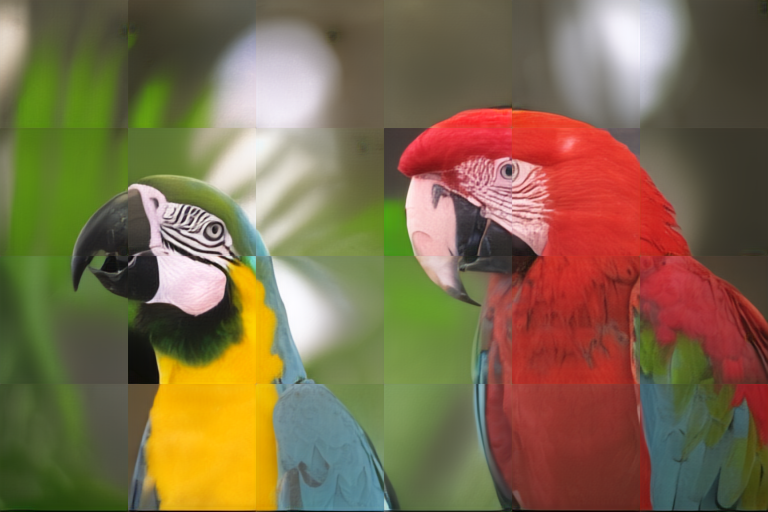} 
  \\
  {\large No mask} & {\large Random (MR=0.5)} & {\large Random (MR=0.4)}& {\large Random (MR=0.3)} & {\large Random (MR=0.2)} \\
  \includegraphics[width=0.28\textwidth]{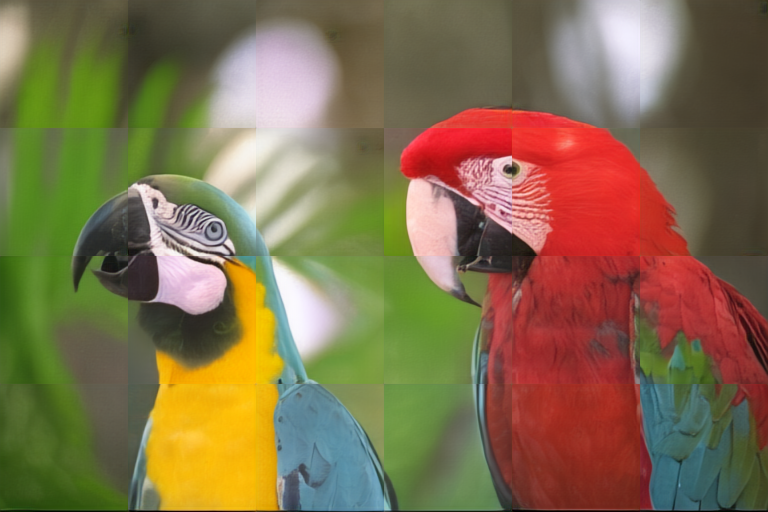}  &
  \includegraphics[width=0.28\textwidth]{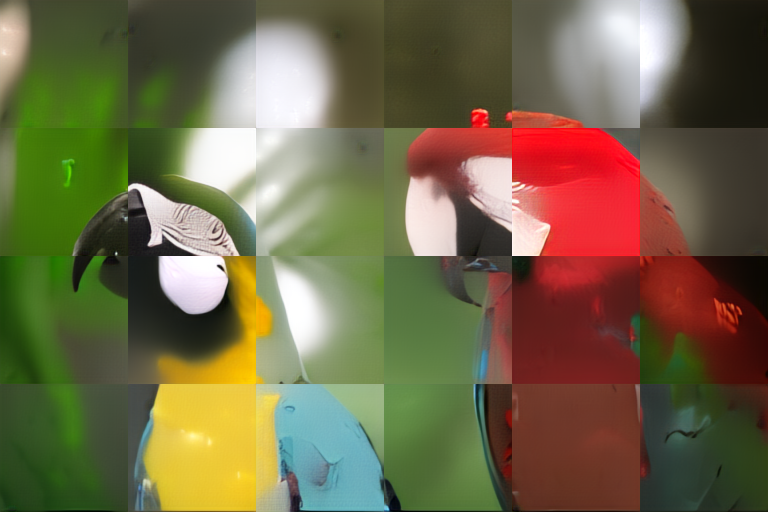}  &
  \includegraphics[width=0.28\textwidth]{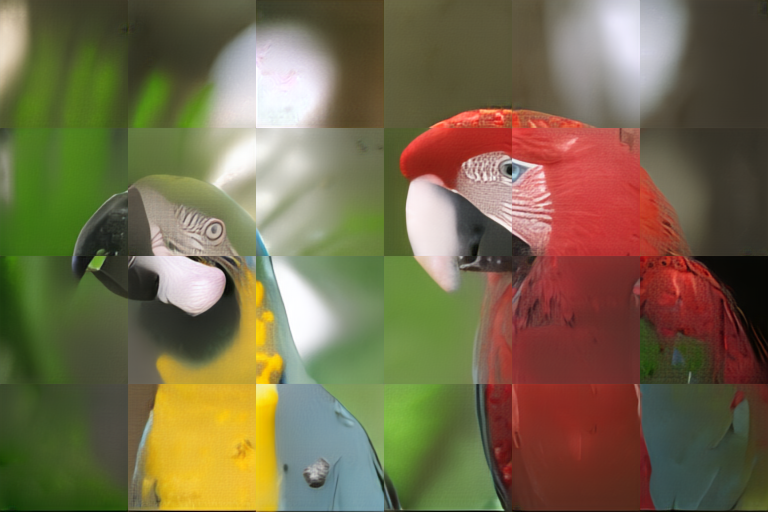}  &
  \includegraphics[width=0.28\textwidth]{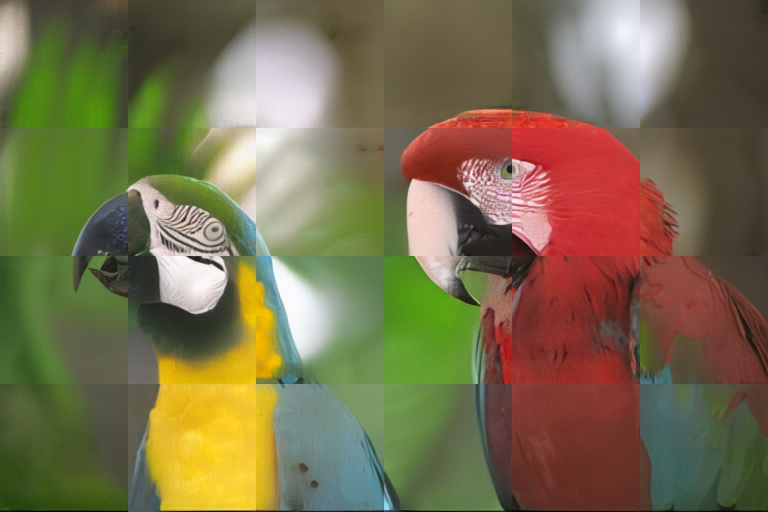}  &
  \includegraphics[width=0.28\textwidth]{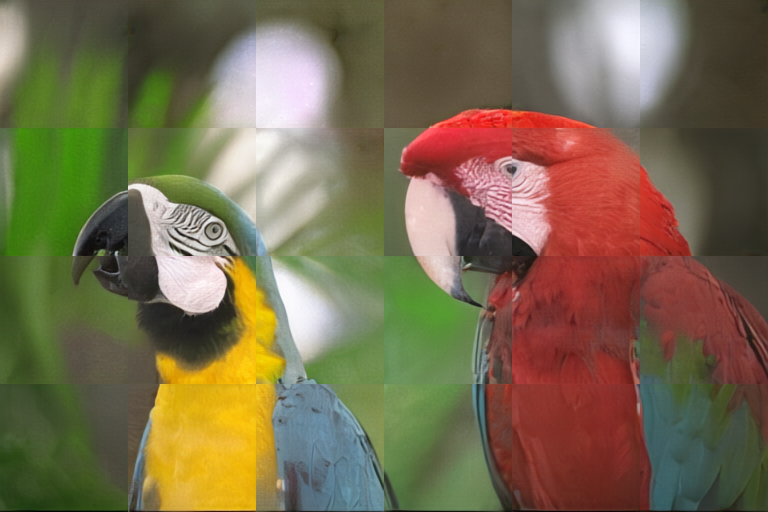}
  \\
\end{tabular}}
   \end{center}
	  \caption{{Examples of reconstructed images under different mask strategies with SNR $=0$ dB. }}
	  \label{fig:outage example images}
	  \vspace{-3mm}
  \end{figure*}

\rv{In this subsection, we analyze the proposed SGD-JSCC scheme in detail, evaluating its performance with different guidance and masking strategies. We adopt the channel configuration described in Section \ref{subsubsec: awgn without snr}, where the SNR information of the AWGN channels is obtained via the proposed SNR estimation module.}
\subsubsection{Performance Evaluation of Different Guidance Schemes}\label{subsubsec: performance evaluation of different guidance schemes}
\rv{
We conduct an ablation study on our diffusion model using four different semantic guidance schemes: no guidance, 
text guidance, edge map guidance, and joint text-edge map guidance. As shown in Fig. \ref{Fig: semantics guided DM ablation study}, in the low SNR regime (i.e., SNR $\leq 5$ dB), both text semantics and edge maps contribute to measurable performance improvements over the unconditional baseline when evaluated by LPIPS, with a combination of text and edge maps achieving further gains. When assessed by CLIP score and FID, the text-guided approach demonstrates the highest performance—largely because text provides coarse yet significant contextual information that helps improve overall image perception. However, purely text-driven guidance struggles to accurately capture fine-grained structural details and may reconstruct incorrect (albeit realistic-looking) content; the edge map plays a crucial role in addressing this shortcoming by preserving clear contour information.
We also observe that the unconditional DM achieves optimal performance in the high SNR regime, as it inherently contains relatively accurate semantic information, making additional guidance for denoising unnecessary. In contrast, the performance of the proposed SGD-JSCC model declines in this high SNR environment, and this can be attributed to two main factors. First, text semantics provide only coarse semantic information. With the introduction of classifier-free guidance, the diffusion model tends to refine image details to enhance realism rather than accurately reconstruct the original image. Second, the edge map semantics are transmitted using a JSCC model, which introduces transmission errors that may mislead the DM into refining incorrect structural information, ultimately leading to performance degradation. In conclusion, while incorporating semantics can enhance performance in low SNR conditions, in high SNR scenarios, unconditional denoising can yield satisfactory results.
}

Exemplary images are provided in Fig. \ref{fig:ablation study} to visualize the role of text and edge map. First, under eq. SNR $=-10$ dB, the reconstructed edge map retains most of the key structural information compared with the original image. For the first row image that comprises a bear and grass, the edge map-guided method successfully reconstructs the bear and grass, whereas the unconditional guided one struggles, reconstructing an unidentified animal instead of a bear. However, as shown in the second row, single structural guidance can also lead to errors: the edge map-guided reconstruction preserves only structural information, missing key semantics such as color. Fortunately, this issue is effectively addressed by adding text guidance, which provides the missing semantic information. These results demonstrate the necessity and effectiveness of hybrid semantic guidance.
\subsubsection{Performance Evaluation of Blind CSI Estimation}\label{subsubsec: csi estimation}
\rv{In this part, we evaluate the accuracy of the estimation module and analyze the robustness of the SGD-JSCC scheme with respect to channel estimation errors. We quantify the performance of the SNR estimation module by calculating the L1 norm of the difference between the predicted signal power scalar, denoted as $\hat{\alpha}$ and the ground truth signal power scalar in (\ref{eq: snr estimation}), $\alpha$. Similarly, we assess phase estimation by measuring the L1 norm of the difference between the predicted phase $\hat{\phi}$ and the actual phase $\phi$.
Furthermore, we investigate the impact of channel estimation errors on overall performance by examining the LPIPS metric across various levels of manually introduced errors. The results are presented in Fig. \ref{Fig: estimation}. Our findings indicate that the estimation errors for both SNR and phase are acceptable. Moreover, the joint estimation of SNR and phase results in only a slight performance degradation compared to single-item estimation when the another is set as the ground truth. 
Interestingly, the phase estimation error decreases monotonically as the SNR increases, while the SNR estimation exhibits a local minimum. This local minimum is found to be correlated with the lowest SNR value encountered during the training phase; more results on SNR estimation errors are provided in \cite{zhang2025semantics}. 
Besides, as we will shown in Table \ref{table: computation time analysis: a}, the computation cost of both SNR estimation and phase estimation is bearable since their lightweight architectures, making them suitable for real-time applications.
As illustrated in Fig. \ref{Fig:estimation :c}, both over-denoising and under-denoising caused by estimation inaccuracies can adversely affect performance; however, the typical estimation accuracy shown in Fig. \ref{Fig:estimation :a} has a negligible impact on the final reconstruction quality. A similar trend is observed for phase estimation, as depicted in Fig. \ref{Fig:estimation :d}. These results validate the effectiveness of our pilot-free scheme, confirming that our channel estimation module is sufficiently accurate and that the SGD-JSCC scheme exhibits robust performance in the presence of channel estimation errors.}
\begin{table*}[t]
	\centering
	\vspace{3mm}
	\caption{\rv{Performance comparison of different schemes under fast fading channels.}}\label{table: fast fading evaluation}
	\label{table:performance of fast fading}
	\vspace{3mm}
	\resizebox{0.9\textwidth}{!}{
	\rv{\begin{tabular}{cc|cccc|cccc}
	\toprule
	\multirow{2}{*}{\parbox{1.2cm}{\centering SNR}} & \multirow{2}{*}{\parbox{1.2cm}{\centering Method}} & \multicolumn{4}{c|}{LPIPS $\downarrow$} & \multicolumn{4}{c}{CLIP Score $\uparrow$}\\ 
	\cmidrule(lr){3-6} \cmidrule(lr){7-10}
	 & & $\rho=1$ & $\rho=128$ & $\rho=512$ & $\rho=1024$    & $\rho=1$ & $\rho=128$ & $\rho=512$ & $\rho=1024$     \\
	\midrule
	\multirow{7}{*}{\centering -5dB} 
	& ADJSCC (w/o fine-tune)     &0.35 &0.36 &0.37 & 0.39&20.78 &20.77 &20.84 & 20.93 \\
	& JSCCformer (w/o fine-tune) &0.35 &0.35 &0.36 &0.37 &21.03 &21.07 &21.13 &21.05  \\
	 & ADJSCC (w/ fine-tune)    &0.34 &0.34 & 0.36&0.36 & 21.47& 21.46&21.45 & 21.44 \\
	 & JSCCformer (w/ fine-tune)&0.32 &0.33 &0.34 &0.35 &22.53 & 22.46&22.29 &22.22  \\
	 & CDDM                      &\textbf{0.28} &\textbf{0.29} &\textbf{0.30} &0.32 &{25.26} &25.22 &25.02 &24.87  \\
	 & SGD-JSCC (w/o filling)    &0.29&0.30 &\textbf{0.30} &\textbf{0.31} & \textbf{25.34}&25.18 & \textbf{25.25}&{25.05}  \\
	 & SGD-JSCC (w/ filling)    &0.29 &0.30 &\textbf{0.30} &\textbf{0.31} & 25.14&\textbf{25.23} &{25.15} &\textbf{25.09}  \\
	\midrule
	\multirow{7}{*}{\centering 0dB} 
	& ADJSCC (w/o fine-tune)     &0.26 &0.26 &0.28 & 0.29&22.18 &20.18 &22.37 &22.44  \\
	& JSCCformer (w/o fine-tune) &0.23 &0.24 &0.26 & 0.27&22.79 &22.78 &22.83 &22.84  \\
	 & ADJSCC (w/ fine-tune)    &0.23 &0.24 &0.25 &0.27 & 24.07&24.03 &23.87 &23.87  \\
	 & JSCCformer (w/ fine-tune)& 0.23& 0.23&0.24 & 0.25&25.07 &24.97 &24.82 &24.61  \\
	 & CDDM                      &\textbf{0.16} &\textbf{0.18} &0.19 &0.21 &{28.23} & 27.91&27.60 &27.28  \\
	 & SGD-JSCC (w/o filling)    &{0.17} &\textbf{0.18} &\textbf{0.18} &0.19 & \textbf{28.40}&28.18 &\textbf{28.08} &\textbf{27.80}  \\
	 & SGD-JSCC (w/ filling)    &\textbf{0.16} &\textbf{0.18} &\textbf{0.18} &\textbf{0.18} & 28.16& \textbf{28.35}& \textbf{28.06}&\textbf{27.79}  \\
	\midrule
	\multirow{7}{*}{\centering 5dB} 
	& ADJSCC (w/o fine-tune)     &0.21 &0.21 &0.23 &0.24 &24.19 &24.27 & 24.35& 24.54 \\
	& JSCCformer (w/o fine-tune) &0.17 & 0.18&0.20 &0.21 &24.63 &24.62 &24.70 & 24.75 \\
	 & ADJSCC (w/ fine-tune)    & 0.15&0.15 &0.16 &0.17 & 26.65&26.48 & 26.34& 26.31 \\
	 & JSCCformer (w/ fine-tune)& 0.16&0.16 & 0.17&0.18 &27.41 &27.28 &27.04 &26.91  \\
	 & CDDM                      &\textbf{0.10} &0.11 &0.12 &0.13 &{29.50} &29.24 &29.16 &28.81  \\
	 & SGD-JSCC (w/o filling)    &0.11 &{0.12} &{0.12} &\textbf{0.12} &\textbf{29.56} &{29.45} &{29.42} &\textbf{29.29}  \\
	 & SGD-JSCC (w/ filling)    &\textbf{0.10} &\textbf{0.11} &\textbf{0.11} & \textbf{0.12}&\textbf{29.57} &\textbf{29.51} &\textbf{29.45} & \textbf{29.29} \\
	\bottomrule
	\end{tabular}}
	}
	\vspace{-6mm}
	\end{table*}
\subsubsection{Performance Evaluation of Mask Strategies}
\rv{
As shown in Fig. \ref{fig: network architecture of diffusion model}, the proposed diffusion framework supports masked latent data, enabling us to achieve rate-adaptive transmission and importance-aware resource allocation by only transmitting partial latent features. 
We evaluate our SGD-JSCC scheme under various masking ratios (MRs) and strategies using the Kodak dataset. Since Kodak images have different resolutions, each image is divided into $128\times128$ patches for independent transmission. For a given patch $\mathbf{x}$, the JSCC encoder outputs a latent feature $\mathbf{f}\in \mathbb{R}^{c\times\frac{h}{8}\times\frac{w}{8}}$, which we concatenate into a matrix $\mathbf{u}\in \mathbb{R}^{c\times \frac{hw}{64}}$ composed of $\frac{hw}{64}$ tokens, each with embedding dimension $c$. We then mask a subset of these tokens to form a reduced latent representation $\hat{\mathbf{f}}\in \mathbb{R}^{c\times \hat{N}}$, defining the mask ratio as ${\rm MR}=1-\frac{\hat{N}}{hw/64}$. Two masking strategies are considered: Random, which randomly drops tokens, and L2-norm, which selects the lowest L2-norm tokens for masking. Fig. \ref{Fig:mask strategies comparison:a} shows that the L2-norm method outperforms the random approach by preserving more critical information, while Fig. \ref{Fig:mask strategies comparison:b} illustrates that reconstruction quality improves as MR decreases, meaning more tokens are retained. 
We also provide examples of reconstructed images under different masking strategies in Fig. \ref{fig:outage example images}. 
The results validate the benefits of selectively transmitting crucial latent features.
}
\begin{figure}[t] 
	\centering
	\captionsetup{skip=2pt}
	\subfigure[Performance comparison of different mask strategies, with edge map and text guidance, under SNR$=0$dB.]{
		\label{Fig:mask strategies comparison:a} 
		\includegraphics[width=0.47\linewidth]{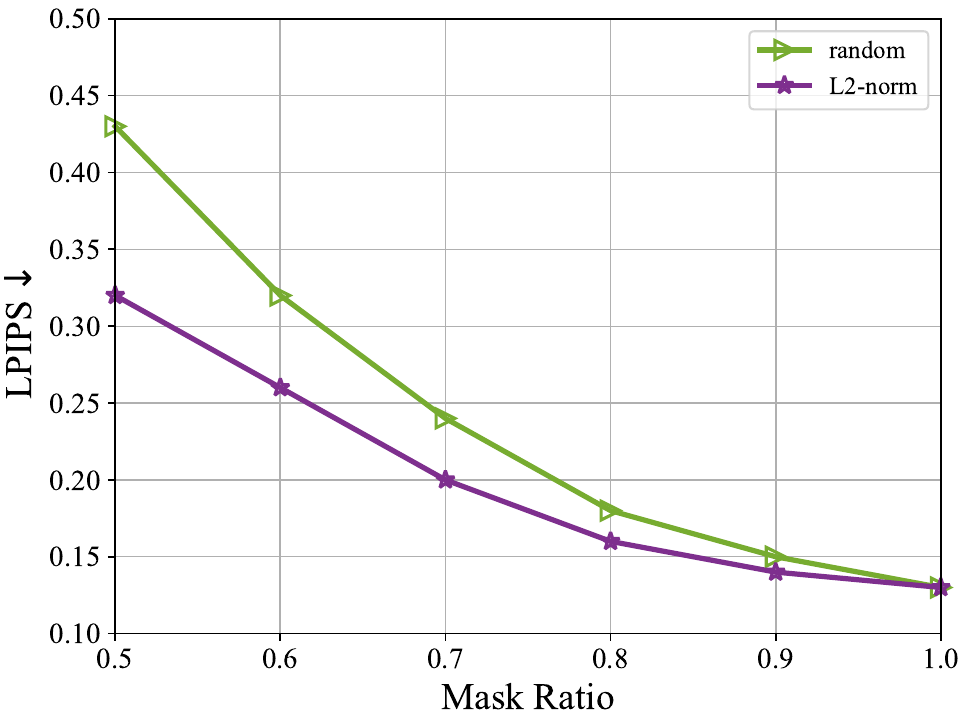}}
		\subfigure[Average Performance of L2-norm mask strategy, with edge map and text guidance.]{
			\label{Fig:mask strategies comparison:b} 
			\includegraphics[width=0.47\linewidth]{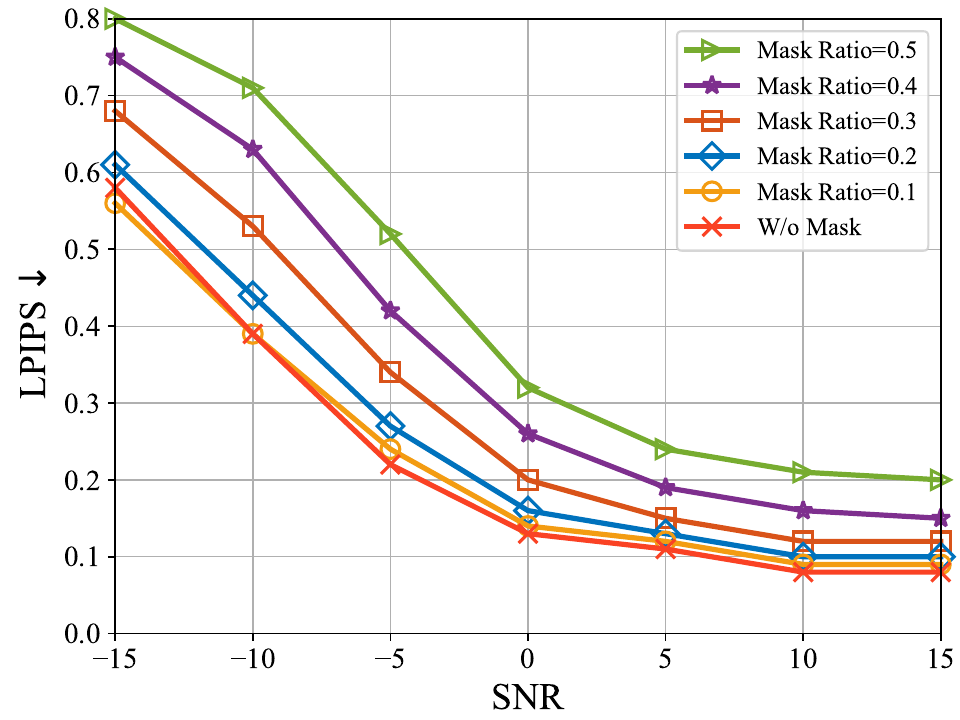}}	
		\vspace{0.5mm}
	\caption{Performance evaluation with mask operation.}
	\vspace{-3mm}
	\label{Fig: outage evaluation}
\end{figure}
\subsection{Performance Evaluation over Fast Fading Channels}
\rv{In this subsection, we evaluate the performance of the proposed SGD-JSCC scheme under fast fading channel conditions. 
We consider that the receiver has perfect CSI for equalization. As discussed in Section \ref{sec: extension to rayleigh fading case}, fast fading channels pose significant challenges because each symbol may experience independent fading, resulting in imbalanced SNR levels. To capture this effect, we define a block length, $\rho$, during which the channel fading scalar remains constant for $\rho$ consecutive symbols. We compare our SGD-JSCC scheme with ADJSCC, JSCCFormer, and CDDM \cite{wu2024cddm} under varying fast fading scenarios. Specifically, we provide results for ADJSCC and JSCCFormer in two configurations: one trained on slow fading channels (without fine-tuning) and one fine-tuned on fast fading channels with $\rho=1$. For CDDM, every channel output element is assigned the same noise level, which is computed as the average. The noisy latent feature is then denoised with our DM. 
For our approach, we examine two variants: the original SGD-JSCC procedure (hereafter referred to as SGD-JSCC with water-filling) and a variant that omits the addition of extra noise to channel symbols at each iteration (SGD-JSCC without water-filling). We consider four different values of 
$\rho$ and evaluate performance using LPIPS and CLIP scores, with the aggregated results summarized in Table \ref{table: fast fading evaluation}.
}

\rv{The results demonstrate that diffusion-based schemes inherently handle fast fading conditions more effectively, consistently outperforming both ADJSCC and JSCCFormer. When $\rho=1$, indicating fully uncorrelated channel states per symbol, CDDM achieves performance comparable to our SGD-JSCC. However, as $\rho$ increases, the performance of CDDM drops substantially, likely due to its reduced adaptability to blocks of symbols sharing the same fading conditions. In contrast, our SGD-JSCC approach shows only a slight decline in performance, underscoring the robustness of its selective update mechanism. Moreover, comparisons between the methods with and without water-filling reveal a modest yet consistent advantage when additional noise is added  into symbols for achieving an equalized SNR level over symbols. This finding reinforces the importance of the water-filling step, which ensures that the diffusion model receives theoretically optimal inputs and, consequently, achieves improved reconstruction quality.
}
\vspace{-3mm}
\begin{table}[]
		\centering
		\caption{Computation Time Analysis}
		\begin{subtable}[Computation time analysis of SGD-JSCC]{\rv{
			\centering
			\label{table: computation time analysis: a}
			\begin{tabular}{llll}
			\toprule
			\multirow{2}{*}{Semantics Extraction/s} & \multicolumn{1}{|l|}{Text} & \multicolumn{1}{l|}{0.241} & \multirow{2}{*}{0.269} \\ 
									   & \multicolumn{1}{|l|}{Edge map}  & \multicolumn{1}{l|}{0.028} &                        \\ \midrule
			\multirow{2}{*}{JSCC Enc. Time/s} & \multicolumn{1}{|l|}{Image} & \multicolumn{1}{l|}{0.004} & \multirow{2}{*}{0.015} \\ 
									   & \multicolumn{1}{|l|}{Edge map}  & \multicolumn{1}{l|}{0.011} &                        \\ \midrule
			\multirow{3}{*}{Blind CSI Estimation/s} & \multicolumn{1}{|l|}{SNR only estimation}  & \multicolumn{1}{l|}{0.0009} & \multirow{3}{*}{0.023} \\ 
									   & \multicolumn{1}{|l|}{Phase only estimation} & \multicolumn{1}{l|}{0.0014} &                        \\ 
									   & \multicolumn{1}{|l|}{Joint estimation} & \multicolumn{1}{l|}{0.0230} &                        \\ \midrule
			\multirow{3}{*}{DM \& JSCC Dec. Time/s} & \multicolumn{1}{|l|}{Edge map}  & \multicolumn{1}{l|}{0.011} & \multirow{3}{*}{1.224} \\ 
									   & \multicolumn{1}{|l|}{DM denoising} & \multicolumn{1}{l|}{1.209} &                        \\ 
									   & \multicolumn{1}{|l|}{Image} & \multicolumn{1}{l|}{0.004} &                        \\ \midrule
			Total Time/s                 & \multicolumn{3}{r}{1.531}                                                           \\ 
			\bottomrule
		\end{tabular}}}
		\end{subtable}
\begin{subtable}[Computation Time Comparison]{
			\label{table: computation time analysis: b}
		\rv{\begin{tabular}{lr}
				\toprule
				Method        & Execution Time/s \\ \midrule
				ADJSCC        & 0.009            \\ 
				JSCCFormer    & 0.021            \\ 
				DiffJSCC      & 2.605            \\ 
				DeepJSCC-Diff & 1.543            \\ 
				SGD-JSCC      & 1.531            \\ \bottomrule
		\end{tabular}}}
		\end{subtable}
		\vspace{0mm}
		\end{table}
	\subsection{Computation Complexity Analysis} 
\rv{
In this subsection, we evaluate the computational performance of the proposed SGD-JSCC scheme using an RTX A5000 GPU, averaging results over 200 images from the COCO dataset. 
Table \ref{table: computation time analysis: a} summarizes the processing time for each component of our system. Notably, the encoding time is dominated by the extraction of text description, 
due to the higher complexity of BLIP2 model. In contrast, the decoding process is mainly bottlenecked by the diffusion denoising step, which currently requires 50 iterations. Table \ref{table: computation time analysis: b} includes a comparison of the conduction times for various diffusion-based schemes, all employing 50 denoising steps. While these diffusion-based approaches offer superior perceptual performance, they do so at the cost of increased decoding time. Fortunately, this limitation can be mitigated through knowledge distillation \cite{pei2024latent}, which trains a student diffusion model capable of achieving comparable denoising quality in far fewer steps, thereby reducing computational load. Moreover, our scheme demonstrates lower overall computation time compared to DiffJSCC and DeepJSCC-Diff, primarily owing to our adoption of a lightweight diffusion model rather than a pre-trained model designed for generation.
}
\section{Conclusion}\label{sec: conclusion}
In this paper, we propose a novel semantics-guided diffusion DeepJSCC scheme, called SGD-JSCC. First, we explored different types of semantics and their corresponding transmission schemes. Then, we designed a DiT model for channel denoising, supporting both text and edge map guidance by integrating a cross-attention mechanism and ControlNet architecture. We made necessary modifications to the original DM and trained it from scratch to seamlessly integrate with DeepJSCC. Furthermore, we introduced a water-filling-inspired scheme to address fading channel scenarios, enabling the use of a DM trained under AWGN conditions without the need for specific fine-tuning. Experimental results demonstrate that the proposed scheme outperforms existing methods. For future work, we aim to extend the proposed scheme to MIMO channels and explore the corresponding CSI-free transmission.

	\bibliographystyle{ieeetr}
	\bibliography{BibDesk_File_v2}

\begin{thebibliography}{10}

\bibitem{gunduz2022beyond}
D.~Gündüz, Z.~Qin, I.~E. Aguerri, H.~S. Dhillon, Z.~Yang, A.~Yener, K.~K.
  Wong, and C.-B. Chae, ``Beyond transmitting bits: Context, semantics, and
  task-oriented communications,'' {\em IEEE Journal on Selected Areas in
  Communications}, vol.~41, no.~1, pp.~5--41, 2023.

\bibitem{bourtsoulatze2019deep}
E.~Bourtsoulatze, D.~B. Kurka, and D.~G{\"u}nd{\"u}z, ``Deep joint
  source-channel coding for wireless image transmission,'' {\em IEEE Trans.
  Cogn. Commun. Netw.}, vol.~5, no.~3, pp.~567--579, 2019.

\bibitem{gunduz2024joint}
D.~G{\"u}nd{\"u}z, M.~A. Wigger, T.-Y. Tung, P.~Zhang, and Y.~Xiao, ``Joint
  source--channel coding: Fundamentals and recent progress in practical
  designs,'' {\em Proceedings of the IEEE}, 2024.

\bibitem{dai2022nonlinear}
J.~Dai, S.~Wang, K.~Tan, Z.~Si, X.~Qin, K.~Niu, and P.~Zhang, ``Nonlinear
  transform source-channel coding for semantic communications,'' {\em IEEE J.
  Sel. Areas Commun.}, vol.~40, no.~8, pp.~2300--2316, 2022.

\bibitem{wu2024transformer}
H.~Wu, Y.~Shao, E.~Ozfatura, K.~Mikolajczyk, and D.~G{\"u}nd{\"u}z,
  ``Transformer-aided wireless image transmission with channel feedback,'' {\em
  IEEE Trans. Wireless Commun.}, early access, 2024.

\bibitem{wu2024deep}
H.~Wu, Y.~Shao, C.~Bian, K.~Mikolajczyk, and D.~G{\"u}nd{\"u}z, ``{Deep joint
  source-channel coding for adaptive image transmission over MIMO channels},''
  {\em IEEE Trans. Wireless Commun.}, 2024.

\bibitem{yang2021deep}
M.~Yang, C.~Bian, and H.-S. Kim, ``{Deep joint source channel coding for
  wireless image transmission with OFDM},'' in {\em Proc. IEEE International
  Conference on Communications(ICC)}, pp.~1--6, 2021.

\bibitem{wu2022channel}
H.~Wu, Y.~Shao, K.~Mikolajczyk, and D.~G{\"u}nd{\"u}z, ``{Channel-adaptive
  wireless image transmission with OFDM},'' {\em IEEE Wireless Commun. Lett.},
  vol.~11, no.~11, pp.~2400--2404, 2022.

\bibitem{bian2024process}
C.~Bian, Y.~Shao, H.~Wu, E.~Ozfatura, and D.~Gunduz, ``Process-and-forward:
  Deep joint source-channel coding over cooperative relay networks,'' {\em
  [Online]. Available: https://arxiv.org/abs/2403.10613}, 2024.

\bibitem{yilmaz2023distributed}
S.~F. Yilmaz, C.~Karamanl{\i}, and D.~G{\"u}nd{\"u}z, ``Distributed deep joint
  source-channel coding over a multiple access channel,'' in {\em Prof. IEEE
  Int'l Conf. on Comms. (ICC)}, pp.~1400--1405, 2023.

\bibitem{zhang2023model}
P.~Zhang, X.~Xu, C.~Dong, K.~Niu, H.~Liang, Z.~Liang, X.~Qin, M.~Sun, H.~Chen,
  N.~Ma, {\em et~al.}, ``Model division multiple access for semantic
  communications,'' {\em Frontiers of Information Technology \& Electronic
  Engineering}, vol.~24, no.~6, pp.~801--812, 2023.

\bibitem{tung2022deepjscc}
T.-Y. Tung, D.~B. Kurka, M.~Jankowski, and D.~G{\"u}nd{\"u}z, ``Deepjscc-q:
  Constellation constrained deep joint source-channel coding,'' {\em IEEE J.
  Sel. Areas Inf. Theory}, vol.~3, no.~4, pp.~720--731, 2022.

\bibitem{bo2022learning}
Y.~Bo, Y.~Duan, S.~Shao, and M.~Tao, ``Joint coding-modulation for digital
  semantic communications via variational autoencoder,'' {\em IEEE Transactions
  on Communications}, vol.~72, no.~9, pp.~5626--5640, 2024.

\bibitem{erdemir2023generative}
E.~Erdemir, T.-Y. Tung, P.~L. Dragotti, and D.~G{\"u}nd{\"u}z, ``Generative
  joint source-channel coding for semantic image transmission,'' {\em IEEE J.
  Sel. Areas Commun.}, vol.~41, no.~8, pp.~2645--2657, 2023.

\bibitem{wu2024cddm}
T.~Wu, Z.~Chen, D.~He, L.~Qian, Y.~Xu, M.~Tao, and W.~Zhang, ``{CDDM: Channel
  denoising diffusion models for wireless semantic communications},'' {\em IEEE
  Trans. Wireless Commun.}, 2024.

\bibitem{xu2021wireless}
J.~Xu, B.~Ai, W.~Chen, A.~Yang, P.~Sun, and M.~Rodrigues, ``Wireless image
  transmission using deep source channel coding with attention modules,'' {\em
  IEEE Trans. Circuits Syst. Video Technol.}, vol.~32, no.~4, pp.~2315--2328,
  2021.

\bibitem{ho2020denoising}
J.~Ho, A.~Jain, and P.~Abbeel, ``Denoising diffusion probabilistic models,''
  {\em Proc. Adv. in Neural Inf. Proc. Sys. (NeurIPS)}, pp.~6840--6851, 2020.

\bibitem{dhariwal2021diffusion}
P.~Dhariwal and A.~Nichol, ``Diffusion models beat gans on image synthesis,''
  {\em Proc. Advances in Neural Information Processing Systems (NeurIPS)},
  vol.~34, pp.~8780--8794, 2021.

\bibitem{rombach2022high}
R.~Rombach, A.~Blattmann, D.~Lorenz, P.~Esser, and B.~Ommer, ``High-resolution
  image synthesis with latent diffusion models,'' in {\em IEEE/CVF Conf. Comp.
  Vision and Pattern Recog. (CVPR)}, pp.~10684--10695, 2022.

\bibitem{chen2024pixart}
J.~Chen, Y.~Wu, S.~Luo, E.~Xie, S.~Paul, P.~Luo, H.~Zhao, and Z.~Li,
  ``Pixart-$\delta$: Fast and controllable image generation with latent
  consistency models,'' {\em [Online]. Available:
  https://arxiv.org/abs/2401.05252}, 2024.

\bibitem{zhang2023adding}
L.~Zhang, A.~Rao, and M.~Agrawala, ``Adding conditional control to
  text-to-image diffusion models,'' in {\em Proc. IEEE/CVF International
  Conference on Computer Vision (ICCV)}, pp.~3836--3847, 2023.

\bibitem{lei2023text+}
E.~Lei, Y.~B. Uslu, H.~Hassani, and S.~S. Bidokhti, ``Text+ sketch: Image
  compression at ultra low rates,'' {\em [Online]. Available:
  https://arxiv.org/abs/2307.01944}, 2023.

\bibitem{qiao2024latency}
L.~Qiao, M.~Mashhadi, Z.~Gao, C.~H. Foh, P.~Xiao, and M.~Bennis,
  ``Latency-aware generative semantic communications with pre-trained diffusion
  models,'' {\em [Online]: https://arxiv.org/abs/2403.17256}, 2024.

\bibitem{wang2024fast}
Y.~Wang, W.~Yang, Z.~Xiong, Y.~Zhao, S.~Mao, T.~Q. Quek, and H.~V. Poor,
  ``Fast-gsc: Fast and adaptive semantic transmission for generative semantic
  communication,'' {\em [Online]. Available: https://arxiv.org/abs/2407.15395},
  2024.

\bibitem{wijesinghe2023diff}
A.~Wijesinghe, S.~Zhang, S.~Wanninayaka, W.~Wang, and Z.~Ding, ``Diff-go:
  Diffusion goal-oriented communications to achieve ultra-high spectrum
  efficiency,'' {\em [Online]: https://arxiv.org/abs/2312.02984}, 2023.

\bibitem{yang2024lossy}
R.~Yang and S.~Mandt, ``Lossy image compression with conditional diffusion
  models,'' {\em Proc. Advances in Neural Information Processing Systems
  (NeurIPS)}, Vancouver, Canada, 2024.

\bibitem{Niu:SPAWC:23}
X.~Niu, X.~Wang, D.~Gündüz, B.~Bai, W.~Chen, and G.~Zhou, ``A hybrid wireless
  image transmission scheme with diffusion,'' in {\em IEEE Int'l Wrks. on Sig.
  Proc. Adv. in Wireless Comms. (SPAWC)}, pp.~86--90, 2023.

\bibitem{yilmaz2023high}
S.~F. Yilmaz, X.~Niu, B.~Bai, W.~Han, L.~Deng, and D.~Gunduz, ``High perceptual
  quality wireless image delivery with denoising diffusion models,'' {\em
  [Online]. Available: https://arxiv.org/abs/2309.15889}, 2023.

\bibitem{chen2024commin}
J.~Chen, D.~You, D.~G{\"u}nd{\"u}z, and P.~L. Dragotti, ``Commin: Semantic
  image communications as an inverse problem with inn-guided diffusion
  models,'' in {\em IEEE Int'l Conf. on Acous., Speech and Sig. Proc.
  (ICASSP)}, pp.~6675--6679, Seoul, Korea, 2024.

\bibitem{yang2024diffusion}
M.~Yang, B.~Liu, B.~Wang, and H.-S. Kim, ``Diffusion-aided joint source channel
  coding for high realism wireless image transmission,'' {\em arXiv preprint
  arXiv:2404.17736}, 2024.

\bibitem{pei2024latent}
J.~Pei, F.~Cheng, P.~Wang, H.~Tabassum, and D.~Shi, ``Latent diffusion
  model-enabled real-time semantic communication considering semantic
  ambiguities and channel noises,'' {\em [Online]. Available:
  https://arxiv.org/abs/2406.06644}, 2024.

\bibitem{xie2024hybrid}
H.~Xie, Z.~Qin, Z.~Han, and K.~B. Letaief, ``Hybrid digital-analog semantic
  communications,'' {\em arXiv preprint arXiv:2405.12580}, 2024.

\bibitem{guo2024diffusion}
L.~Guo, W.~Chen, Y.~Sun, B.~Ai, N.~Pappas, and T.~Quek, ``Diffusion-driven
  semantic communication for generative models with bandwidth constraints,''
  {\em [Online], https://arxiv.org/abs/2407.18468}, 2024.

\bibitem{wang2024diffcom}
S.~Wang, J.~Dai, K.~Tan, X.~Qin, K.~Niu, and P.~Zhang, ``Diffcom: Channel
  received signal is a natural condition to guide diffusion posterior
  sampling,'' {\em [Online]. Available: https://arxiv.org/abs/2406.07390},
  2024.

\bibitem{pan2022extreme}
Z.~Pan, X.~Zhou, and H.~Tian, ``Extreme generative image compression by
  learning text embedding from diffusion models,'' {\em [Online]. Available:
  https://arxiv.org/abs/2211.07793}, 2022.

\bibitem{careil2023towards}
M.~Careil, M.~J. Muckley, J.~Verbeek, and S.~Lathuili{\`e}re, ``Towards image
  compression with perfect realism at ultra-low bitrates,'' in {\em in Proc.
  International Conference on Learning Representations (ICLR)}, 2023.

\bibitem{li2023blip}
J.~Li, D.~Li, S.~Savarese, and S.~Hoi, ``Blip-2: Bootstrapping language-image
  pre-training with frozen image encoders and large language models,'' in {\em
  Proc. International conference on machine learning(ICML)}, pp.~19730--19742,
  Honolulu, USA, 2023.

\bibitem{zhou2024muge}
C.~Zhou, Y.~Huang, M.~Pu, Q.~Guan, R.~Deng, and H.~Ling, ``Muge: Multiple
  granularity edge detection,'' in {\em IEEE/CVF Conf. on Computer Vision and
  Pattern Recog. (CVPR)}, pp.~25952--25962, 2024.

\bibitem{song2020denoising}
J.~Song, C.~Meng, and S.~Ermon, ``Denoising diffusion implicit models,'' {\em
  [Online]. Available: https://arxiv.org/abs/2010.02502}, 2020.

\bibitem{chen2023importance}
T.~Chen, ``On the importance of noise scheduling for diffusion models,'' {\em
  [Online]. Available: https://arxiv.org/abs/2301.10972}, 2023.

\bibitem{peebles2023scalable}
W.~Peebles and S.~Xie, ``Scalable diffusion models with transformers,'' in {\em
  Proc. IEEE/CVF International Conference on Computer Vision (CVPR)},
  pp.~4195--4205, 2023.

\bibitem{chen2023pixart}
J.~Chen, J.~Yu, C.~Ge, L.~Yao, E.~Xie, Y.~Wu, Z.~Wang, J.~Kwok, P.~Luo, H.~Lu,
  {\em et~al.}, ``Pixart-$\alpha$: Fast training of diffusion transformer for
  photorealistic text-to-image synthesis,'' {\em [Online]. Available:
  https://arxiv.org/abs/2310.00426}, 2023.

\bibitem{gao2023mdtv2}
S.~Gao, P.~Zhou, M.-M. Cheng, and S.~Yan, ``{MDTv2: Masked Diffusion
  Transformer is a Strong Image Synthesizer},'' {\em [Online]. Available:
  https://arxiv.org/abs/2303.14389}, 2023.

\bibitem{yang2019snr}
K.~Yang, Z.~Huang, X.~Wang, and F.~Wang, ``An snr estimation technique based on
  deep learning,'' {\em Electronics}, vol.~8, no.~10, p.~1139, 2019.

\bibitem{kirillov2023segment}
A.~Kirillov, E.~Mintun, N.~Ravi, H.~Mao, C.~Rolland, L.~Gustafson, T.~Xiao,
  S.~Whitehead, A.~C. Berg, W.-Y. Lo, {\em et~al.}, ``Segment anything,'' in
  {\em in Proc. IEEE/CVF International Conference on Computer Vision (CVPR)},
  pp.~4015--4026, 2023.

\bibitem{sun2024journeydb}
K.~Sun, J.~Pan, Y.~Ge, H.~Li, H.~Duan, X.~Wu, R.~Zhang, A.~Zhou, Z.~Qin,
  Y.~Wang, {\em et~al.}, ``Journeydb: A benchmark for generative image
  understanding,'' {\em Proc. Advances in Neural Information Processing Systems
  (NeurIPS)}, vol.~36, Vancouver, Canada, 2024.

\bibitem{changpinyo2021conceptual}
S.~Changpinyo, P.~Sharma, N.~Ding, and R.~Soricut, ``Conceptual 12m: Pushing
  web-scale image-text pre-training to recognize long-tail visual concepts,''
  in {\em IEEE/CVF Conf. on Computer Vision and Pattern Recog. (CVPR)},
  pp.~3558--3568, 2021.

\bibitem{gadre2024datacomp}
S.~Y. Gadre, G.~Ilharco, A.~Fang, J.~Hayase, G.~Smyrnis, T.~Nguyen, R.~Marten,
  M.~Wortsman, D.~Ghosh, J.~Zhang, {\em et~al.}, ``Datacomp: In search of the
  next generation of multimodal datasets,'' {\em Advances in Neural Inf. Proc.
  Systems (NeurIPS)}, vol.~36, 2024.

\bibitem{zhu2022celebv}
H.~Zhu, W.~Wu, W.~Zhu, L.~Jiang, S.~Tang, L.~Zhang, Z.~Liu, and C.~C. Loy,
  ``Celebv-hq: A large-scale video facial attributes dataset,'' in {\em
  European Conf. on Comp. Vision (ECCV)}, pp.~650--667, 2022.

\bibitem{lin2014microsoft}
T.-Y. Lin, M.~Maire, S.~Belongie, J.~Hays, P.~Perona, D.~Ramanan,
  P.~Doll{\'a}r, and L.~Zitnick, ``Microsoft coco: Common objects in context,''
  in {\em European Conf. on Comp. Vision (ECCV)}, pp.~740--755, 2014.

\bibitem{zhang2025semantics}
M.~Zhang, H.~Wu, G.~Zhu, R.~Jin, X.~Chen, and D.~G{\"u}nd{\"u}z,
  ``Semantics-guided diffusion for deep joint source-channel coding in wireless
  image transmission,'' {\em arXiv preprint arXiv:2501.01138}, 2025.

\end{thebibliography}

	\end{document}